\documentclass[fleqn,usenatbib]{mnras}

\usepackage{newtxtext,newtxmath}
\usepackage{xcolor}
\usepackage[T1]{fontenc}
\usepackage{scalerel}
\usepackage{pdflscape} 
\usepackage{graphicx}
\usepackage{amsmath}
\usepackage{indentfirst}
\usepackage{orcidlink}
\usepackage{fontawesome5}
\usepackage{array}
\usepackage{multirow}
\usepackage{booktabs}
\usepackage{threeparttable}
\usepackage{lineno}
\newcommand{\Mtwohc}{M_{\rm 200c}}

\newcommand{\Rfivehc}{R_{\rm 500c}}
\newcommand{\Mfivehc}{M_{\rm 500c}}

\newcommand{\Rtwoh}{R_{\rm 200c}}        
        
\newcommand{\mtwoh}{M_{\rm 200c}}        
        
\newcommand{\mfiveh}{M_{\rm 500c}}             
\newcommand{\LX}{L_{\rm X}}

\newcommand{\Mgas}{M_{\rm gas}}
\newcommand{\fgas}{f_{\rm gas}}
\newcommand{\fstar}{f_{\star}}
\newcommand{\fstarICL}{f_{\star, \rm ICL}}
\newcommand{\fbh}{f_{\rm BH}}
\newcommand{\fb}{f_{\rm b}}

\newcommand{\YSZ}{Y_{\rm SZ}}

\newcommand{\MstarBCG}{M_{\star,\,\rm BCG}}

\newcommand{\Mbh}{M_{\rm BH}}
\newcommand{\Mstarsat}{M_{\star,\,\rm sat}}
\newcommand{\Mstargap}{M_{\star,\,\rm 14}}
\newcommand{\MstarICL}{M_{\star,\,\rm ICL}}
\newcommand{\Mbhcent}{M_{\rm BH,\,cent}}
\newcommand{\vcircdm}{v_{\rm c,\,DM}}

\newcommand{\mdelta}{M_{\Delta}} 
\newcommand{\rdelta}{R_{\Delta}}

\newcommand{\rhocrit}{\rho_{\rm crit }}
\newcommand{\Mfid}{M_{\rm fid}}

\newcommand{\msol}{\ensuremath{\, {\rm M}_\odot}}    
\newcommand{\msun}{\ensuremath{\, {\rm M}_\odot}} 
\newcommand{\kpc}{\ensuremath{\, {\rm kpc}}}         
\newcommand{\mpc}{\ensuremath{\, {\rm Mpc}}}

\newcommand{\Nsat}{\ensuremath{N_{\rm sat}}}

\newcommand{\MstarsatThresh}{\ensuremath{M_{\rm \star,\,sat}^{\rm th}}}
\newcommand{\Mstartot}{\ensuremath{M_{\rm \star,\,tot}}}

\newcommand{\sigmaSat}{\ensuremath{\sigma_{\rm sat,\,1D}}}

\newcommand{\sigmaeight}{\ensuremath{\sigma_8}}
\newcommand{\ns}{\ensuremath{n_{\rm s}}}
\newcommand{\omegam}{\ensuremath{\Omega_{\rm m}}}
\newcommand{\omegab}{\ensuremath{\Omega_{\rm b}}}

\newcommand{\omegaLambda}{\ensuremath{\Omega_{\Lambda}}}

\newcommand{\FLAMLm}{\rm FLAMINGO-L1\_m8}
\newcommand{\FLAMLmbold}{\rm \textbf{FLAMINGO-L1\_m8}}
\newcommand{\TNGCluster}{TNG-\textsc{cluster}}

\defcitealias{Planck2014}{Planck Collab. (2014)}

\newcommand{\github}[1]{%
   \href{#1}{\faGithub}%
}

\definecolor{bleudefrance}{rgb}{0.19, 0.55, 0.91}
\definecolor{forestgreen}{rgb}{0.13, 0.55, 0.13}
\definecolor{violet}{RGB}{128, 0, 128}

\title[Mass Proxy Quality of Galaxies, Stars, BHs]{Mass Proxy Quality of Halo Properties in {\textsc{illustris}}TNG and \textsc{flamingo} Simulations: II. Galaxies, Stars, and Black Holes}

\author[Eddie Aljamal et al.]{
Eddie Aljamal$^{\orcidlink{0009-0002-2253-4583}}$, $^{1}$\thanks{E-mail: ealjamal@umich.edu}
August E. Evrard$^{\orcidlink{0000-0002-4876-956X}}$,$^{1}$
Arya Farahi$^{\orcidlink{0000-0003-0777-4618}}$,$^{2}$
Dylan Nelson$^{\orcidlink{0000-0001-8421-5890}}$,$^{3,4}$
\\
$^{1}$Department of Physics and Leinweber Institute for Theoretical Physics, University of Michigan\\
$^{2}$Department of Statistics and Data Science, The University of Texas at Austin\\
$^{3}$Universität Heidelberg, Zentrum für Astronomie, Institut für Theoretische Astrophysik, Albert-Ueberle-Str. 2, D-69120 Heidelberg, Germany\\
$^{4}$Universität Heidelberg, Interdisziplinäres Zentrum für Wissenschaftliches Rechnen, INF 205, 69120 Heidelberg, Germany\\
}

\date{}
\pubyear{2026}

\begin{document}

\label{firstpage}
\pagerange{\pageref{firstpage}--\pageref{lastpage}}
\maketitle

\begin{abstract}
We study the stellar, black hole, and satellite galaxy populations of group- and cluster-scale halos in the {\textsc{illustris}}TNG, TNG-{\textsc{cluster}}, and {\textsc{flamingo}} cosmological hydrodynamical simulations. We characterize the mass-conditioned scaling relations, covariance, conditional likelihoods, and mass proxy quality (MPQ) of integral properties within $\Rfivehc$: total stellar mass ($\Mstartot$), brightest central galaxy stellar mass ($\MstarBCG$), intracluster light mass ($\MstarICL$), total black hole mass ($\Mbh$), satellite occupation ($\Nsat$), and satellite velocity dispersion ($\sigmaSat$), at $z\in\{0,0.5,1,2\}$. While normalization differences of up to a factor of two emerge among simulations, many properties exhibit remarkably similar mass-localized slopes, intrinsic scatters, and MPQs across halo samples and redshifts. Correlations among stellar, satellite, and gas properties are generally consistent in sign and their dependence on halo mass and redshift, whereas those involving $\Mbh$ and $\Nsat$ show greater variation, particularly in TNG-Cluster. Conditional likelihoods are nearly lognormal for many properties, including $\Mstartot$, $\Nsat$, and $\sigmaSat$, while $\MstarBCG$ and $\MstarICL$ are negatively skewed. Among the properties considered here, total stellar mass is the best mass proxy, with a halo mass scatter of $7$--$14\%$, followed by $\Nsat$ with a mass scatter of $9$--$18\%$. The consistent correlation behavior of $\MstarICL$ highlights its potential for next-generation observational studies. Finally, $\sigmaSat$ exhibits highly consistent scaling, with mass-independent slope and scatter and a largely neutral correlation structure, suggesting that it is relatively insensitive to astrophysical complexity.
\end{abstract}


%%%%%%%%%%%%%%%%%%%%%%%%%%%%%%%%%%%%%%%%%%%%%%%%%%

\section{Introduction}
Galaxy groups and clusters occupy the high-mass tail of the collapsed matter distribution and provide powerful probes of both cosmology and galaxy formation physics. Their space-time abundance is sensitive to the growth of structure and the expansion history of the Universe, while their internal baryonic contents encode cumulative effects of gas cooling, star formation, hierarchical assembly, and feedback from both active galactic nuclei (AGN) and supernovae (SN) \citep{Voit2005, Kravtsov2012}. The statistical relationship between observable cluster properties and total halo mass is therefore central to both cosmological analyses and studies of the galaxy-halo connection.

Scaling relations between cluster observables and halo mass provide a fundamental connection between observations and the underlying matter distribution. Multi-wavelength observables such as the X-ray luminosity \citep{Mantz2010XrayScaling, Mahdavi2013, Mantz2016b, Sereno2020, Chiu2022, Bahar2022}, thermal Sunyaev--Zel'dovich (tSZ) signal \citep{Planck2011, Planck2014, Shirasaki2024}, brightest central galaxy (BCG) stellar mass \citep{Zhang2016, Kravtsov2018, GoldenMarx2018, GoldenMarx2019, Bradshaw2020}, and optical richness \citep{Farahi2016, Mulroy2019} are widely used as tracers of halo mass and assembly history. These observables probe distinct components of the baryon cycle and are sensitive, to varying degrees, to processes such as radiative cooling, star formation, mergers, tidal stripping, and AGN feedback. Consequently, their scaling relations, scatter, and covariance contain important information about the astrophysical processes governing cluster evolution.

The first paper of this series \citep[][hereafter Paper I]{Aljamal2025} features the mass--property relations (MPRs) of intrinsic hot gas observables in the {\textsc{illustris}}TNG and {\textsc{flamingo}} simulations, including the X-ray luminosity, gas mass, X-ray temperature, pressure, and thermal Sunyaev--Zel'dovich (tSZ) signal. While differences in normalization are often evident, the slope and scatter behaviors are more robust, with many gas scaling relations approaching self-similar slopes and minimum variance at the highest halo masses. A key consistent finding is that gas mass and total thermal energy ($\YSZ$) constitute the best mass proxies across group and cluster scales, with a total mass precision approaching $4\%$ for the most massive halos.

In this work, we shift focus to the stellar and galaxy components of massive halos, specifically the total stellar mass, BCG stellar mass, intracluster light (ICL) mass, number of satellite galaxies, satellite velocity dispersion, and total black hole (BH) mass. The scaling relations of these properties and covariance structure with the gas properties encode important information about halo assembly histories and the baryonic processes that shape galaxy clusters. For example, the total stellar mass reflects the integrated efficiency of star formation and baryon conversion; the abundance and kinematics of satellite galaxies—closely related to the halo occupation distribution framework \citep{Hearin2016, Zehavi2018, Wechsler2018}—probe hierarchical merging and dynamical evolution; while the diffuse ICL component is strongly influenced by mergers and tidal stripping \citep{Murante2007, Contini2014, Brown2024, Contreras-Santos2025}.

The growing statistical power of current and upcoming multiwavelength surveys further motivates a detailed understanding of these scaling relations and their covariance structure. Optical and near-infrared surveys such as LSST \citep{IvezicLSST2019} and Euclid \citep{BlanchardEuclid2020} will identify and characterize enormous samples of galaxy clusters across a broad redshift range using galaxy richness, weak lensing, and stellar light measurements. At the same time, next-generation X-ray and millimeter-wave surveys including eROSITA \citep{Predehl2021eROSITA}, South Pole Telescope \citep[SPT;][]{Carlstrom2011SPT}, Simons Observatory \citep{Galitzki2018Simons}, CMB-S4 \citep{Abazajian2019CMBS4}, and CMB-HD \citep{Sehgal2019CMBHD} will provide complementary measurements of the hot intracluster medium through X-ray emission and the tSZ effect. Together, these surveys will enable joint multiwavelength analyses that combine stellar, gas, weak lensing, and dynamical observables to improve cluster mass calibration and constrain the covariance between different tracers of halo mass. The full scientific potential of these datasets would be unleashed if consistent (method-independent) theoretical predictions for cluster scaling relations could be demonstrated across a wide range of masses and redshifts.

In the last decade, cosmological hydrodynamical and magnetohydrodynamical simulations have advanced in numerical methodology, dynamic range, and the treatment of the nonlinear baryonic processes that govern galaxy and cluster formation \citep{McCarthy2017, Cui2018, Pillepich2018TNGmethods, Schaye2023Flamingo}. Modern simulation suites now simultaneously achieve large cosmological volumes and sufficiently high resolution to statistically sample the rare high-mass end of the halo population while moderately resolving the stellar, gaseous, and BH components of constituent galaxies. These advances now enable detailed characterization of how the normalization, slope, scatter, and covariance of cluster scaling relations vary with halo mass and redshift \citep{Farahi2016, Barnes2017a, Truong2018, Pop2022, Braspenning2024, Braspenning2026}, while also allowing systematic comparisons between different numerical methods and baryonic feedback models \citep{Anbajagane2020stellar, Rasia2025fractions, Aljamal2025}. Such comparisons are essential for identifying which statistical properties of massive halos are robustly predicted and which remain sensitive to specific choices for implementing baryonic physics.

This work is organized as follows. In \S\ref{sec:methods-sims}, we describe the two simulation suites used in this work, emphasizing the similarities and differences in their treatments of baryonic physics, and define the galaxy, stellar, and BH properties analyzed throughout this study. We briefly review the Kernel Localized Linear Regression \citep[KLLR][]{Farahi2022KLLR} framework employed to extract continuous, mass-conditioned scaling relation parameters from discrete halo populations. In \S\ref{sec:star-bh-MPRs} and \S\ref{sec:mpr-Nsat-vel-disp}, we present the normalization, slope, and scatter of the MPRs for the stellar, BH, and galaxy properties. Section~\ref{sec:correlations} examines their mass-conditioned correlation structure, including correlations between optical\footnote{We use the term ``optical'' as a generic reference to the stellar, galactic, and BH properties examined here, many of which are observable at optical/IR wavelengths.} and hot gas properties. We additionally investigate the shapes of the residual likelihood distributions and assess their consistency with the commonly assumed log-normal form in \S\ref{sec:kernel-shapes}. In \S\ref{sec:MPQ}, we study the implications of these statistics for the mass proxy quality (MPQ) of individual optical observables. We defer the discussion of the combined MPQs of gas and optical properties to a subsequent letter (Aljamal \textit{et al.}, in prep., Paper III). Finally, we summarize our results and discuss their broader implications in \S\ref{sec:conclusion}.

\begin{table*}
\centering
    \caption{Cosmological and simulation parameters for the three simulations; the dimensionless Hubble constant $h \equiv H_0/100 \text{ km s}^{-1} \text{Mpc}^{-1}$, the normalized density of matter $\omegam$, the cosmic mean baryon fraction $\fb = \omegab/\omegam$, the normalization of the power spectrum $\sigmaeight$, the power-law index of the primordial matter power spectrum $\ns$, the box side-length $L$, the number of dark matter particles $N_{\rm DM}$, the dark matter particle mass $m_{\rm DM}$, the baryonic particle mass $m_{\rm baryon}$, and the gravitational softening length at $z = 0$ $\epsilon^{z = 0}$.}
    \label{tab:sim-constants}
\begin{threeparttable}
\begin{tabular}{ccccccccccc}
    \toprule & \\[-9pt]
    Simulation & $h$ & $\omegam$ & $\fb$ & $\sigmaeight$ & $\ns$ & $L$ [\mpc/h] & $N_{\rm DM}$ & $m_{\rm DM}$ [$\msol$] &  
        $m_{\rm baryon}$ [$\msol$] & $\epsilon^{z = 0}$ [\kpc]\\ 
    \midrule
        \text{TNG300} & $0.6774$ & $0.3089$ & $0.157$ & $0.8159$ & $0.9667$ & $205$ & $2500^3$ & $5.9 \times 10^7$ & $1.1 \times 10^7$ & $1.48$\\
        \text{TNG-Cluster}\tnote{*} & $0.6774$ & $0.3089$ & $0.157$ & $0.8159$ & $0.9667$ & $680$ & $8192^3$\tnote{$\dagger$} & $6.1 \times 10^7$ & $1.2 \times 10^7$ & $1.48$\\
        \text{FLAM-L1\_m8} & $0.681$ & $0.306$ & $0.159$ & $0.807$ & $0.967$ & $1000$ & $3600^3$ & $7.06 \times 10^8$ & $1.34 \times 10^8$ & $2.85$\\
    \bottomrule
\end{tabular}
\begin{tablenotes}
    \item[*] Zoom simulation
    \item[$\dagger$] Effective resolution
\end{tablenotes}
\end{threeparttable}
\end{table*}

%%%%%%%%%%%%%%%%%%%%%%%%%%%%%%%%%%%%%%%%%%%%%%%%%%%%%%%%%%%%

\begin{figure*}
    \centering
    \includegraphics[width = 0.75\textwidth]{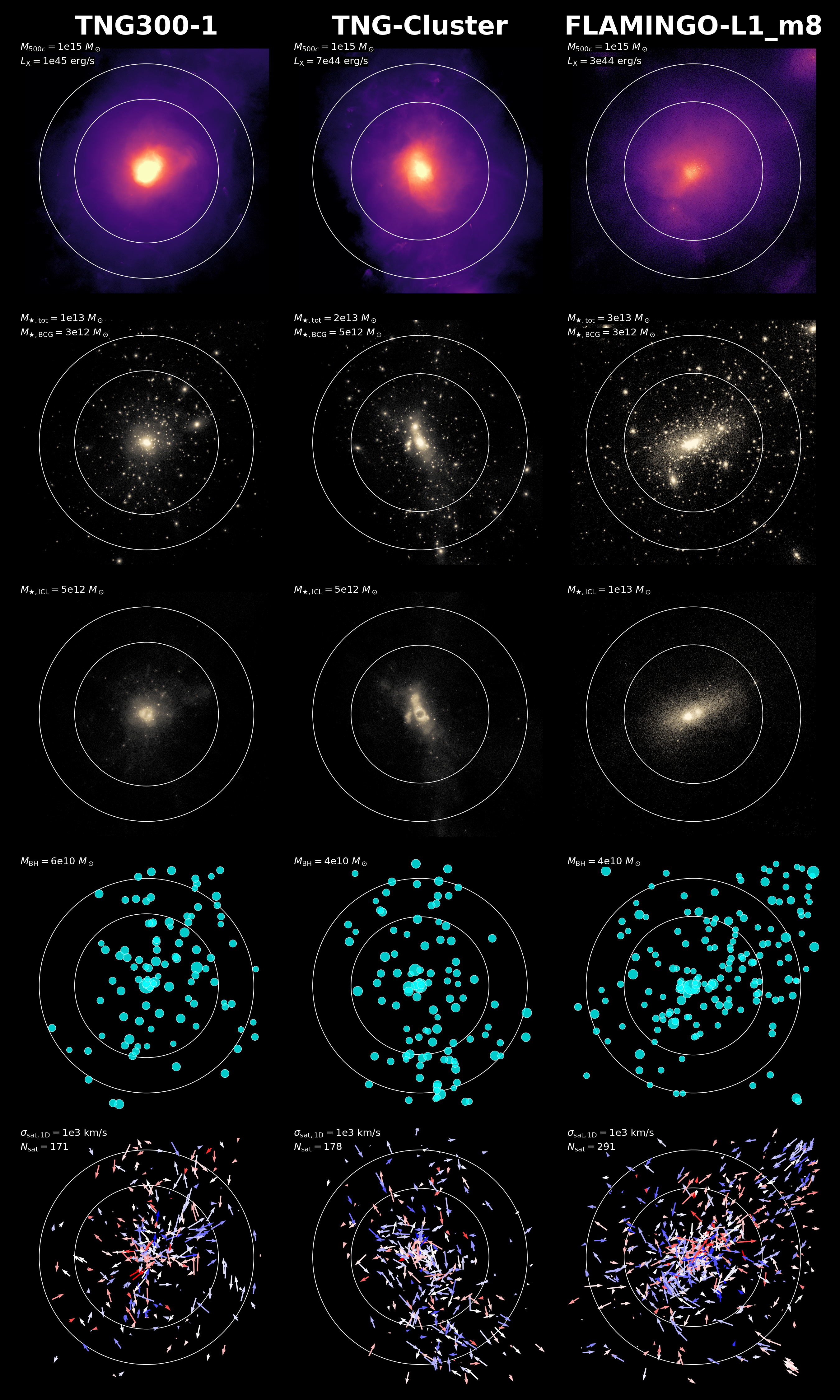}
    \caption{Example massive halos with $\Mfivehc \simeq 10^{15}\msun$ in TNG300-1 (left), TNG-Cluster (middle), and {\FLAMLm} (right) samples.  Images are projected along the $z$-axis of the simulated volume, and each row uses a common scaling across columns. The rows show, from top to bottom, soft-band X-ray surface brightness, stellar distribution, intracluster light (ICL), BHs with masses above $10^7\msun$, and satellite galaxy velocities. The values in the upper left of each panel give the corresponding property values for the halo shown, including the total soft-band luminosity $\LX$ and the other properties summarized in Table~\ref{tab:prop-definitions}. Black hole marker radii are proportional to the logarithm of their masses. Red and blue arrows indicate velocity components out of and into the page, respectively. The two circles indicate $\Rfivehc$ and $\Rtwoh$.
}
    % \label{fig:cool-plot}
\end{figure*}

%%%%%%%%%%%%%%%%%%%%%%%%%%%%%%%%%%%%%%%%%%%%%%%%%%%%%%%%%%%%

\section{Simulations and Halo Sample Properties}\label{sec:methods-sims}

Here we first describe the simulations used (\S\ref{sec:simulations}), specifically highlighting differences in treatments of compact objects and feedback, summarized in Table~\ref{tab:sim-comparison}.  In \S\ref{sec:halo-properties} we define the stellar, BH, and galactic properties employed in our population analysis (Table~\ref{tab:prop-definitions}).

\subsection{Simulations}\label{sec:simulations}

We analyze halos drawn from the hydrodynamical simulations TNG300-1, {\TNGCluster}, and {\FLAMLm}. TNG300 and TNG-Cluster are part of the {\textsc{illustris}}TNG project \citep{Nelson2018, Marinacci2018, Pillepich2018, Springel2018, Naiman2018, Nelson2024}, while {\FLAMLm} is part of the {\textsc{flamingo}} project \citep{Schaye2023Flamingo, Kugel2023, Helly2026FLAMrelease}. The simulations assume a flat $\Lambda$CDM cosmology with parameters listed in Table \ref{tab:sim-constants}. The simulations differ in their hydrodynamic solvers, AGN feedback implementations, star formation prescriptions, calibration strategies, and subhalo-finding algorithms. Since these differences directly affect the stellar, BH, and satellite properties studied in this work, we summarize the most relevant methodological distinctions here, listed in Table \ref{tab:sim-comparison}.

TNG300 and TNG-Cluster are evolved using the moving-mesh magnetohydrodynamics code AREPO \citep{Springel2010Arepo, Weinberger2020Arepo}, whereas {\FLAMLm} employs the SPHENIX smoothed-particle hydrodynamics scheme implemented in SWIFT \citep{Borrow2022, Schaller2024}. TNG300 is a periodic-volume simulation with side length $302.6\,{\rm Mpc}$, while TNG-Cluster is a suite of zoom-in simulations taken from a $(1\,{\rm Gpc})^3$ box, targeting massive cluster halos at equal mass resolution \citep{Nelson2024}, providing $204$ clusters with $10^{14.5} < \Mfivehc/\msol < 10^{15}$ and $31$ clusters with $\Mfivehc > 10^{15} \msol$ at $z = 0$. The TNG-Cluster sample is volume complete for $\Mtwohc \ge 10^{15}\msol$ and randomly samples lower mass halos such that the number is uniform in log-mass in the combined TNG300-1 and TNG-Cluster samples. 

{\FLAMLm} evolves structure in a $(1\,{\rm Gpc})^3$ cosmological volume optimized for large statistical cluster  samples. Initial conditions are created using \texttt{monofonic} \citep{Hahn2021, Elbers2022} that include neutrinos. Though {\FLAMLm} is run with purely thermal AGN feedback, the lower resolution FLAMINGO-L1\_m9 provides many AGN and supernova feedback variations (including only a kinetic feedback mode) calibrated to variations in the observed cluster gas fractions and stellar mass function, respectively \citep{Schaye2023Flamingo, Helly2026FLAMrelease}.
Paper IV in this series will explore MPR and MPQ sensitivity to these variations.

\subsubsection{AGN feedback}\label{sec:AGNfeedback}

The simulation methods differ significantly in their approaches to implementing BH growth and AGN feedback. As we will see, these approaches yield significant differences in population-level  normalizations for stars and BHs within groups and clusters within $\Rfivehc$.  Consistent with our findings of Paper I, higher order scaling relation features, such as slope and intrinsic scatter, are less sensitive to specific AGN treatment, particularly at the highest halo masses. 

In {\textsc{illustris}}TNG, the supermassive black hole model is described in detail in \citet{Weinberger2017TNGmethods} (see also the second TNG methods paper \citet{Pillepich2018TNGmethods}). Black holes are seeded with an initial mass of $6.2 \times 10^6\msun$ in halos above a threshold mass of $7.3 \times 10^{10}\msun$ and accrete gas at the Bondi rate limited by the Eddington limit,
\begin{equation}
    \dot{M}_{\rm BH}
    =
    \min\left(
        \dot{M}_{\rm Bondi},
        \dot{M}_{\rm Edd}
    \right).
\end{equation}
The Bondi and Eddington rates \citep{HoyleLyttleton1939, BondiHoyle1944, Bondi1952} are given by:
\begin{align}
    \dot{M}_{\rm Edd} &= \frac{4\pi G M_{\rm BH}m_{\rm p}}{\epsilon_{\rm r}\sigma_{\rm T}c}\\
    \dot{M}_{\rm Bondi} &= \frac{4\pi G^2 M_{\rm BH}^2\rho}{c_{\rm s}^3}
\end{align}
where $G$, $c_{\rm s}$, $M_{\rm BH}$, $\rho$ are the universal gravitational constant, the speed of sound, the black hole's mass and the gas density, respectively.

The AGN feedback mode is determined by the Eddington ratio,
\begin{equation}
    f_{\rm Edd}
    \equiv
    \frac{\dot{M}_{\rm BH}}{\dot{M}_{\rm Edd}}.
\end{equation}
and a transition threshold that is dependent on the BH mass
\begin{equation}
    \chi
    =
    \min\left[
        0.002
        \left(
            \frac{M_{\rm BH}}{10^8\msun}
        \right)^2,
        0.1
    \right].
\end{equation}
At high accretion rates $f_{\rm Edd} \ge \chi$, feedback proceeds through thermal mode—or quasar mode—for which thermal energy is continuously injected into the surrounding gas \citep{ShakuraSunyaev1976LowAccretion}. At low accretion rates, the model transitions to a kinetic mode in which momentum is stochastically injected into surrounding gas cells.

Integrated over their evolutionary histories, the total feedback energy is dominated by the thermal mode in lower-mass systems, where kinetic injection events are relatively infrequent. The kinetic contribution becomes increasingly important for systems hosting black holes above ${\sim} 10^{8.3}\msun$ and galaxies with stellar masses above ${\sim}10^{10.5}\msun$, such that the cumulative kinetic energy can exceed that injected through the thermal mode \citep{Zinger2020}. At high halo masses, however, the instantaneous feedback energy injection rate evolves from being dominated by the thermal mode at early times to the kinetic mode at late times \citep{Weinberger2018SMBH}.

In contrast, the fiducial \textsc{flamingo} simulations employ the single-mode thermal AGN feedback model of \citet{BoothSchaye2009}. Black holes with an initial subgrid mass of $10^{5}\msun$ are seeded in halos with masses greater than $2.8 \times 10^{11}\msun$ at the location of the densest gas particle. To prevent excursions from the bottom of the potential well, BHs are subsequently repositioned to the location of the gas particle within the SPH kernel support radius and three softening lengths that has the lowest gravitational potential energy, following the prescriptions of \citet{Springel2005Nature} and \citet{BoothSchaye2009} (see also \citet{Bahe2022BHreposition} for further details on BH repositioning and its effects).

\begin{table*}
    \centering
    \caption{Summary of the simulation methodologies most relevant for this work.}
    \label{tab:sim-comparison}
    \renewcommand{\arraystretch}{1.2}
    \begin{tabular}{p{3.2cm} p{5.8cm} p{5.8cm}}
        \toprule
        Methodology & {\textsc{illustris}}TNG (TNG300/TNG-Cluster) & {\textsc{flamingo}} ({\FLAMLm}) \\
        \midrule
        
        Hydrodynamics
        & AREPO moving-mesh magnetohydrodynamics \citep{Springel2010Arepo}.
        & SWIFT + SPHENIX smoothed-particle hydrodynamics \citep{Borrow2022, Schaller2024}. \\ 
        
        AGN feedback
        & Two-mode AGN feedback: thermal and kinetic \citep{Weinberger2017TNGmethods, Pillepich2018TNGmethods}.
        & Single-mode thermal AGN feedback \citep{BoothSchaye2009, Schaye2023Flamingo}. \\
        
        BH accretion
        & Bondi accretion limited by the Eddington rate. Seed subgrid mass $6.2 \times 10^6 \msun$.
        & Modified Bondi-Hoyle accretion with density-dependent boost factor. Seed subgrid mass $10^{5}\msun$. \\
        
        BH repositioning
        & No explicit repositioning.
        & Repositioned toward local potential minima \citep{Springel2005Nature, BoothSchaye2009, Bahe2022BHreposition}. \\
        
        Star formation
        & Multiphase ISM model and galactic winds driven by SN \citep{Pillepich2018TNGmethods}
        & Pressure-dependent star formation with stochastic kinetic SN feedback \citep{SchayeDallaVecchia2008, Chaikin2023} \\
        
        Calibration strategy
        & Calibrated primarily to observed GSMF, stellar-to-halo mass relation, and gas mass in virial radius.
        & Calibrated to the observed low-redshift GSMF and cluster gas fractions.\\
        
        Subhalo finder
        & \textsc{SUBFIND} \citep{Springel2001Subfind, Dolag2009}.
        & \textsc{HBT-HERONS} \citep{ForouharHBTHERONS}. \\
        
        Subhalo methodology
        & Instantaneous density-based structure identification.
        & History-based particle membership tracking. \\
        
        \bottomrule
    \end{tabular}
\end{table*}

Black holes accrete using a modified Bondi-Hoyle prescription,
\begin{equation}
    \dot{M}_{\rm mBondi}
    =
    \alpha
    \frac{
        4\pi G^2 M_{\rm BH}^2 \rho
    }
    {
        (c_s^2 + v^2)^{3/2}
    },
\end{equation}
where $v$ is the relative velocity between the black hole and surrounding gas. The boost factor $\alpha$ compensates for unresolved multiphase gas structure and depends on the gas density \citep{Schaye2023Flamingo}.

Feedback energy accumulates in a reservoir and is released in discrete heating events that raise the temperature of the nearest gas particle by a fixed amount,
\begin{equation}
    \Delta T_{\rm AGN}
    \simeq
    10^{8.07}\,{\rm K}.
\end{equation}
Unlike TNG, the fiducial \textsc{flamingo} model does not include a kinetic low-accretion feedback channel, although alternative kinetic-feedback runs exist within the broader simulation suite \citep{Husko2022}. 

\subsubsection{Star formation and calibration}\label{sec:starFormCal}

The simulations also differ in their treatment of star formation and in the observables used to calibrate their subgrid models. 

In {\textsc{illustris}}TNG, star formation occurs in cold, dense gas following the Kennicutt--Schmidt relation and is regulated by a subgrid multiphase interstellar medium (ISM) model \citep{SpringelHernquist2003}. Stellar feedback is implemented through kinetic galactic winds driven by supernovae, with wind velocities and energetics that scale with the local dark matter velocity dispersion \citep{Pillepich2018TNGmethods}. Stellar evolution returns mass and metals to the ISM while tracking the enrichment of nine chemical elements. The TNG model parameters were calibrated to reproduce several galaxy-scale observables, including the galaxy stellar mass function (GSMF), stellar-to-halo mass relation, galaxy sizes, and gas fractions within halos \citep{Pillepich2018TNGmethods, Nelson2019release}. The resulting galaxy population also reproduces the observed spatial clustering of galaxies and its dependence on stellar mass and color \citep{Springel2018}.

Though TNG300-1 and TNG-Cluster adopt the same physical model and mass resolution, TNG-Cluster includes one modification to the star formation prescription that directly affects the stellar mass content and, indirectly, several other halo properties. The change was originally introduced in TNG50-1 following the inclusion of magnetic fields, where very dense star-forming gas required timesteps as short as ${\sim}10$ yr to satisfy the numerical integration criteria. To avoid prohibitively small timesteps, the dependence of the star formation timescale on gas density was modified \citep{Nelson2019}.

In the \citet{SpringelHernquist2003} star formation model, the star formation timescale is
\begin{equation}\label{eq:Springel-SFR-SFtime}
t_{\star}(\rho) = t_0^{\star} \left(\frac{\rho}{\rho_{\rm th}}\right)^{-\alpha_{\star}},
\end{equation}
where $t_0^{\star}$ sets the gas consumption timescale, $\rho_{\rm th}$ is the threshold density for star formation, and $\rho$ is the local gas density. In TNG100 and TNG300, the exponent is fixed to $\alpha_{\star} = 1/2$. For TNG50-1, however, a steeper dependence with $\alpha_{\star} = 1$ was adopted for gas above the runaway density threshold \citep{Nelson2019}. Since the star formation rate scales as $\rho^{\alpha_{\star}}$, this modification accelerates star formation in the densest gas, allowing the gas to be consumed before the timestep becomes prohibitively small. The same modification was subsequently adopted in TNG-Cluster, introducing a difference in the star formation model relative to TNG300. As we show below, this leads to non-negligible differences in stellar and BH properties. 

\textsc{flamingo}, in contrast, adopts a pressure-dependent star formation prescription, which reproduces the observed Kennicutt-Schmidt relation \citep{SchayeDallaVecchia2008}. Feedback from star formation is implemented through isotropic, stochastic kinetic energy injection \citep{Chaikin2022, Chaikin2023}, while stellar mass loss and chemical enrichment are tracked using the same nine elements as TNG \citep{Wiersma2009}. The stellar and AGN feedback parameters are jointly calibrated to reproduce both galaxy- and cluster-scale observables at low redshift, including the GSMF and cluster gas fractions \citep{Kugel2023}. The simulations have also been shown to reproduce the observed cosmic star formation history \citep{Schaye2023Flamingo}.

These differences in star formation, feedback, and calibration strategy can significantly influence the efficiency of star formation and quenching in massive halos, as well as the partitioning of stellar mass between the BCG, satellite population, and ICL (Aljamal \textit{et al.}, in prep., Paper IV).

\begin{table*}
    \centering
    \caption{Galaxy, stellar, and BH properties measured within a 3D aperture of $\Rfivehc$. Satellite galaxy properties impose a bound stellar mass threshold of $\MstarsatThresh = 10^{9.5}\msun$.  MPRs of the first six properties are described in Sections \ref{sec:star-bh-MPRs} and \ref{sec:mpr-Nsat-vel-disp} while the last two are used in \S\ref{sec:correlations}.  }
    \label{tab:prop-definitions}
    \begin{tabular}{p{2cm} p{11cm}}
         \toprule
         Property & Definition\\
         \midrule
         $\Mstartot$ & Total stellar mass (sum of stellar particle masses)\\
         $\MstarBCG$ & Stellar mass bound to the central subhalo within $100$ physical kpc of its center\\
         $\Mbh$ & Total BH mass (sum of BH subgrid masses)\\
         $\MstarICL$ & Stellar mass not bound to satellites and located beyond $100$ physical kpc of the BCG center\\
         $\Nsat$ & Number of satellite galaxies above $\MstarsatThresh$ \\
         $\sigmaSat$ & Gapper velocity dispersion for halos with $\Nsat \geq 10$ \\
         $\Mstarsat$ & Total stellar mass bound to any satellite subhalo (no threshold imposed)\\
         $\Mstargap$ & Stellar mass gap between the most and fourth most massive galaxies for halos with $\Nsat \geq 10$ \\
         \bottomrule
    \end{tabular}
\end{table*}

\subsubsection{Subhalo identification}\label{sec:subhaloIDmethods}

The simulations also differ in their subhalo-finding methodologies. In TNG, halos are first identified using a friends-of-friends (FoF) algorithm with linking length $b=0.2$, after which substructure is identified using \textsc{SUBFIND} \citep{Springel2001Subfind, Springel2005Nature, Dolag2009}. \textsc{SUBFIND} identifies locally overdense regions within halos and iteratively removes gravitationally unbound particles to construct self-bound subhalos. The method therefore relies primarily on the instantaneous density and velocity fields at a given simulation snapshot.

\textsc{flamingo} instead uses \textsc{HBT-HERONS} \citep{ForouharHBTHERONS}, a history-based extension of HBT+ \citep{Han2012HBT, Han2018}. The algorithm identifies subhalos across successive snapshots using the particle membership of their progenitors, allowing subhalo identities to remain temporally continuous through mergers and tidal stripping. Candidate substructures are identified, and their particle memberships refined, through iterative unbinding, while satellite particles are inherited from their progenitors rather than reconstructed independently from the instantaneous density field at each snapshot. 

These methodological differences are particularly relevant for satellite galaxy populations, which are summarized in \citet{ForouharHBTHERONS}, and ICL measurements, since they directly affect how stellar particles are partitioned between central galaxies, satellites, diffuse stellar components, and unresolved substructure.

\subsection{Halo Properties}\label{sec:halo-properties}

Consistent with \citet{Aljamal2025}, we compute intrinsic properties using all particles and cells within $\Rfivehc$ of the halo center, adopting the corresponding SOAP (Spherical Overdensity and Aperture Processor) properties for FLAMINGO when available \citep{McGibbonSOAP2025}.

For \textbf{true halo mass}, we adopt the standard spherical overdensity mass definition, $M_{\Delta}$, for which the total mass enclosed within a sphere of radius $R_{\Delta}$ has a mean density equal to $\Delta$ times the critical density, $\rhocrit(z) = 3H(z)^2/(8\pi G)$. The halo mass and radius are therefore linked by the definition
\begin{equation}
    \mdelta = \frac{4\pi}{3}\Delta \rhocrit (z) \rdelta^3. \label{eq:SO-true-mass}
\end{equation}
Throughout this work we adopt $\Delta = 500$, a convention commonly used in X-ray and SZ analyses. We have verified that our results are qualitatively and quantitatively similar when using $\Delta = 200$.

We study intrinsic halo properties listed in Table~\ref{tab:prop-definitions}: the total stellar mass $\Mstartot$, bright central galaxy (BCG) stellar mass $\MstarBCG$, total BH mass $\Mbh$, intracluster light (ICL) mass $\MstarICL$, number of satellite galaxies $\Nsat$, and one-dimensional satellite velocity dispersion $\sigmaSat$. We also consider two additional properties---the total satellite stellar mass $\Mstarsat$ and the stellar mass gap $\Mstargap$---as well as the mean total baryon fraction, $\fb$, within $\Rfivehc$.  

The centers (positions) of halos and subhalos are defined as the locations of their most bound particles, corresponding to the minima of their gravitational potentials, in all three simulations. We use the halo center when computing all quantities within $\Rfivehc$. In \textsc{flamingo}, the central subhalo position coincides with the halo center by construction, whereas in {\textsc{Illustris}}TNG the halo and central subhalo centers are determined independently. For the halo samples considered in this work, however, the central subhalo and halo centers coincide at all redshifts.

\textbf{Total and BCG stellar masses.} The total stellar mass, $\Mstartot$, is defined as the sum of the masses of all stellar particles within $\Rfivehc$. We define the BCG stellar mass, $\MstarBCG$, as the total stellar mass bound to the central subhalo—identified by the subhalo finder—within a fixed aperture of $100$ physical kpc of the center. This aperture lies within $\Rfivehc$ for all halos in our samples. 

\textbf{Black hole mass.} The total black hole mass, $\Mbh$, is defined as the sum of the subgrid masses of all black hole particles within $\Rfivehc$. 

\textbf{Intracluster light mass.} The intracluster light stellar mass, $\MstarICL$, is defined as the total stellar mass within $\Rfivehc$ that is either unbound to any subhalo or gravitationally bound to the central subhalo while located more than $100\kpc$ from the halo center. Together, $\MstarBCG$, $\Mstarsat$, and $\MstarICL$ comprise the total stellar mass budget of the halo. This definition is motivated by \citet{Zhang2019DESyr1} which suggests that $100\kpc$ is the aperture dividing the stellar content of the BCG and the ICL in massive systems.

\textbf{Number of satellite galaxies.} 
We impose a lower limit on the bound stellar mass of satellite galaxies of $\MstarsatThresh = 10^{9.5}\msun$, corresponding to a minimum number of stellar particles  of roughly $360$ for TNG and $20$ for {\FLAMLm}. The number of satellite galaxies, $\Nsat$, is defined as the number of satellite galaxies with centers inside $\Rfivehc$ and stellar masses above $\MstarsatThresh$.

\textbf{One-dimensional velocity dispersion.}  For a given halo, we use the $x$-component of all satellite galaxy velocities to estimate the one-dimensional velocity dispersion using the gapper estimator, an outlier-resistant scale estimator based on ordered velocity gaps \citep{Beers1990}. For each halo, we adopt the gapper velocity dispersion measured from the original satellite sample and apply the gapper estimator to $1000$ bootstrap realizations of the satellite population to estimate its uncertainty.  We use a single component, rather than a quadrature sum of all components, to better align with the observationally accessible line-of-sight velocities. 

\begin{table*}
    \centering
      \caption{
    Halo mass ranges and sample sizes used for the KLLR analyses at each redshift. ``Stellar/BH'' corresponds to properties defined for the full halo sample, including the total stellar mass, BCG stellar mass, ICL mass, BH mass, and satellite occupation. ``Satellite'' corresponds to properties involving computations on satellite galaxies, including the velocity dispersion and stellar mass gap, and is therefore restricted to higher halo masses. Mass ranges are given in $\log_{10}(\Mfivehc/\msun)$. At $z = 0$, the mass ranges for TNG300-1 and TNG-Cluster, together with the number of halos remaining after the $\Nsat \geq 10$ cut, are insufficient for reliable KLLR analyses, so these cases are excluded.
    }
    \small
    \setlength{\tabcolsep}{2.5pt}
    \begin{tabular}{c cccc cccc cccc}
        \toprule
        & \multicolumn{4}{c}{\textbf{TNG300-1}}
        & \multicolumn{4}{c}{\textbf{TNG-Cluster}}
        & \multicolumn{4}{c}{\textbf{\FLAMLmbold}} \\
        \cmidrule(lr){2-5} \cmidrule(lr){6-9} \cmidrule(lr){10-13}
        \textbf{$z$}
        & \multicolumn{2}{c}{\textbf{Stellar/BH}}
        & \multicolumn{2}{c}{\textbf{Satellite}}
        & \multicolumn{2}{c}{\textbf{Stellar/BH}}
        & \multicolumn{2}{c}{\textbf{Satellite}}
        & \multicolumn{2}{c}{\textbf{Stellar/BH}}
        & \multicolumn{2}{c}{\textbf{Satellite}} \\
        & Range & $N$
        & Range & $N$
        & Range & $N$
        & Range & $N$
        & Range & $N$
        & Range & $N$ \\
        \midrule
        $\mathbf{2}$ 
        & $[13.00, 13.57]$ & $279$
        & --- & ---
        & $[13.00, 13.94]$ & $300$
        & --- & ---
        & $[13.00, 13.97]$ & $11\,067$
        & $[13.44, 13.97]$ & $938$ \\

        $\mathbf{1}$ 
        & $[13.00, 14.03]$ & $1\,264$
        & $[13.67, 14.03]$ & $79$
        & $[13.43, 14.45]$ & $311$
        & $[13.61, 14.45]$ & $258$
        & $[13.00, 14.50]$ & $47\,372$
        & $[13.47, 14.50]$ & $8\,059$ \\

        $\mathbf{0.5}$ 
        & $[13.00, 14.26]$ & $1\,993$
        & $[13.67, 14.26]$ & $225$
        & $[13.78, 14.81]$ & $312$
        & $[13.78, 14.81]$ & $307$
        & $[13.00, 14.78]$ & $74\,280$
        & $[13.52, 14.78]$ & $13\,845$ \\

        $\mathbf{0}$ 
        & $[13.00, 14.45]$ & $2\,528$
        & $[13.72, 14.45]$ & $347$
        & $[14.22, 15.04]$ & $312$
        & $[14.22, 15.04]$ & $312$
        & $[13.00, 14.98]$ & $93\,795$
        & $[13.61, 14.98]$ & $17\,046$ \\
        \bottomrule
    \end{tabular}
    \label{tab:halo-counts}
\end{table*}

We restrict analyses involving $\sigmaSat$ and $\Mstargap$ to halos satisfying $\Nsat \geq 10$. We set the minimum mass used in the KLLR analysis to the lowest halo mass above which at least $90\%$ of halos in the full sample are retained after applying the $\Nsat \geq 10$ cut. Rather than requiring $100\%$ completeness, this definition prevents a small number of massive halos with unusually few satellite galaxies from imposing an unnecessarily conservative lower mass limit.

\textbf{Satellite stellar mass.} The total satellite stellar mass, $\Mstarsat$, is defined as the total mass of stellar particles within $\Rfivehc$ that are bound to satellite galaxies, irrespective of the stellar mass of the host satellite. Consequently, satellite galaxies with bound stellar masses below the threshold $\MstarsatThresh$ still contribute to this quantity.

\textbf{Stellar mass gap.} The stellar mass gap, $\Mstargap$, is the difference in bound stellar mass between the most massive\footnote{At $z=0$, the central galaxy is the most massive galaxy for $99\%$ of halos with $\Mfivehc \geq 10^{13}\msun$.} (in bound stellar mass) and fourth-most massive galaxy. This quantity is motivated by the observational magnitude gap, which has been shown to correlate with halo formation history \citep{Jones2003, DOnghia2005, Dariush2010, Deason2013, GoldenMarx2018, Vitorelli2018, Farahi2020}, as well as with the stellar mass of the BCG \citep{GoldenMarx2018, Farahi2020, GoldenMarx2022} and ICL \citep{GoldenMarx2025, MontenegroTaborda2025}. Inclusion of this quantity in modeling of the BCG stellar mass-halo mass relation has also been shown to reduce the intrinsic scatter \citep{GoldenMarx2018, GoldenMarx2019}.

\subsection{Kernel Localized Linear Regression (KLLR)}\label{sec:kllr}

As in Paper I, we use the Kernel Localized Linear Regression (KLLR; \citealt{Farahi2022KLLR}) method, a flexible framework for characterizing scaling relations that are scale-dependent.  We refer the reader to that paper for more detail, and include here only a brief description of equations that define our notation.  
We use $\mathbf{s}=\ln(\mathbf{S}/S_{\rm fid})$ and $\mu=\ln(\Mfivehc/\Mfid)$, where $S_{\rm fid}$ and $\Mfid$ are fiducial scales. The local mean scaling relation of a chosen property $a$ is
\begin{equation}\label{eq:KLLR-scaling}
    \langle s_a \ \vert \  \mu, z \rangle = \pi_a(\mu, z) + \alpha_a(\mu, z)\mu,
\end{equation}
where $\pi_a(\mu, z)$ and $\alpha_a(\mu, z)$ are the mass- and redshift-dependent normalization and slope, respectively. At fixed redshift, the normalization and slope of property $a$ centered on mass scale $\mu_c$ are estimated by minimizing the weighted residual sum of squares, 
\begin{equation}
    \epsilon_a^2(\mu_c) = \sum_{i = 1}^n w^2(\mu_i-\mu_c)[s_{a, i} - \pi_a(\mu_c) - \alpha_a(\mu_c)(\mu_i-\mu_c)]^2,
\end{equation}
where the index $i$ runs over all halos and the weights are given by a Gaussian kernel centered on $\mu_c$,
\begin{equation}
    w(\mu_i - \mu_c) = \frac{1}{\sqrt{2\pi}\sigma_{\text{KLLR}}} \exp\bigg[-\frac{(\mu_i-\mu_c)^2}{2\sigma^2_{{\rm KLLR}}} \bigg],
\end{equation}
Here, $\sigma_{\rm KLLR}$ is the kernel width, chosen to be $0.2\ln(10)$ in this work. 

\begin{table}
    \centering
    \caption{
    Mean baryon, gas, stellar, ICL, and BH mass fractions within $\Rfivehc$ at $\Mfivehc = 10^{14.5}\msun$ for TNG300-1, TNG-Cluster, and {\FLAMLm} at $z=0$. The fractions are estimated from the KLLR scaling relations evaluated at this halo mass.
    }
    \label{tab:mean-fractions}
    \begin{tabular}{lccc}
         \toprule
         \textbf{Fraction} & \textbf{TNG300-1} & \textbf{TNG-Cluster} & \FLAMLmbold \\
         \midrule
         $\fb$ & $0.153$ & $0.155$ & $0.138$\\
         $\fgas$ & $0.139$ & $0.137$ & $0.109$ \\
         $\fstar$ & $0.013$ & $0.018$ & $0.029$ \\
         $\fstarICL$ & $3.39 \times 10^{-3}$ & $3.82 \times 10^{-3}$ & $8.96 \times 10^{-3}$\\
         $\fbh$ & $5.70\times10^{-5}$ & $3.67\times10^{-5}$ & $4.11\times10^{-5}$ \\
         \bottomrule
    \end{tabular}
\end{table}

The mass-conditioned covariance between two properties, $a$ and $b$, is then computed in a similar weighted manner
\begin{equation}\label{eq:KLLR-covariance}
    C_{a, b} (\mu, z) = A\sum_{i = 1}^n \ w_i \ \delta s_{a, i} \ \delta s_{b, i},
\end{equation}
where $\delta s_{x, i}$ is the residual of property $x$ from the KLLR mean of the $i$th halo and $A$ is a normalization factor that only depends on the weights, $w_i$. 

\begin{figure*}
    \centering
    \includegraphics[width = 0.8\textwidth]{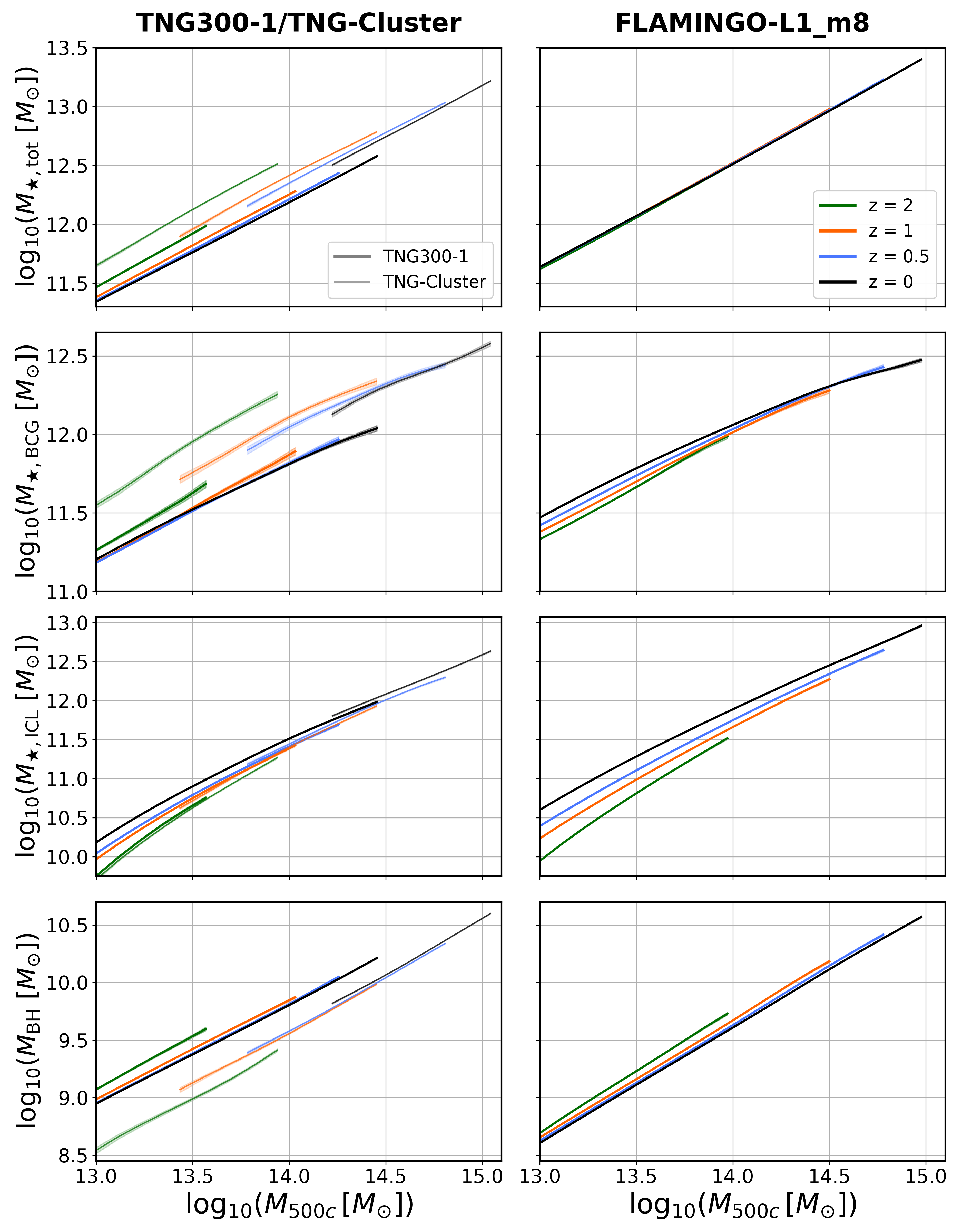}
    \caption{Mean stellar and BH properties as a function of halo mass at four redshifts $z = 0$ (black), $z = 0.5$ (blue), $z = 1$ (red) and $z = 2$ (green) for the three simulations TNG300-1 (left column), TNG-Cluster (left column, thin line) and {\FLAMLm} (right column).
}
    \label{fig:stellar-norms}
\end{figure*}

The corresponding correlation coefficient is 
\begin{equation}\label{eq:KLLR-correlation}
    r_{a, b}(\mu, z) = \frac{C_{a, b}(\mu,z)}{\sqrt{C_{aa}(\mu,z)} \sqrt{{C_{bb}(\mu,z)}}}.
\end{equation}
which lies in the range $[-1,1]$.

The scale-dependent intrinsic scatter is given by the square root of the diagonal covariance terms,
\begin{equation}\label{eq:KLLR-scatter}
    \sigma_{a} (\mu, z)= \sqrt{C_{a, a}(\mu,z)} .
\end{equation}

To examine the shapes of mass-conditioned likelihoods in \S\ref{sec:kernel-shapes}, we compute \textit{normalized residuals},
\begin{equation}\label{eq:KLLR-normalized-res}
    \widehat{\delta s}_{a, i} \equiv \frac{\delta s_{a, i}}{\sigma_a(\mu_i,z)}.
\end{equation}

%%%%%%%%%%%%%%%%%%%%%%%%%%%%%%%%%%%%%%%%%%%%%%%%%%%%%%%%%%%%%%%%%%%%
%%%%%%%%%%%%%%%%%%%%%%%%%%%%%%%%%%%%%%%%%%%%%%%%%%%%%%%%%%%%%%%%%%%%
\section{Stars, Black Holes, and Total Baryons}\label{sec:star-bh-MPRs}

In this section, we examine the mass and redshift dependence of stellar and BH properties. We also discuss the baryon fraction as a function of halo mass and redshift in each simulation, using it as a framework to interpret differences in the scaling parameters. The numbers of halos used in the KLLR analyses of the stellar, BH, and satellite scaling relations are reported in Table~\ref{tab:halo-counts}, including the reduced samples used for satellite-based observables that require $\Nsat \geq 10$. 

For additional context, Table~\ref{tab:mean-fractions} reports the mean baryon, gas, stellar, ICL, and BH mass fractions inside $\Rfivehc$ at $z=0$ evaluated at $\Mfivehc = 10^{14.5}\msun$ for each simulation. At $z = 0$, FLAMINGO contains more than twice the total stellar mass of TNG300-1 and about $1.6$ times that of TNG-Cluster, with the same trend being reflected in the ICL fraction. This excess is partially offset by a higher gas fraction in TNG300, which exceeds that of FLAMINGO by $28\%$, while TNG300-1 and TNG-Cluster differ by only $2\%$ in gas content. The BH mass fraction is highest in TNG300-1 and lowest in TNG-Cluster. Overall, the gas and stellar mass fractions are broadly consistent with values reported in observational and simulation studies \citep{Tremmel2019, Rasia2025fractions}, with TNG tending toward higher gas and lower stellar fractions, and FLAMINGO toward lower gas and higher stellar fractions.

The scaling relations are derived using kernel centers beginning
at the minimum masses reported in Table~\ref{tab:halo-counts} and increasing in intervals of ${\sim} 0.1$ dex up to the mass of the $21$st most massive halo at each redshift. Unless otherwise stated, shaded regions in all figures represent the estimated $1\sigma$ uncertainties obtained from $1000$ bootstrap realizations.

\subsection{MPR Normalizations}\label{sec:mpr-star-bh-norm}
Figure~\ref{fig:stellar-norms} presents the MPR normalizations for the total stellar mass $\Mstartot$, the BCG stellar mass $\MstarBCG$, the ICL stellar mass $\MstarICL$, and the total BH mass $\Mbh$. We discuss each in turn as follows.

\subsubsection{$\Mstartot-\Mfivehc$}\label{sec:Mstartot-scaling}
The $\Mstartot-\Mfivehc$ relation in the FLAMINGO halo population is remarkably insensitive to redshift across the full halo mass range studied. In contrast, TNG300-1 exhibits mild redshift evolution for $z \leq 1$, with the total stellar mass increasing slightly with redshift at fixed halo mass and a more pronounced increase emerging by $z = 2$. TNG-Cluster shows qualitatively similar behavior to TNG300, although with a larger amplitude of evolution at fixed halo mass.  The lack of redshift evolution in FLAMINGO results from a cancellation between the negative evolution of both $\MstarBCG$ and $\MstarICL$, discussed in the following sections, and the strong positive evolution of $\Mstarsat$ shown in Figure~\ref{fig:M-star-sat-kllr-params}. 

These trends can be compared with both observational and simulation studies in the literature. Using a sample of $91$ tSZ-selected clusters from SPT, \citet{Chiu2018} found no statistically significant redshift evolution in stellar mass conditioned on SZ-estimated mass over the range $0.2 < z < 1.25$, broadly consistent with the weak evolution observed in both simulations. However, the observational uncertainties are large, and larger cluster samples will be needed to distinguish between the predicted redshift evolution in the simulations. Similarly, the simulation study of \citet{Battaglia2013} reported that the stellar mass fraction of massive halos remains approximately constant with redshift. In contrast, using a KLLR analysis of the BAHAMAS and MACSIS simulations, \citet{Farahi2018} found that the normalization of the total stellar mass scaling relation decreases mildly with increasing redshift for both low- and high-mass halos out to $z = 1$.

At the present epoch, halos in FLAMINGO contain approximately twice the total stellar mass of halos in TNG300, consistent with the larger stellar mass fractions reported in Table~\ref{tab:mean-fractions}. The differences between the simulations decrease with increasing redshift, although a systematic offset remains between the two TNG simulations. In particular, TNG-Cluster exhibits a nearly constant enhancement of ${\sim} 0.12$ dex in $\Mstartot$ relative to TNG300-1 at $z = 0$, increasing mildly with redshift to ${\sim} 0.2$ dex by $z = 2$. In Appendix~\ref{sec:Mstartot-discrepancy}, we show that the excess stellar mass in TNG-Cluster is driven primarily by the BCG, with only minor contributions from the other stellar components.

\subsubsection{$\MstarBCG-\Mfivehc$}
Similar behavior is present in the $\MstarBCG-\Mfivehc$ relation. At $z = 0$ and $\Mfivehc = 10^{14.5}\msun$, FLAMINGO predicts BCG stellar masses that are larger than those in TNG300-1 by ${\sim} 0.25$ dex, while remaining comparable to those in TNG-Cluster. Toward the high-mass end, however, the FLAMINGO relation flattens, leading to more massive BCGs in TNG-Cluster for the most massive halos. The offset between TNG300-1 and TNG-Cluster persists across all redshifts and remains approximately ${\sim} 0.2$ dex in magnitude. Overall, the normalization of the TNG300-1 relation is in closer agreement with observational constraints from \citet{Kravtsov2018} and \citet{GoldenMarx2022}.

The redshift evolution of the $\MstarBCG-\Mfivehc$ relation in TNG300-1 and TNG-Cluster broadly mirrors that of the total stellar mass relation, with BCG stellar masses exhibiting no significant evolution out to $z = 1$, in agreement with the observational results of \citet{GoldenMarx2019} and \citet{GoldenMarx2022}. In FLAMINGO, however, the evolution is qualitatively different for lower-mass systems ($\Mfivehc \lesssim 10^{14}\msun$), where BCG stellar masses decrease with increasing redshift at fixed halo mass, while the relation for more massive halos remains approximately redshift independent.

\subsubsection{$\MstarICL-\Mfivehc$}
The differences among the simulations are more modest for the $\MstarICL-\Mfivehc$ relation. At $z = 0$, FLAMINGO contains approximately twice as much stellar mass in the ICL as both TNG simulations, reflective of its larger total stellar mass content. This offset decreases with increasing redshift. Unlike the total and BCG stellar masses, the difference between TNG300-1 and TNG-Cluster is comparatively small for the ICL component.

All three simulations exhibit similar redshift evolution, with the ICL mass decreasing with increasing redshift at fixed halo mass. This trend is broadly consistent with theoretical expectations of a declining ICL fraction toward higher redshift \citep{Rudick2011, Contini2014, Kimmig2025}, while still retaining a non-negligible ICL component even at early times \citep{Joo2023}.

\subsubsection{$\Mbh-\Mfivehc$}
For the $\Mbh-\Mfivehc$ relation, the simulations exhibit trends opposite to those seen for the total stellar mass. FLAMINGO halos contain systematically less total BH mass than TNG300, while remaining broadly consistent with TNG-Cluster. A clear mass-dependent offset is also present between the two TNG simulations, with TNG-Cluster containing less BH mass than TNG300-1 at fixed halo mass. Interestingly, the magnitude of this offset is comparable, in fractional terms, to that observed in the $\Mstartot-\Mfivehc$ relation, but with the opposite sign. This may indicate that, in TNG's AGN model, more efficient star formation is associated with gas conditions that result in lower SMBH accretion rates.

The redshift evolution of the $\Mbh-\Mfivehc$ relation is similar in TNG300-1 and FLAMINGO, with little evolution out to $z = 1$ and a modest increase in BH mass at fixed halo mass by $z = 2$. In contrast, TNG-Cluster halo populations exhibit a mild decrease in BH mass with increasing redshift at fixed halo mass over the same interval.

\subsection{Mean baryon fraction}\label{sec:baryon-fraction}

\begin{figure}
    \centering
    \includegraphics[width = \linewidth]{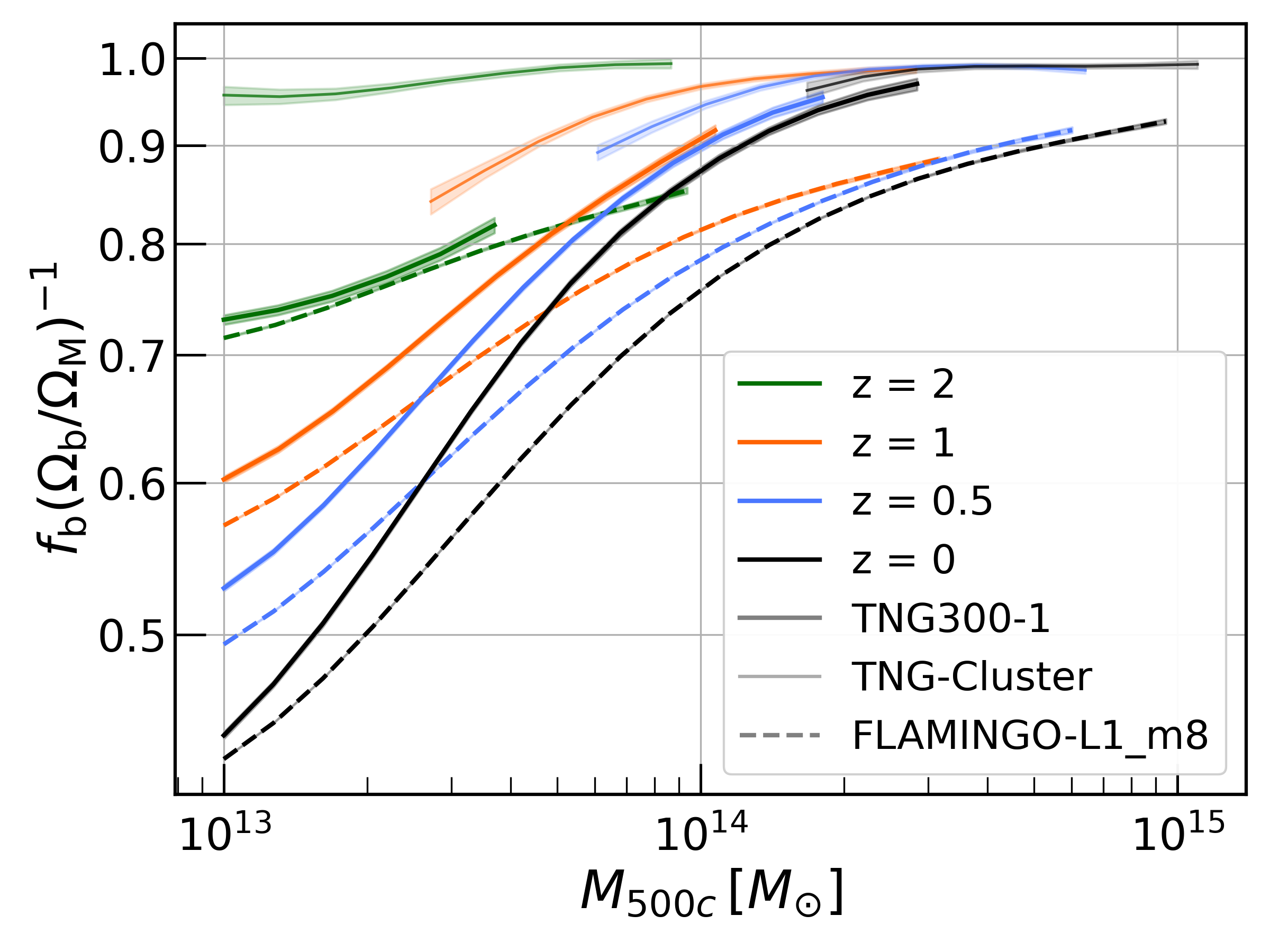}
    \vspace{-18truept}
    \caption{Mean baryon fraction within $\Rfivehc$ for halo populations in TNG300-1 (solid line), TNG-Cluster (solid, thin) and {\FLAMLm} (dashed) as a function of mass and redshift.  Values are normalized by the cosmic mean baryon fraction of each simulation (Table~\ref{tab:sim-constants}).
}
    \label{fig:baryon-fraction}
\end{figure}

Figure~\ref{fig:baryon-fraction} shows the mass and redshift dependence of the KLLR normalization of the baryon fraction within $\Rfivehc$, normalized by the cosmic mean baryon fraction listed in Table~\ref{tab:sim-constants}.

At all halo masses and for $z<2$, halos in the TNG simulations retain systematically more baryons within $\Rfivehc$ than halos in FLAMINGO, with the discrepancy increasing toward higher halo masses. Both simulation suites exhibit a strong mass dependence in the total baryon fraction, with more massive halos retaining a larger fraction of their baryons within the halo aperture, while lower-mass systems more efficiently expel baryons beyond $\Rfivehc$ through feedback processes, consistent with trends previously reported for the gas fraction in hydrodynamical simulations \citep{McCarthy2017, LeBrun2017, Henden2018, Farahi2018, Rasia2025fractions}. For the most massive halos, however, TNG-Cluster approaches an approximately constant baryon fraction across all redshifts, remaining at $99.3\%$ of the cosmic mean, whereas FLAMINGO reaches only $92.7\%$. At fixed halo mass, TNG-Cluster also retains a systematically larger baryon fraction than TNG300-1 at all redshifts. The higher baryon fraction in TNG-Cluster is consistent with its higher stellar mass fraction, while the gas fractions of the two TNG simulations remain in closer agreement (see Appendix~\ref{sec:fgas-fstar}). 

\begin{figure*}
    \centering
    \includegraphics[width = 0.8\textwidth]{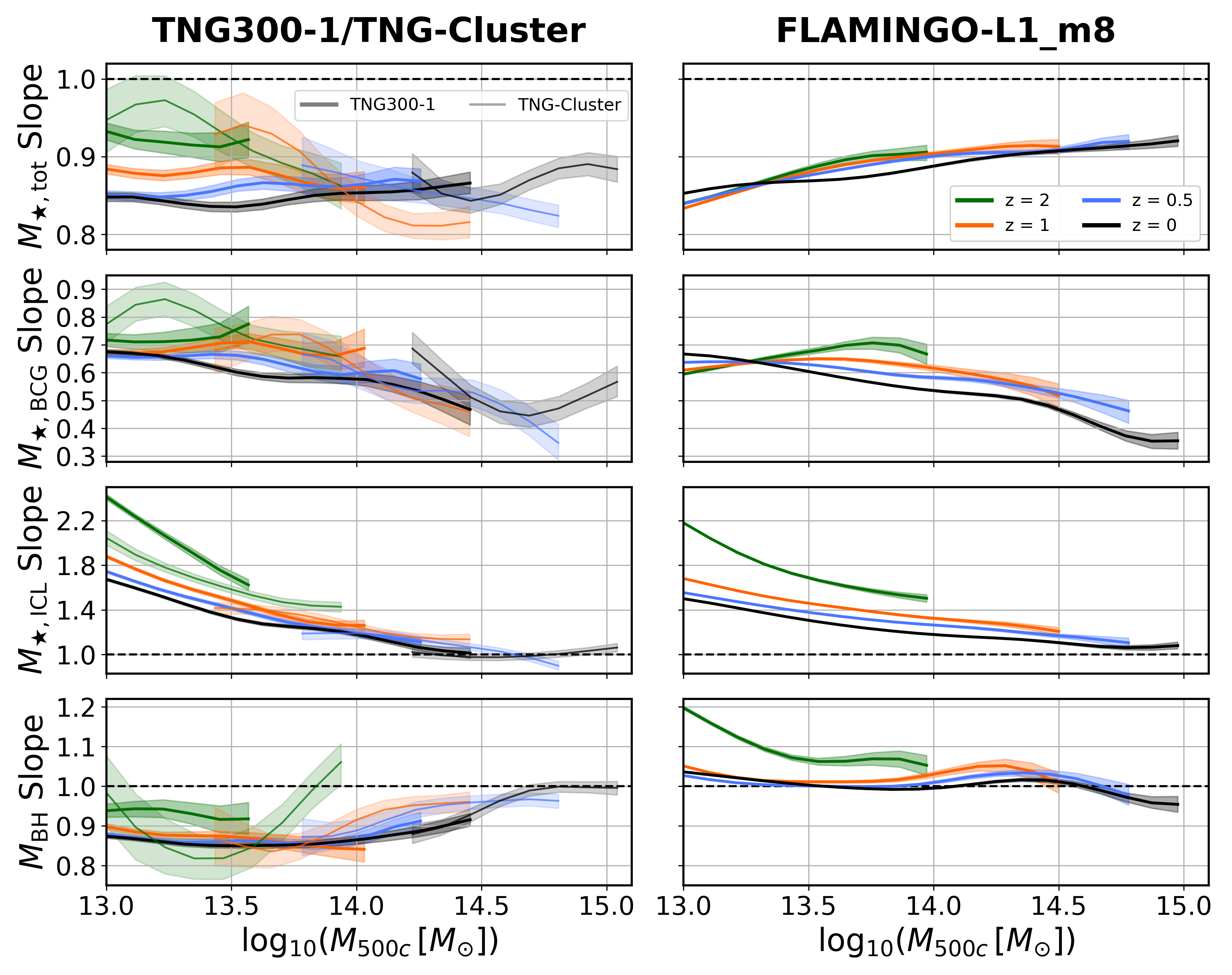}
    \caption{Halo mass and redshift dependence of the MPR slopes for (top to bottom) $\Mstartot$, $\MstarBCG$, $\MstarICL$, and $\Mbh$ at the same redshifts and color scheme used in Figure~\ref{fig:stellar-norms}. The dashed lines denote the self-similar expectation of unity.
}
    \label{fig:stellar-slopes}
\end{figure*}

At fixed halo mass, the baryon fraction generally decreases with time, particularly for halos with $\Mfivehc \lesssim 10^{14.5}\msun$, in agreement with the results of \citet{Rasia2025fractions}. Toward higher masses, this redshift dependence weakens and the baryon fraction becomes nearly redshift independent, asymptotically approaching the cosmic baryon fraction in all simulations. Interestingly, at $z=2$, TNG300-1 and FLAMINGO exhibit comparable baryon fractions across much of the mass range, resulting from a near cancellation between the higher $\fgas$ and lower $\fstar$ in TNG300-1 (Appendix~\ref{sec:fgas-fstar}). In contrast, TNG-Cluster retains systematically more baryons, by ${\sim} 10$--$20\%$, and remains close to the cosmic baryon fraction across the full halo mass range studied.

\begin{figure*}
    \centering
    \includegraphics[width = 0.8\textwidth]{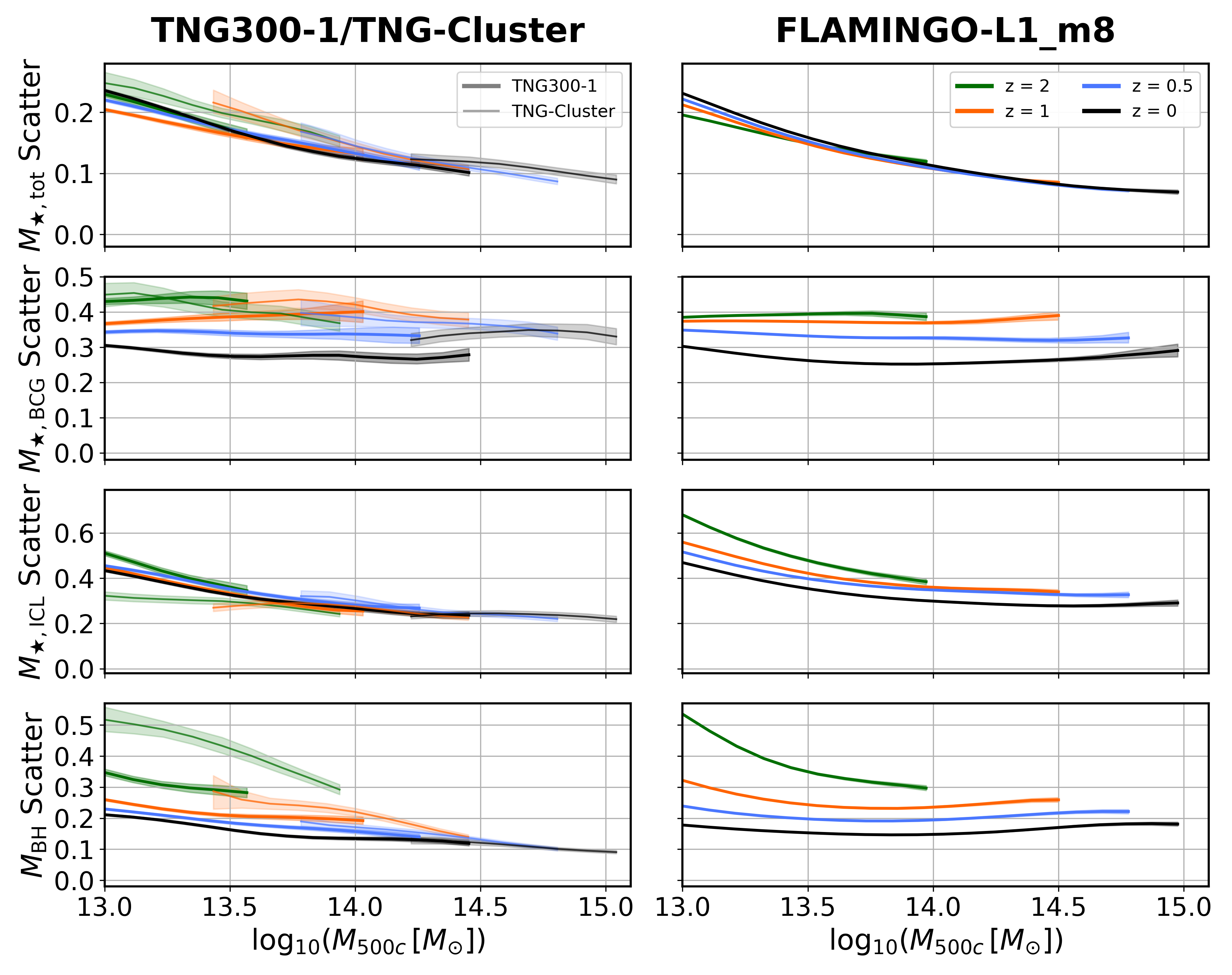}
    \caption{Halo mass and redshift dependence of MPR intrinsic scatter, equation~\eqref{eq:KLLR-scatter}. The format is identical to that of Figure~\ref{fig:stellar-slopes}.  
}
    \label{fig:stellar-scatters}
\end{figure*}

\subsection{MPR slopes and intrinsic variance}\label{sec:mpr-star-bh-slope-scatter}
In Figures~\ref{fig:stellar-slopes} and \ref{fig:stellar-scatters}, we present the mass and redshift dependence of the MPR slopes and intrinsic scatter values (in natural logarithmic units) for the stellar and BH properties. Despite the differences in normalization discussed above, the simulations exhibit broad quantitative and qualitative agreement in the behavior of both the slope and scatter as functions of halo mass and redshift.

\subsubsection{$\Mstartot-\Mfivehc$}
As anticipated from early cooling timescale arguments, the $\Mstartot-\Mfivehc$ relation at $z = 0$ is sublinear in all three simulations. In TNG300-1 and TNG-Cluster, the slope remains nearly constant with halo mass, taking values of ${\sim} 0.84$--$0.89$ across the full mass range studied. In contrast, FLAMINGO exhibits a systematic increase in slope with halo mass, rising from ${\sim} 0.85$ at group scales to ${\sim} 0.92$ for the most massive clusters, approaching the self-similar expectation of unity. Similar sublinear slopes have been reported in previous simulation studies \citep{Barnes2017b, Bahe2017}.

The redshift evolution of the slope is weak overall. In FLAMINGO, the slope increases mildly with redshift over the range $10^{13.5}\msun \lesssim \Mfivehc \lesssim 10^{14.5}\msun$, while in TNG300-1 the increase is largely confined to halos with $\Mfivehc \lesssim 10^{13.5}\msun$, with TNG-Cluster showing a similar trend.

The intrinsic scatter in the $\Mstartot-\Mfivehc$ relation shows notable agreement among the simulations and little redshift evolution. In all cases, the scatter decreases monotonically with halo mass, falling from ${\sim} 0.25$ at group scales to ${\sim} 0.10$ in the TNG simulations and to ${\sim} 0.08$ in FLAMINGO for the most massive clusters. These results are broadly consistent with observational estimates from \citet{Chiu2018}, who reported a slope of $0.80 \pm 0.12$ and an intrinsic scatter of $0.22 \pm 0.03$, as well as previous simulation studies \citep{Farahi2018, Anbajagane2020stellar} which find both similar slopes and a decline in scatter toward higher halo masses.  As discussed in \S\ref{sec:MPQ}, total stellar mass offers the best MPQ among the properties studied here. 

\subsubsection{$\MstarBCG-\Mfivehc$}
The combination of AGN feedback and longer cooling times in the core of massive halos produces sublinear slopes for the $\MstarBCG-\Mfivehc$ relation. The slope generally declines with increasing mass, approaching values of ${\sim} 0.4$ for halos with $\Mfivehc \gtrsim 10^{14.5}\msun$. An exception occurs in TNG-Cluster, where the slope shows a mild upturn for the most massive systems, consistent with the behavior reported for BAHAMAS clusters. Overall, the inferred slopes agree well with values reported from multiple simulations by \citet{Anbajagane2020stellar}. 

The measured slopes are also broadly consistent with observational studies. \citet{Kravtsov2018} found a slope of ${\sim} 0.6$ for nearby clusters, while \citet{GoldenMarx2018} reported a similar value. However, our slopes are generally steeper than those reported in several other observational analyses \citep{Zhang2016, Mulroy2019, GoldenMarx2019, GoldenMarx2022}, particularly at the massive end.

The redshift evolution of the slope broadly mirrors that of the $\Mstartot$ relation, although the effect is somewhat stronger for the BCG component. In TNG300, the slope increases mildly with redshift for lower-mass halos, while FLAMINGO shows a stronger increase for systems with $\Mfivehc \gtrsim 10^{13.5}\msun$. This trend is qualitatively consistent with the observational studies of \citet{GoldenMarx2019} and \citet{GoldenMarx2022}, which found steeper BCG scaling relations at earlier times. At $z = 2$, FLAMINGO additionally shows a weak increase of slope with halo mass, differing from the flatter behavior seen at lower redshifts.

Compared to the slope, the intrinsic scatter of the $\MstarBCG-\Mfivehc$ relation depends more strongly on redshift than on halo mass. All three simulations are in close agreement, exhibiting an almost mass-independent scatter of slightly below ${\sim} 0.3$ across most of the mass range. The scatter increases modestly with redshift in all simulations. Observational studies report intrinsic scatters spanning ${\sim} 0.2$--$0.5$ dex \citep{Zhang2016, Kravtsov2018, GoldenMarx2018, Mulroy2019, GoldenMarx2019, GoldenMarx2022}, consistent with the values found here, although most observational analyses report little evidence for redshift evolution \citep{GoldenMarx2019, GoldenMarx2022}.

\subsubsection{$\MstarICL-\Mfivehc$}
The slope of the $\MstarICL-\Mfivehc$ relation exhibits a strong dependence on halo mass, steepening at the group scale ($\Mfivehc \lesssim 10^{14}\msun$), and the scatter is also larger at lower masses.  These features reflect the decreased effectiveness of stripping, due to lower velocity dispersions in groups, as well as the enhanced stochasticity due to a small number of available satellites to strip.  At cluster-scale masses, the slope approaches unity and becomes nearly constant in TNG-Cluster. Likewise, the scatter declines from ${\sim} 0.4$ at low masses to ${\sim} 0.25$ for massive clusters, where it subsequently flattens.

\begin{figure*}
    \centering
    \includegraphics[width = 0.7\textwidth]{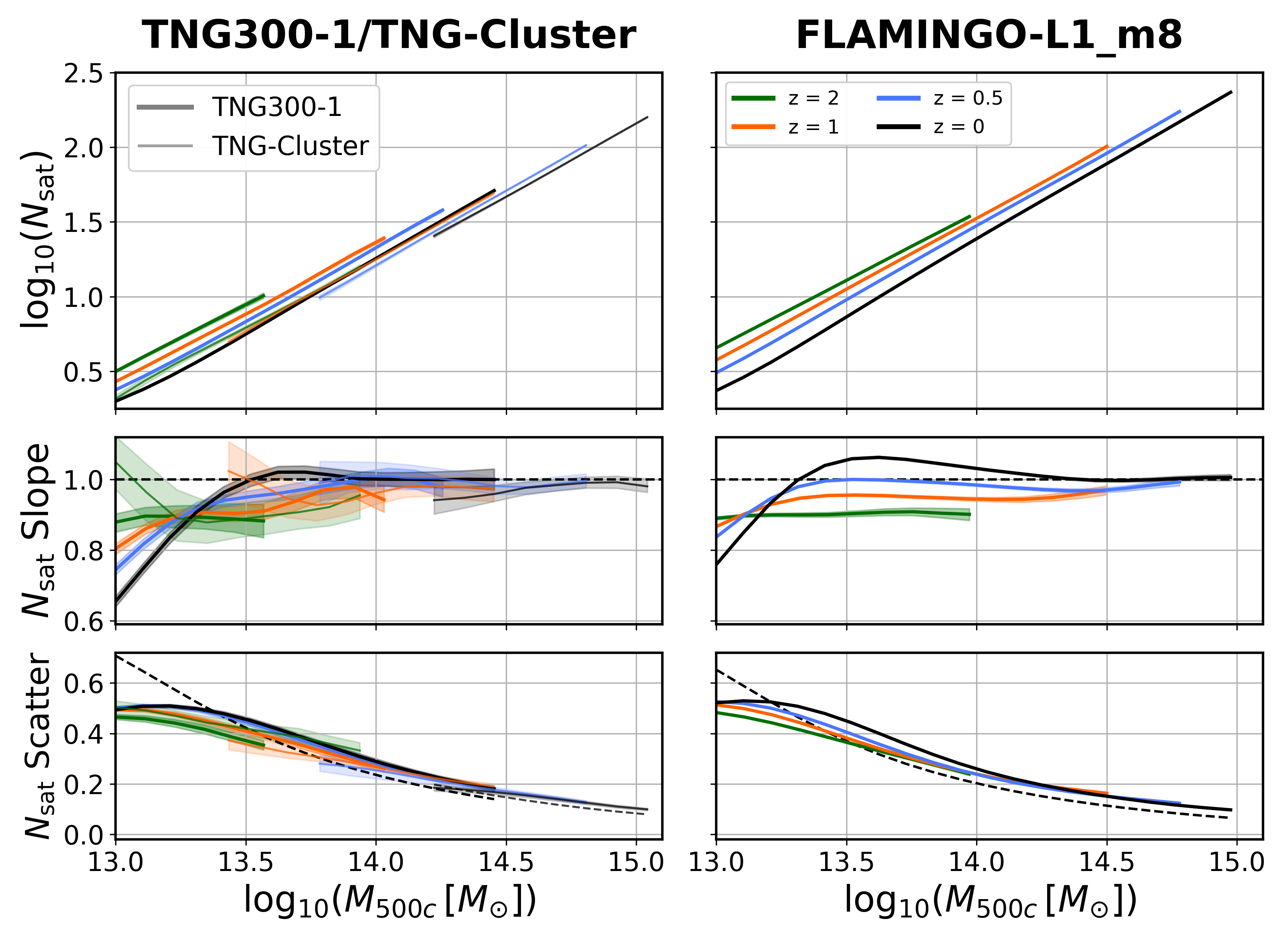}
    \caption{Satellite occupation, the number of satellite galaxies with stellar mass above $10^{9.5} \msol$ within $\Rfivehc$, scaling parameters: normalization (top panel), slope (middle) and intrinsic logarithmic scatter (bottom). In the latter, the dashed lines show the $z=0$ Poisson fractional scatter, $\langle \Nsat \vert \Mfivehc,z\rangle^{-1/2}$, derived from the top panel. The dashed line in the middle panel indicates the self-similar slope of $1$.
}
    \label{fig:N-sat-kllr-params}
\end{figure*}

The simulations also show consistent redshift evolution in the slope, particularly at lower masses, where the slope increases toward higher redshift, most noticeably by $z = 2$. The redshift dependence of the scatter differs between the simulation suites. In the TNG simulations, the scatter remains largely unchanged with redshift, whereas FLAMINGO exhibits a systematic increase toward higher redshift.

The strong mass and redshift dependence of the $\MstarICL-\Mfivehc$ relation presents an interesting target for future observational studies. Wide-field surveys such as Euclid will enable measurements of diffuse intracluster light for large cluster samples \citep{BlanchardEuclid2020, EuclidBellhouseICLdetectionlimits, KlugeEuclidICLPerseus}, while targeted surveys such as LoVoCCS offer complementary opportunities to characterize the ICL in nearby well-studied systems \citep{Fu2022LoVoCCS, Fu2024LoVoCCS, Englert2026LoVoCCS}. Measurements spanning a broad range of halo masses and redshifts could test the predicted mass and redshift-dependence of the relation, particularly the steep slope at group scales, its convergence toward unity in massive halos, and the increase in both the slope and intrinsic scatter toward higher redshift. 

\subsubsection{$\Mbh-\Mfivehc$}
The slope of the $\Mbh-\Mfivehc$ relation is close to unity at lower halo masses in both TNG300-1 and FLAMINGO, although TNG300-1 remains slightly sublinear while FLAMINGO is slightly superlinear over much of the mass range, with typical slopes of ${\sim} 0.85$ and ${\sim} 1.05$, respectively. Toward higher halo masses, however, the simulations converge. In TNG300-1 and TNG-Cluster, the slope gradually increases with halo mass, approaching unity for the most massive systems, while the FLAMINGO slope decreases slightly, remaining close to unity and reaching ${\sim} 0.95$ at the high-mass end.

The redshift evolution of the slope is weak in all simulations. Both TNG300-1 and TNG-Cluster show little variation with redshift across the full mass range, while FLAMINGO exhibits a noticeable steepening for lower-mass halos at $z = 2$.

The intrinsic scatter of the $\Mbh-\Mfivehc$ relation also shows only weak mass dependence. In the TNG simulations, the scatter is nearly constant with a minimum value of ${\sim} 0.1$ for the most massive halos, while FLAMINGO exhibits a somewhat larger minimum of ${\sim} 0.15$. The redshift evolution of the scatter is broadly similar between the simulation suites. In the TNG simulations, the scatter increases only slightly with redshift, primarily for lower-mass halos and mainly at $z < 2$. At $z = 2$, however, both FLAMINGO and TNG-Cluster show a stronger increase in scatter that extends across most of the halo mass range.

The $\Mbh-\Mfivehc$ relation also presents an opportunity for future observational tests of the simulations. As a first step, measurements of central SMBH masses in groups and clusters, combined with independent halo mass estimates from weak lensing or X-ray observations, could test the predicted relation over a broad range of halo masses. Extending these measurements to include SMBHs hosted by satellite galaxies would provide an estimate of the total BH content and enable a more direct comparison with the relation studied here. Such measurements could help distinguish between the different predictions of the simulations and determine which treatments of black hole growth and AGN feedback more accurately describe the buildup of black holes in massive halos.

%%%%%%%%%%%%%%%%%%%%%%%%%%%%%%%%%%%%%%%%%%%%%%%%%%%%%%%%%%%%%%%%%%%%
%%%%%%%%%%%%%%%%%%%%%%%%%%%%%%%%%%%%%%%%%%%%%%%%%%%%%%%%%%%%%%%%%%%%
\section{Satellite occupation and velocities}\label{sec:mpr-Nsat-vel-disp}

We turn next to the statistics of satellite galaxies having stellar masses above the threshold value, $\MstarsatThresh = 10^{9.5} \msol$. 
We first examine scaling of mean satellite counts (or occupation), then unpack the detailed shape of the the galaxy stellar function in massive halos before closing with virial scaling analysis of satellite velocity dispersion.  For the latter, the $\Nsat \ge 10$ criterion for robust velocity dispersion measures discussed in \S\ref{sec:halo-properties} implies a minimum halo mass that is both simulation and redshift dependent (Table~\ref{tab:halo-counts}).  At $z=2$, the combination of limited sample sizes and narrow mass ranges for halos satisfying this threshold precludes application of KLLR for TNG300-1 and TNG-Cluster samples, so we restrict analysis of $\sigmaSat$ to $z \le 1$.

\subsection{Satellite Occupation}\label{sec:Nsatstats} 

Log-mean counts of satellite galaxies with stellar mass above $\MstarsatThresh$ are shown in the top panel of Figure \ref{fig:N-sat-kllr-params}, with scale-dependent slopes and scatter values in the middle and lower panels, respectively.  

At $z=0$ and $\Mfivehc \simeq 10^{14.5}\msun$, the mean satellite occupation is ${\sim} 58$ in TNG300, ${\sim}48$ in TNG-Cluster, and ${\sim} 78$ in FLAMINGO. The higher satellite occupation in FLAMINGO is broadly consistent with its larger total stellar mass content, as halos contain approximately twice the stellar mass of their TNG counterparts while maintaining a comparable satellite-to-total stellar mass fraction. These normalization offsets persist across redshift, with all simulations exhibiting increasing satellite occupation toward higher redshift at fixed halo mass.

\begin{figure}
    \centering
    \includegraphics[width = \linewidth]{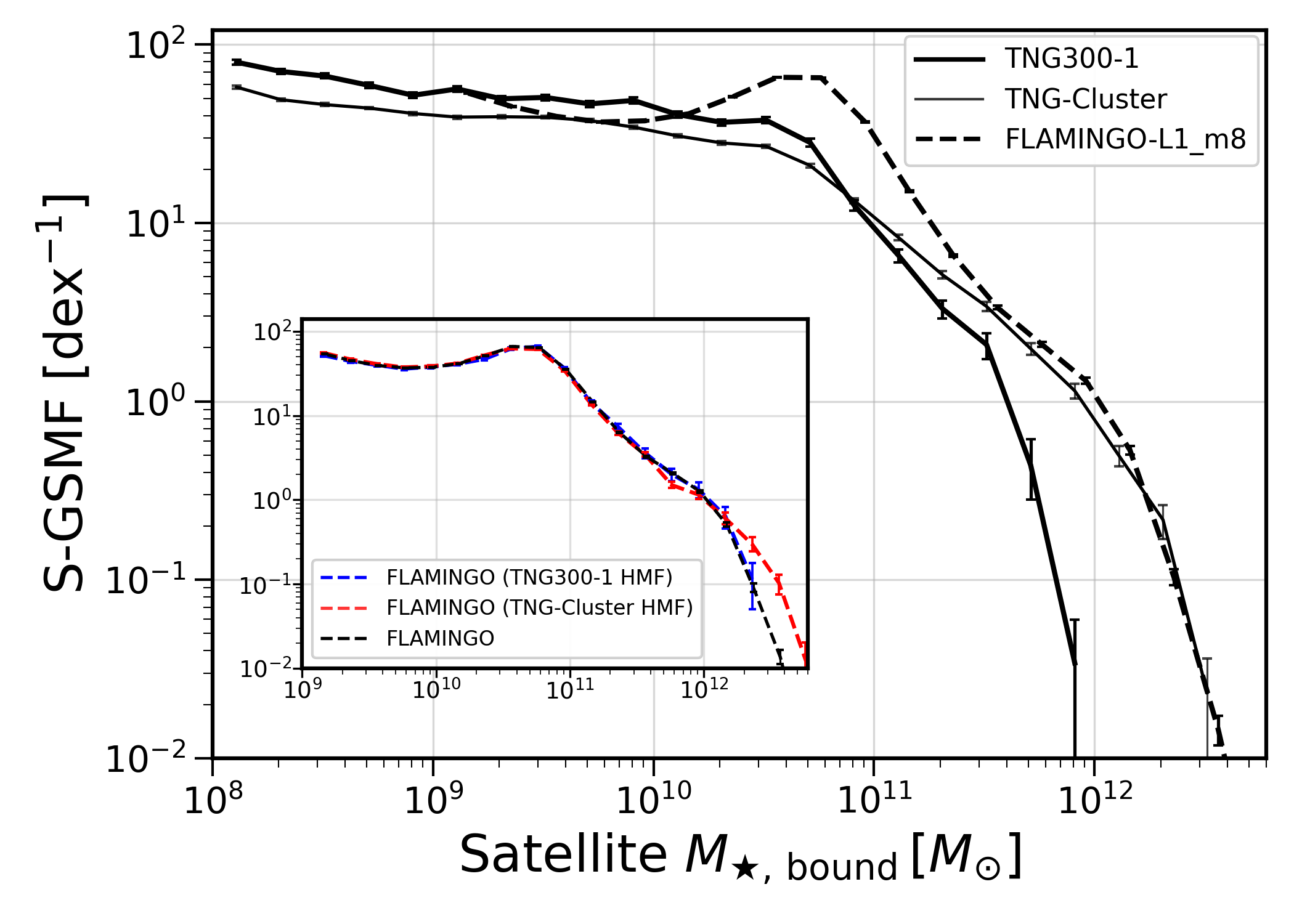}
    \vspace{-18truept}
    \caption{The mean satellite galaxy stellar mass function (S-GSMF) at $z = 0$ for a $\Mfivehc \geq 10^{14}\msun$ halo, derived from cluster-scale halos from the samples of TNG300-1 (solid line), TNG-Cluster (solid, thin line) and {\FLAMLm} (dashed line).  Each is derived via inversely weighting by halo mass the counts of satellites in halos above $10^{14}\msun$, equation~\eqref{eq:SGSMF}. The inset shows the S-GSMF for the original FLAMINGO halo sample (black) along with those derived from subsamples of the full FLAMINGO set designed to match the mass distributions of TNG300-1 (blue) and TNG-Cluster (red). The close agreement demonstrates that the distinct halo mass selections of the two TNG samples is unlikely to strongly bias the S-GSMF shape, except at the highest stellar masses, members of which emerge only in the most massive halos. The error bars represent the $16$th--$84$th percentile range from $1000$ bootstrap realizations of the satellite galaxy samples.
    } 
    \label{fig:S-GSMF}
\end{figure}

While the satellite occupation studied here is not directly equivalent to observational estimators such as redMaPPer richness \citep{Rykoff2014redMaPPerI}, the normalizations provide a useful point of comparison with observational richness--mass relations. For example, \citet{Mulroy2019} report a richness of ${\sim}125$ at $M_{{\rm WL},500c} = 7.4 \times 10^{14}\msun$, while the corresponding satellite occupations at similar halo masses in our simulations are ${\sim}107$ for TNG-Cluster and ${\sim}182$ for FLAMINGO.  

Surprisingly, TNG-Cluster halos contain \emph{fewer} satellite galaxies, while at the same time \emph{more} stars, than similar mass halos of TNG300. A greater fraction of TNG-Cluster halos' stellar mass resides in the central galaxy, whose BCGs are systematically more massive, leaving $\Mstarsat$ to be more consistent between TNG300-1 and TNG-Cluster at fixed halo mass (Appendix~\ref{sec:TNG-discrepancy}).  The different shapes of the galaxy stellar mass function within clusters help explain this conundrum, a topic we address in the next section.

Despite the disagreement in satellite count normalization, the slope and scatter are largely consistent across the simulations, in keeping with our Paper I findings for the hot gas features. At $z=0$, the slope of $\Nsat$ with halo mass is close to the self-similar expectation and consistent with unity in TNG for halos with $\Mfivehc > 10^{13.5}\msun$, and in FLAMINGO for $\Mfivehc > 10^{14.3}\msun$. At lower masses, the FLAMINGO relation becomes somewhat superlinear, while both simulations exhibit flattening as the satellite counts become small where small number statistics confound an approach based on logarithmic properties. There is a similar redshift dependence in both simulations, more prominent in FLAMINGO owing to its larger halo sample, with the slope decreasing to sublinear values with increasing redshift and reaching ${\sim}0.9$ at $z=2$.

In all simulations, the scatter decreases with halo mass, from ${\sim}40\%$ at low masses to ${\sim}10\%$, broadly following the fractional Poisson scatter at $z=0$ shown as the dashed line. At $\Mfivehc > 10^{13.5}\msun$, the scatter remains slightly above the Poisson expectation, with qualitative and quantitative agreement between the simulations.  Such trends in simulations are also reported at the $\mtwoh$ mass scale by  \citet{Anbajagane2020stellar}. The redshift dependence is confined to halos with $\Mfivehc < 10^{14}\msun$, particularly in FLAMINGO, where the scatter decreases with increasing redshift.

At high halo masses, the agreement of FLAMINGO and TNG-Cluster sample measures for the $\Nsat$ scatter is not expected under Poisson statistics, since the log-mean counts differ by nearly a factor of two.   The concordance reflects a fortuitously larger boost above Poisson statistics in massive FLAMINGO halos. 

\subsection{Cluster-scale Galaxy Stellar Mass Function}\label{sec:GSMF}

To further detail the stellar contents of massive halos, particularly the discrepancy in the two TNG models, we turn to the satellite galaxy stellar mass function (S-GSMF), the differential mean number of satellite galaxies within $\Rfivehc$ as a function of their stellar mass.  Since the $\Nsat$ slope at $z=0$ is nearly linear at cluster scales, $\Mfivehc > 10^{14} \msol$, we construct the mean S-GSMF expected at that epoch for a fiducial halo mass scale, $M_{\rm fid} = 10^{14.5} \msol$, using halo counts inversely weighted by total halo mass,
\begin{equation}\label{eq:SGSMF}
\frac{dn(M_{\rm fid})}{d\ln M_{\star, \rm bound}} \ = \ N_{\rm sam}^{-1} \ \sum_{i=1}^{N_{\rm sam}} \ \frac{M_{\rm fid}}{M_i}  \frac{dn (M_i)}{d\ln M_{\star, \rm bound}} . 
\end{equation}
Here $N_{\rm sam}$ is the number of halos in each sample with total mass above $10^{14} \msol$.  The minimum satellite stellar mass used for each sample is set by requiring at least ten stellar particles.

The result, shown in Figure~\ref{fig:S-GSMF}, illustrates how different approaches to galaxy astrophysics can yield different expectations for the galactic contents of massive halos.  Since the underlying astrophysics is genuinely complex, with multi-scale, fast and slow components coupled in complex ways,  some level of deviation in this emergent function is the natural expectation. 

The FLAMINGO sample exhibits a broad peak near $M_{\star,\rm bound}\sim5\times10^{10.5}\msun$, corresponding roughly 500 stellar particles.  Such a bump is not seen in XMM-XXL clusters at low redshifts \citep{Guglielmo2018_XXLGSMF}.  Above this stellar mass, the S-GSMFs of FLAMINGO and TNG-Cluster converge and become nearly identical for the most massive satellites, which rival their BCGs in heft.  

For bound stellar masses $M_{\star,\rm bound}\lesssim 10^{10.5}\msun$, there is a deficit in TNG-Cluster counts relative to TNG300, with $30\%$ fewer low-mass satellites in the former.  The result is highly statistically significant (${\sim} 8\sigma$), driven by a shift in chaotic astrophysical dynamics caused by the change from one-half to one of the $\alpha_{\star}$-parameter controlling star formation.\footnote{We stress that we cannot at this point describe this shift in terms of the astrophysical processes at play, we merely measure the impact.} 
In \S\ref{sec:correlations}, we show that the difference in low-mass galaxy counts is also reflected in differing signs of the correlation between total stellar mass and satellite galaxy count.  

Compensating the TNG-Cluster deficit at low stellar masses is an abundance of more massive galaxies with $M_{\star,\rm bound} > 10^{11} \msol$.  The compensation brings the total satellite stellar mass nearly in line with that of TNG300-1 (see Appendix~\ref{M-star-sat-MPRs}).  

Halo sample selection by mass could potentially affect the shape of the S-GSMF.  We show a test of selection in the inset of Figure~\ref{fig:S-GSMF} where the S-GSMFs of FLAMINGO sub-samples designed to match the differential halo counts in TNG300-1 and TNG-Cluster are compared to the result for the full sample.  Very similar S-GSMFs result at low stellar masses, indicating that total mass selection is not the main driver of the difference seen in the TNG realizations.  A boost in the S-GSMF at the highest stellar masses for the TNG-Cluster equivalent is expected because the most massive halos will have had more massive progenitor halos hosting more massive central  galaxies.  

\begin{figure*}
    \centering
    \includegraphics[width = 0.7\textwidth]{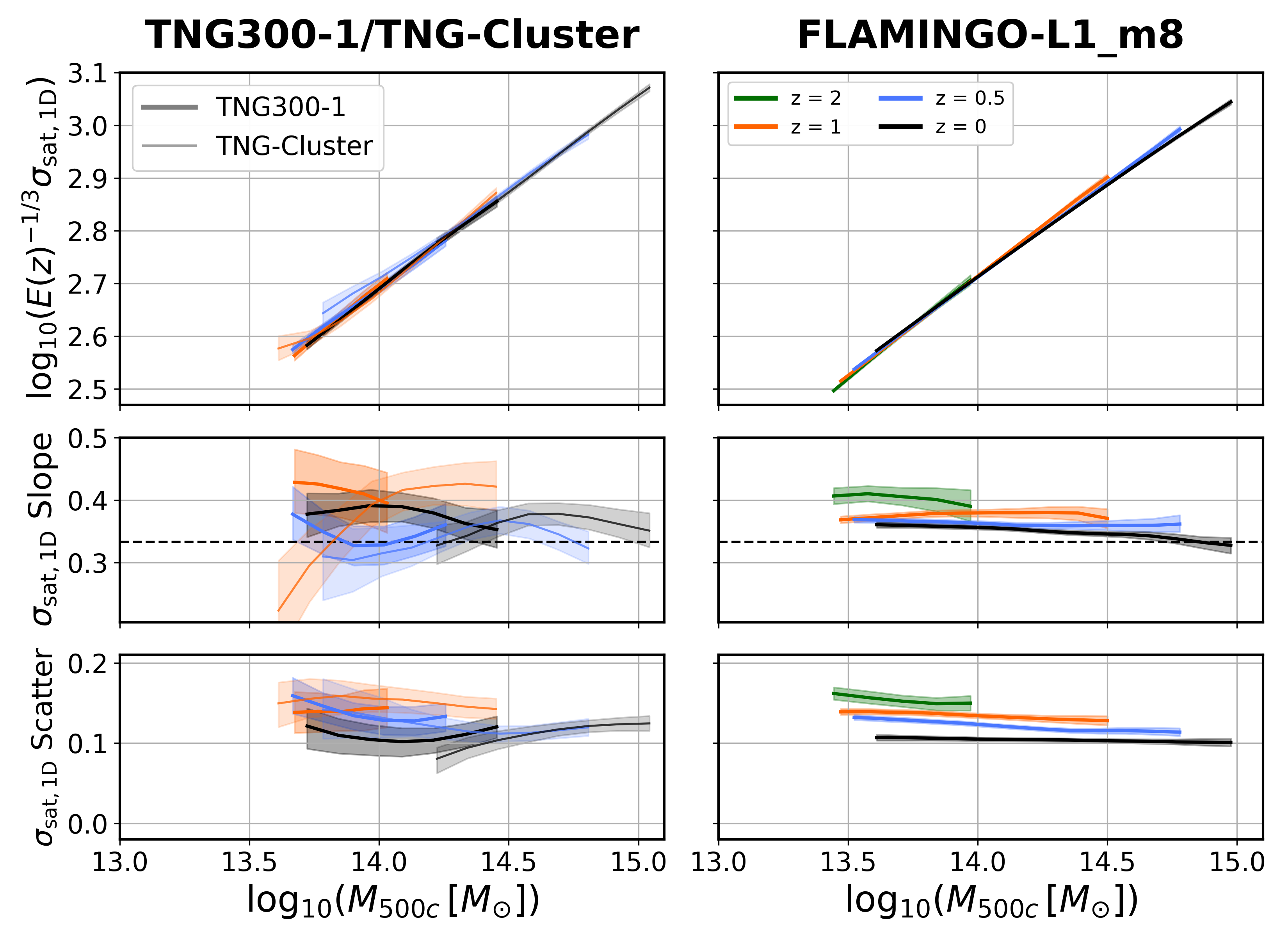}
    \caption{One-dimensional satellite galaxy velocity dispersion, as defined in Table~\ref{tab:prop-definitions}, scaling parameters. The top panel shows the normalization, including the self-similar evolution factor $E(z)^{-1/3}$, where $E(z) = \sqrt{\omegam (1 + z)^3 + \omegaLambda}$; the middle panel shows the slope; and the bottom panel shows the intrinsic scatter. The dotted line in the middle panel indicates the self-similar slope of $1/3$.  
}
    \label{fig:vel-disp-kllr-params}
\end{figure*}

\subsection{Satellite Galaxy Velocity Dispersion}\label{sec:SigmaSatstats}

Figure~\ref{fig:vel-disp-kllr-params} displays the scaling relations of the one-dimensional satellite velocity dispersion using the same format as Figure \ref{fig:N-sat-kllr-params}.  Strong agreement is seen across the simulation samples, with all closely adhering to self-similar redshift evolution, $E(z)^{-1/3}$, in normalization consistent with virial theorem expectations.

The slopes display weak mass dependence, with values slightly above the self-similar expectation, consistent with most observations \citep{Han2015, Viola2015, Rana2022, Zhang2024}.  Slopes reach ${\sim}0.35$--$0.38$ for halos with $\Mfivehc < 10^{14.5}\msun$, in agreement with \citet{Munari2013}, before decreasing toward the self-similar value of $1/3$ at the highest masses in FLAMINGO.  FLAMINGO also shows a weak increase in slope at higher redshifts, behavior mirrored with lower significance in TNG300.

The intrinsic scatter\footnote{We subtract in quadrature the measurement uncertainty of each halo's gapper velocity dispersion derived by bootstrap resampling of its satellites.} in the $\sigmaSat-\Mfivehc$ relation is remarkably mass-independent, particularly in FLAMINGO, with a value of ${\sim}0.1$ at $z=0$. The scatter gently increases with redshift in all simulations, with both the scatter amplitude and its redshift dependence broadly consistent with other simulations \citep{Munari2013, Anbajagane2021velbias}.

In pure dark matter simulations, the virial theorem is precisely respected.  The DM velocity dispersion scales with halo mass with a self-similar slope of $0.3361 \pm 0.015$ and a scatter of $(4.3 \pm 0.2)\%$ \citep{Evrard2008}. The observable satellite galaxy velocity dispersion is a biased tracer of this behavior, with heavier stellar mass galaxies being cooler than lighter ones \citep{Anbajagane2021velbias}. Such trends with galaxy photometric magnitude have been observed \citep{Biviano1992, Adami1998, Biviano2002, Lares2004, Ribeiro2010, Barsanti2016} and recently measured at high significance using more than $600$ clusters in the \textsc{SPIDERS} sample (Aljamal \textit{et al.}, in prep.).

%%%%%%%%%%%%%%%%%%%%%%%%%%%%%%%%%%%%%%%%%%%%%%%%%%%%%%%%%%%%%%%%%%%%%%%%%%%%%%%%%%%%%%%%%%%%%%%%%%%
%%%%%%%%%%%%%%%%%%%%%%%%%%%%%%%%%%%%%%%%%%%%%%%%%%%%%%%%%%%%%%%%%%%%%%%%%%%%%%%%%%%%%%%%%%%%%%%%%%%
\section{Halo Mass-conditioned Correlations}\label{sec:correlations}

In contrast to the often substantial normalization offsets across simulation samples, discussed in \S\ref{sec:mpr-star-bh-norm}, the scatter in individual properties (Figures~\ref{fig:stellar-scatters}, \ref{fig:N-sat-kllr-params} and \ref{fig:vel-disp-kllr-params}) often shows good quantitative agreement. The diagonals of the halo mass-conditioned covariance, $C_{a,a}(\mu,z)$, thus are relatively robust to the modeling of astrophysical processes.  We now turn to addressing the off-diagonal components, $C_{a,b}(\mu,z)$, framed in terms of the correlation coefficient, Equation~\eqref{eq:KLLR-correlation}.  

Mass-conditioned correlations between halo properties provide important constraints on the physical processes governing the co-assembly and co-evolution of the key baryonic components of groups and clusters.  The correlation structure  encodes how astrophysical processes orchestrate the arrangement of hot gas, stars, black holes, and diffuse ICL.  Spatial scales are coupled in complex ways by the algorithms that express the heating, cooling, enrichment and transport associated with star formation, black hole accretion and AGN feedback.  This chemo-thermal activity plays out within a background of dynamics driven largely by gravitation, including mergers and associated shocks and tidal interactions. 

In Paper I, we examined the covariance structure of hot gas observables in TNG and FLAMINGO halo samples, finding strong mass-conditioned correlations among X-ray, thermal, and gas mass properties, with broad agreement across simulations.  By and large, the mainly positive correlations reflect basic physics: in a halo of fixed total mass, more hot gas yields both a higher thermal pressure and increased X-ray luminosity.  Similarly, the correlation of hot gas temperature with thermal pressure is also high, while temperature is only weakly correlated with the amount of hot gas.  

While a few combinations of the stellar properties treated here may have simple expectations---for example, in groups where BCG stellar mass is dominant we can expect a strongly positive correlation between BCG and total stellar masses---many properties, such as BH mass, emerge in ways that, to a large degree, are not directly dependent on the stellar properties we examine here.  

Halo assembly history may also drive correlations among stellar and galaxy properties.  Simulation studies have shown that earlier-forming halos tend to host more massive central galaxies, larger stellar mass gaps, and fewer surviving satellites \citep{Deason2013, Zehavi2018, Artale2018, Contreras2019, Anbajagane2020stellar, Farahi2020}.

Our treatment of correlations is not intended to be comprehensive, since 78 unique property pairs arise from joining the five gas properties of Paper I with the eight core properties of this work (Table~\ref{tab:prop-definitions}).  We structure this section by first presenting a somewhat surprising result: the consistent simulation views of how the amount of ICL correlates with other stellar and BH properties (\S\ref{sec:ICL-corrs}).  Given that hot gas is a key property from Paper I, we analyze correlations of stellar properties with $\Mgas$ in \S\ref{sec:star-gas-corrs}. We close with an examination of the stellar and BH correlations in the five-dimensional space of $\{ \Mstartot, \MstarBCG, \Mbh, \Mstarsat, \Nsat\}$.  
As in Paper I, we overplot the correlation coefficient trends from all simulations in a single panel. For clarity, we show only $z=0$ and $z=1$, while providing the data for all other redshifts in our repository.  

\begin{figure}
    \centering
    \includegraphics[width = 0.6\linewidth]{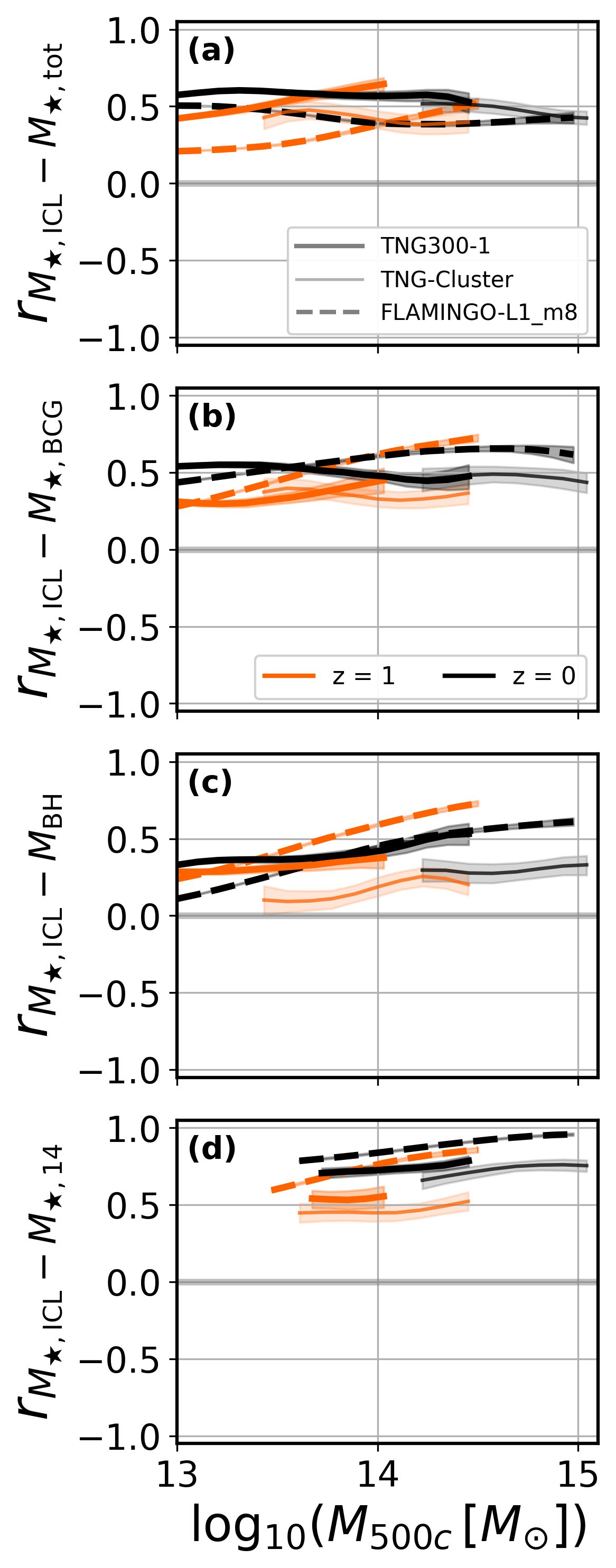}
    \caption{Mass-conditioned correlations of the intracluster light stellar mass with (a) total stellar mass, (b) BCG stellar mass, and (c) the stellar mass gap between the first and fourth stellar mass-ranked galaxies for TNG300-1 (solid lines), TNG-Cluster (thin solid lines), and {\FLAMLm} (dashed lines) at redshifts $z = 0$ (black) and $z = 1$ (orange).
}
    \label{fig:M-star-ICL-correlations}
\end{figure}

%%%%%%%%%%%%%%%%%%%%%%%%%%%%%%%%%%%%%%%%%%%%%%%%%%%%%%%%%%%%%%%%%%%%%%%%%%%%%%%%

\subsection{ICL stellar mass correlations}\label{sec:ICL-corrs}

The intracluster light, built from stars stripped during mergers and tidal interactions, is expected to correlate with satellite disruption, BCG growth, and the redistribution of baryons within the halo \citep{Contini2014, Joo2023}. Both simulations and observations suggest that the ICL retains memory of halo assembly history and the dynamical state of clusters \citep{DeMaio2015, MontesTrujillo2018, DeMaio2020, Contini2021}. 
While the slope and scatter of the ICL mass scaling relation are similar in FLAMINGO and TNG simulations, the normalization of the former is a factor 2--3 larger than the latter, reflecting the overall larger star formation efficiency in FLAMINGO halos seen in Figure~\ref{fig:stellar-norms}.  

Despite this difference, Figure~\ref{fig:M-star-ICL-correlations} shows good agreement in the correlations of ICL stellar mass with four other properties.  In all cases, halo mass and redshift dependencies are weak, and the three simulation samples provide moderately consistent numerical values.  

In particular, panel (d) of Figure~\ref{fig:M-star-ICL-correlations} displays a very strong positive correlation between $\MstarICL$ and the stellar mass gap $\Mstargap$, with the correlation in FLAMINGO increasing with halo mass and approaching near-degeneracy at $z=0$ for the most massive systems. In FLAMINGO, this correlation exhibits only weak redshift dependence, whereas both TNG simulations show a stronger evolution with redshift, with good agreement between TNG300-1 and TNG-Cluster. The stellar mass gap is closely related to the observational “magnitude gap,” which has been widely used as a proxy for halo formation time \citep{Jones2003, DOnghia2005, Dariush2010, Deason2013, GoldenMarx2018, Vitorelli2018, Farahi2020}. In this context, the strong $\MstarICL$--$\Mstargap$ correlation---consistent with the positive correlation between the magnitude gap and BCG+ICL stellar mass found by \citet{GoldenMarx2025}, as well as that between the magnitude gap and ICL fraction found by \citet{MontenegroTaborda2025}---suggests that both quantities are linked to the formation time of the host halo, reflecting the same underlying process of hierarchical satellite processing. A large $\Mstargap$ arises when an overly massive central galaxy is built through merger-driven growth that simultaneously depletes the number of available massive satellites.

Panel (b) shows a strong correlation between ICL and BCG stellar masses, with typical correlation coefficients of $0.4$--$0.7$. Part of this trend may arise because stellar particles originating from the outskirts of the BCG beyond the adopted $100\kpc$ aperture contribute directly to the ICL component. Since both the BCG and ICL are primarily assembled through hierarchical accretion and satellite disruption \citep{DeLucia2007, Contini2014, Contini2018}, the positive correlation reflects a common dependence on halo assembly history.

While $\MstarICL$ constitutes a minority contributor to $\Mstartot$ (see Appendix~\ref{sec:Mstartot-discrepancy}), panel (a) shows that the two quantities exhibit a strong positive correlation of ${\sim} 0.5$ that is approximately mass independent at $z=0$ in all simulations.

\begin{figure*}
    \centering
    \includegraphics[width = 0.9\textwidth]{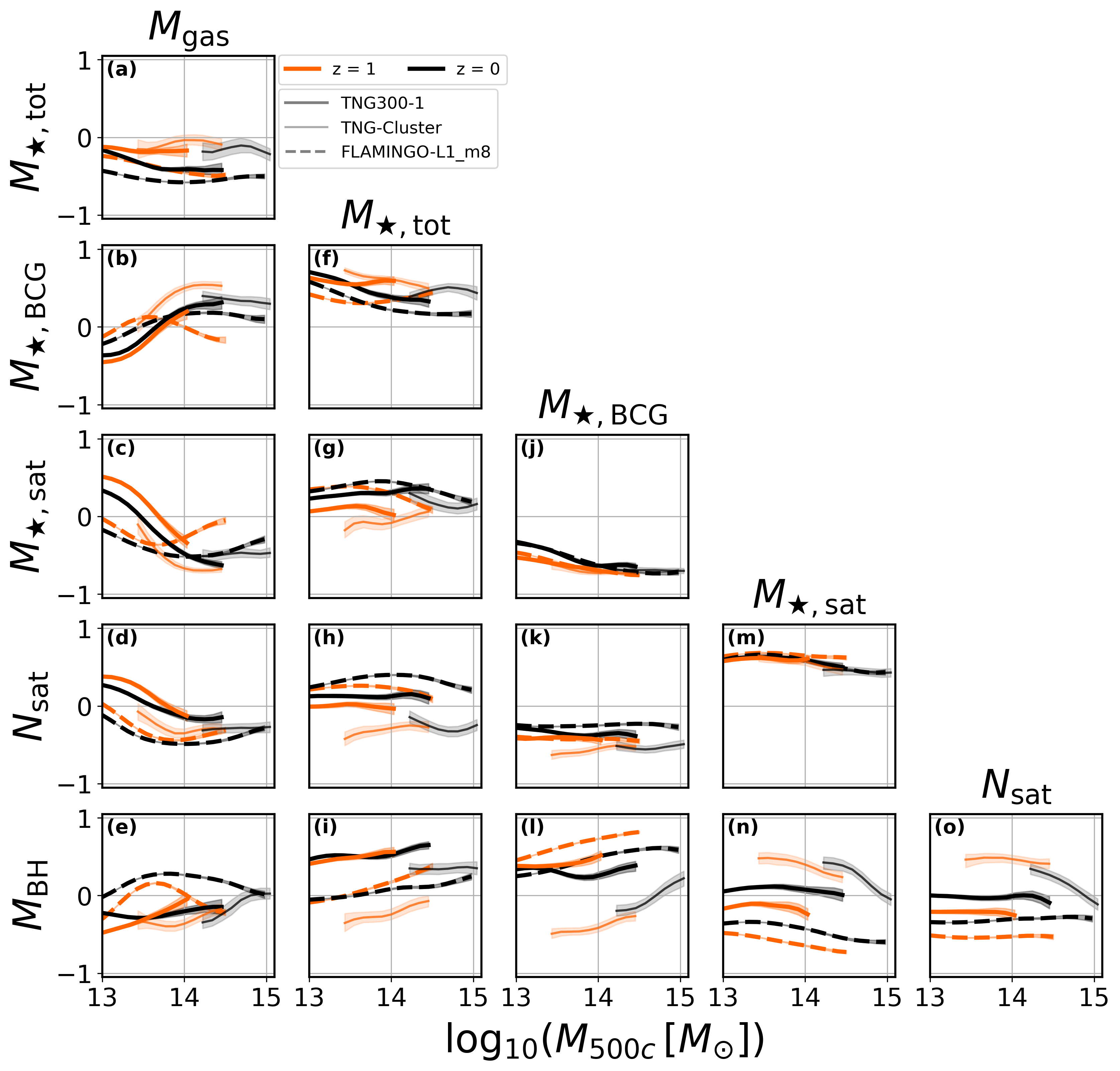}
    \caption{Mass-conditioned correlations for hot gas mass $\Mgas$ (a - e), total stellar mass $\Mstartot$ (f - i), BCG stellar mass $\MstarBCG$ (j - l), and total satellite stellar mass $\Mstarsat$ (m - n) and satellite occupation $\Nsat$ (o) for TNG300-1 (solid lines), TNG-Cluster (thin solid lines), and {\FLAMLm} (dashed lines) at redshifts $z = 0$ (black) and $z = 1$ (red). 
    }
    \label{fig:6-prop-corrs}
\end{figure*}

By contrast, the $\MstarICL-\Mbh$ correlation in panel (c) is relatively weak in lower-mass halos but strengthens toward higher halo masses, particularly in FLAMINGO. In TNG300, the correlation is quantitatively consistent with FLAMINGO, but the mass dependence is weaker, while in TNG-Cluster the correlation is weaker than in both TNG300-1 and FLAMINGO and remains approximately independent of halo mass. 

The consistency of expectations from different simulations motivates measurement in observed samples.  The especially strong correlation between the ICL mass and the magnitude gap should be testable by current and future Euclid and LSST observations, as well as existing low-redshift samples such as LoVoCCS \citep{Fu2022LoVoCCS}.    

%%%%%%%%%%%%%%%%%%%%%%%%%%%%%%%%%%%%%%%%%%%%%%%%%%%%%%%%%
\subsection{Hot gas mass correlations}\label{sec:star-gas-corrs}

We next explore how the amount of hot gas, $\Mgas$, correlates with various stellar properties and total BH mass as a function of system total mass.  The measurements are displayed in the left column of Figure~\ref{fig:6-prop-corrs}, panels (a) through (e).  The remaining panels of this figure are discussed in the next section.  

\subsubsection{Gas and bulk stellar measures}
Panels (a), (c) and (d) of Figure~\ref{fig:6-prop-corrs} show mass-dependent trends of the correlation between hot gas mass and three related measures of the overall stellar content of the halo: $\Mstartot$, $\Mstarsat$, and $\Nsat$, respectively.  At cluster mass scales $>10^{14} \msol$, generally negative values provide support for a simple ``closed box'' interpretation: more stars within $\Rfivehc$ implies less gas, and vice-versa.  The trends at $z=0$ are consistent with those reported in previous simulations \citep{Wu2015, Farahi2018}. 

Other features at cluster mass scales introduce caution into this interpretation. The correlation with total stellar mass at $z=0$ (panel a) from TNG300-1 is marginally inconsistent with that of TNG-Cluster and is more consistent with the expectations of FLAMINGO, which takes on a value close to $-0.5$ nearly independent of mass.  Another concern is a  seeming logical inconsistency in TNG-Cluster: high mass halos retain nearly all of their cosmic baryons within $\Rfivehc$ (Figure~\ref{fig:baryon-fraction}) yet deviations around the means of gas and total stellar masses are uncorrelated, in defiance of a simple closed box model.  
It is worth noting that the magnitude of the gas mass fluctuations in physical units is a factor four\footnote{At high halo masses for TNG-Cluster, the fractional scatter in gas is $0.05$ while it is larger, $0.1$, for stellar mass.  But the gas mass is roughly eight times larger than the stellar mass (Appendix~\ref{sec:fgas-fstar}), hence absolute density fluctuations in gas are a factor four larger than stellar mass fluctuations.} times larger than that of stellar mass.  

At the group scale below $10^{14} \msol$, FLAMINGO halos continue to display anticorrelations between hot gas mass and all of the bulk stellar measures, though often with smaller magnitude relative to cluster scales.  Given the low  baryon fractions in groups, this result does not reflect a ``closed box'' interpretation within $\Rfivehc$.

Group-scale halos in TNG300-1 display complex behaviors, with total stellar mass, $\Mstartot$, and satellite galaxy count, $\Nsat$, weakly correlated with gas mass, while satellite stellar mass, $\Mstarsat$, is more strongly positively correlated for halos below ${\sim} 10^{13.5} \msol$. The overall baryon fraction in these halos is about half the cosmic value, and there is significant scatter in both stellar and gas components. The positive correlation may reflect a small population of recently formed systems built from two or more gas-rich systems with reasonably high central stellar masses.

\subsubsection{Gas with BCG stellar mass and BH mass}
Panel (b) of Figure~\ref{fig:6-prop-corrs} illustrates a complex and generally weak correlation between gas mass and BCG stellar mass.  In FLAMINGO, the correlation amplitude lies below 0.2 at all mass scales and both redshifts, with a positive correlation in the cluster regime at $z=0$.  That a larger BCG stellar mass implies larger gas mass is also seen in both TNG-Cluster and TNG300-1 samples at cluster scales.  At group scales, TNG300-1 demonstrates the opposite trend below ${\sim} 10^{13.5} \msol$.  For these low mass groups, a halo of fixed mass with a bigger BCG contains less gas within $\Rfivehc$.  

The correlations of total BH mass with gas mass, panel (e), somewhat mimic the behavior seen for BCG stellar mass. In FLAMINGO halos, a higher BH mass generally implies more hot gas at $z=0$, but this trend does not persist to $z=1$. This is consistent with the positive correlation between central BH mass and gas fraction at fixed halo mass found in FLAMINGO by \citet{Costello2026}, which they attribute to halo formation history: halos that form earlier grow their BHs earlier, expel gas earlier, and consequently have more time for the gas to re-accrete into the halo. In the TNG samples, this correlation tends to be weakly negative, similar to the correlation between BH mass and gas fraction found in Magneticum \citep{Marini2025}, which employs a two-mode AGN feedback model similar to TNG. As discussed by \citet{Costello2026}, the difference between the Magneticum and FLAMINGO correlations may reflect differences in the timing and efficiency of AGN feedback, with later feedback in Magneticum potentially allowing less time for previously expelled gas to re-accrete. The precise origin of this difference, however, remains unclear. More broadly, as discussed below, other correlations also show no consensus across the simulations in how total BH mass correlates with other stellar properties.

%%%%%%%%%%%%%%%%%%%%%%%%%%%%%%%%%%%%%%%%%%%%%%%%%%%%%%%%%%%%%
\subsection{Stellar, BH, and Satellite correlations}\label{sec:stellarCorrelations}

We now turn to the correlation structure of the total stellar mass and its constituent components, together with the total BH mass and satellite occupation, shown in panels (f) through (o) of Figure~\ref{fig:6-prop-corrs}. We organize the discussion into correlations that show robust, simulation-independent trends (\S\ref{sec:corr-universal}) and those that exhibit significant differences between simulations (\S\ref{sec:corr-divergent}). 

\subsubsection{Robust behaviors}\label{sec:corr-universal}

Like the case of hot gas properties in Paper I, several property pairs have clear mathematical relationships that engender simple expectations for their statistical correlation.  For example, panel (m) shows that the stellar mass in satellites, $\Mstarsat$, is strongly positively correlated with satellite count, $\Nsat$.  All simulations see a very positive correlation, with the value gradually declining from values of 0.8 at group scales to 0.5 for massive clusters.  The departure from unity suggests that there are non-trivial shifts in the shape of the S-GSMF on a per halo basis.  

Another simple expectation is that satellite accretion drives up BCG stellar mass growth at the expense of satellite metrics \citep{Cui2014, Bahe2017, Tremmel2019, Bahe2019, Anbajagane2020stellar, MontenegroTaborda2025}. The $\MstarBCG$ and $\Mstarsat$ behavior, panel (j), demonstrates a strong, consistent anti-correlation at cluster scales, with a value of $-0.8$ weakening toward $-0.5$ at group scales.  The consistent magnitude of this anti-correlation is perhaps surprising given the different fractional contributions of central and satellite contributions to the total stellar mass discussed in Appendix~\ref{sec:Mstartot-discrepancy}.  The behavior with satellite count, panel (k), also demonstrates anti-correlation, but at smaller magnitude, particularly for FLAMINGO at $z=0$.  

Finally, the BCG and total stellar mass correlation, panel (f) is largely positive because the BCGs contribute a significant fraction of the overall stellar mass. For example, even at the cluster scale of $\Mfivehc = 10^{14.5}\msun$, the BCG contributes approximately $30\%$, $36\%$, and $21\%$ of the total stellar mass in TNG300, TNG-Cluster, and FLAMINGO at $z=0$, respectively (see Figure~\ref{fig:Mstartot-components-scaling-fractions}). The correlation strength increases toward lower halo masses, as expected from the higher fractional contribution of the BCG to the total stellar mass of the halo.

\subsubsection{Inconsistent behaviors}\label{sec:corr-divergent}

Several pairs of stellar property display inconsistent correlations across samples, with TNG-Cluster halos frequently displaying behaviors different from TNG300-1 and FLAMINGO populations.  For example, given that $\Mstarsat$ contributes linearly to $\Mstartot$, one would anticipate a positive correlation for this pair, as well as for the combination of $\Nsat$ and $\Mstartot$.  Panels (g) and (h) show that positive behavior is seen in both TNG300-1 and FLAMINGO populations, though the former is weaker than the latter. 

In TNG-Cluster, however, the $\Mstartot$--$\Nsat$ correlation is consistently negative, meaning that halos of fixed mass with above average total stars have \textit{fewer} satellite galaxies than average.  Yet by $z=0$, such halos have above average values of the stars contained in satellite galaxies.  The differing shapes of the GSMF in high mass halos (Figure~\ref{fig:S-GSMF}) and the discrepant BCG and total stellar mass scalings in the two TNG realizations (Figure~\ref{fig:stellar-norms}) conspire in this anomaly. These differences ultimately stem from a shift in the parameter controlling the star formation efficiency at high gas densities. 

In TNG-Cluster, the $\Nsat$ anti-correlation with $\Mstartot$ weakens in amplitude as the satellite galaxy mass threshold is raised, and changes sign for satellites with stellar mass above $10^{11} \msol$. We further note that a similar weak anti-correlation between $\Mstartot$ and $\Nsat$ was also found in the Magneticum simulation by \citet{Anbajagane2020stellar}.  Similar to the case of TNG-Cluster, that model also produced very massive BCGs in high mass halos. 

Correlations involving total BH mass, shown in the bottom row of Figure~\ref{fig:6-prop-corrs}, are generally inconsistent across the three simulations.  One factor complicating interpretation of correlations is the fact that the simulations show different normalizations in the $\Mbh$--$\Mfivehc$ scaling relation, most notably between TNG300-1 and TNG-Cluster (discussed in Section~\ref{sec:mpr-star-bh-norm}), in which similar seeding and feedback prescriptions are applied but star formation efficiency differs.

Panel (i) shows that the $\Mstartot$–$\Mbh$ correlation is strongly positive in TNG300-1 but extremely weak in FLAMINGO, with only mild redshift dependence. In TNG-Cluster, the correlation shifts from mildly negative at $z=1$ to mildly positive at $z=0$.  

The $\MstarBCG$--$\Mbh$ correlation, panel (l), also shows that TNG-Cluster halos exhibit behaviors different from those of TNG300-1 and FLAMINGO, particularly at $z=1$ where total BH mass is anti-correlated with BCG stellar mass.  However, the correlation between \emph{central} BH mass, meaning the BH contained within the central galaxy, and BCG stellar mass displays much better agreement between the simulations, with consistently positive correlation values. Details are provided in Appendix~\ref{sec:cent-BH-corrs}. This positive correlation between $\MstarBCG$ and $\Mbhcent$ is expected. A halo with an oversized central BH mass likely experienced enhanced BH growth through a combination of gas accretion and BH mergers, processes that are themselves linked to the assembly history of the central galaxy. 

How total BH mass correlates with satellite measures is complex. In all simulations, $\Mstarsat$ and $\Nsat$ are strongly correlated, so behaviors are similar in panels (n) and (o). In FLAMINGO halos, both correlations are negative and stronger at high redshift. TNG300-1 halos exhibit weak anti-correlations at $z=1$ that transition toward zero or weakly positive at $z=0$.  TNG-Cluster shows positive correlations that decrease rapidly with halo mass at $z=0$ to values near zero.  

These correlations can be considered in terms of the previously discussed $\MstarBCG$--$\Mbh$ and $\MstarBCG$--$\Mstarsat$ correlations. In TNG300-1 and FLAMINGO, halos with larger total BH masses tend to host more massive BCGs, while the robust anti-correlation between $\MstarBCG$ and $\Mstarsat$ implies that these systems preferentially contain less stellar mass and fewer satellite galaxies. In TNG-Cluster, however, the $\MstarBCG$--$\Mbh$ correlation is weak or negative, reversing this trend and producing positive correlations between $\Mbh$ and both $\Mstarsat$ and $\Nsat$. Clearly more work is needed to understand the interplay between a halo's BH and stellar contents, including the role of AGN feedback in central versus satellite galaxies. 

\subsection{Satellite Galaxy Velocity Dispersion}\label{correlationDispersion}

We do not show in Figure~\ref{fig:6-prop-corrs} correlations involving the satellite galaxy velocity dispersion because we find none that are significantly different from zero.  Upper limits typically are at the level of $0.1$.  Prior TNG300-1 analysis by \citet{Anbajagane2022baryon} showed a positive correlation, with magnitude $\sim 0.5$, between $\sigmaSat$ and halo concentration for halos of mass $\Mtwohc = 10^{14} \msol$.  Somewhat surprisingly, zero correlation was found at this mass scale between $\sigmaSat$ and formation redshift, defined as the epoch at which the primary progenitor reached half the mass of the final system.  We suspect that this finding implies that the properties examined here have weak dependence on halo concentration.  

\subsection{Discussion: empirical measurements for  discrimination}\label{correlationDiscussion}

While galaxy group and cluster samples are selected using features in a chosen wavelength range (millimeter, optical/IR, or X-ray wavelengths), the improving sensitivity of astronomical observations broadly paints a bright future for multi-property measurements of massive halos in the sky.  Bayesian approaches for inferring scaling parameters (including covariance of pairs of observables) in pure power-law models that account for selection and property covariance are in place \citep{Mantz2010cosmo, Sereno2016, Mulroy2019, Mantz2019}, but applications to date have been limited to small sample sizes.   

As noted in Paper I, the Weighing the Giants sample of 139 clusters, 27 with lensing masses, provided an early measurement of intrinsic correlation between X-ray luminosity and temperature \citep{Mantz2016b}.  Extending to optical and SZ properties for $41$ X-ray-selected clusters in the LoCuSS sample, \citet{Farahi2019} and \citet{Mulroy2019} found statistically significant anti-correlations between gas and stellar properties conditioned on a cluster's mass, particularly between total $K$-band luminosity of cluster galaxies and hot gas mass. These results support the expectations shown in panel (a) of Figure~\ref{fig:6-prop-corrs}, but the measurement is not sensitive enough to distinguish between the cases of low- (TNG-Cluster) or high-amplitude (TNG300-1 and FLAMINGO) anti-correlation.

Measuring intrinsic property correlations for cluster samples remains challenging because of observational uncertainties, projection effects, and sample selection.  The correlation coefficient predictions of Figure~\ref{fig:6-prop-corrs} generally carry small statistical uncertainties.  The significant disagreements in value often seen for property pairs motivate observing teams to assemble multi-wavelength data for thousands of clusters, enabling empirical measurements with similar precision.

\begin{figure}
    \centering
    \includegraphics[width = 0.95\linewidth]{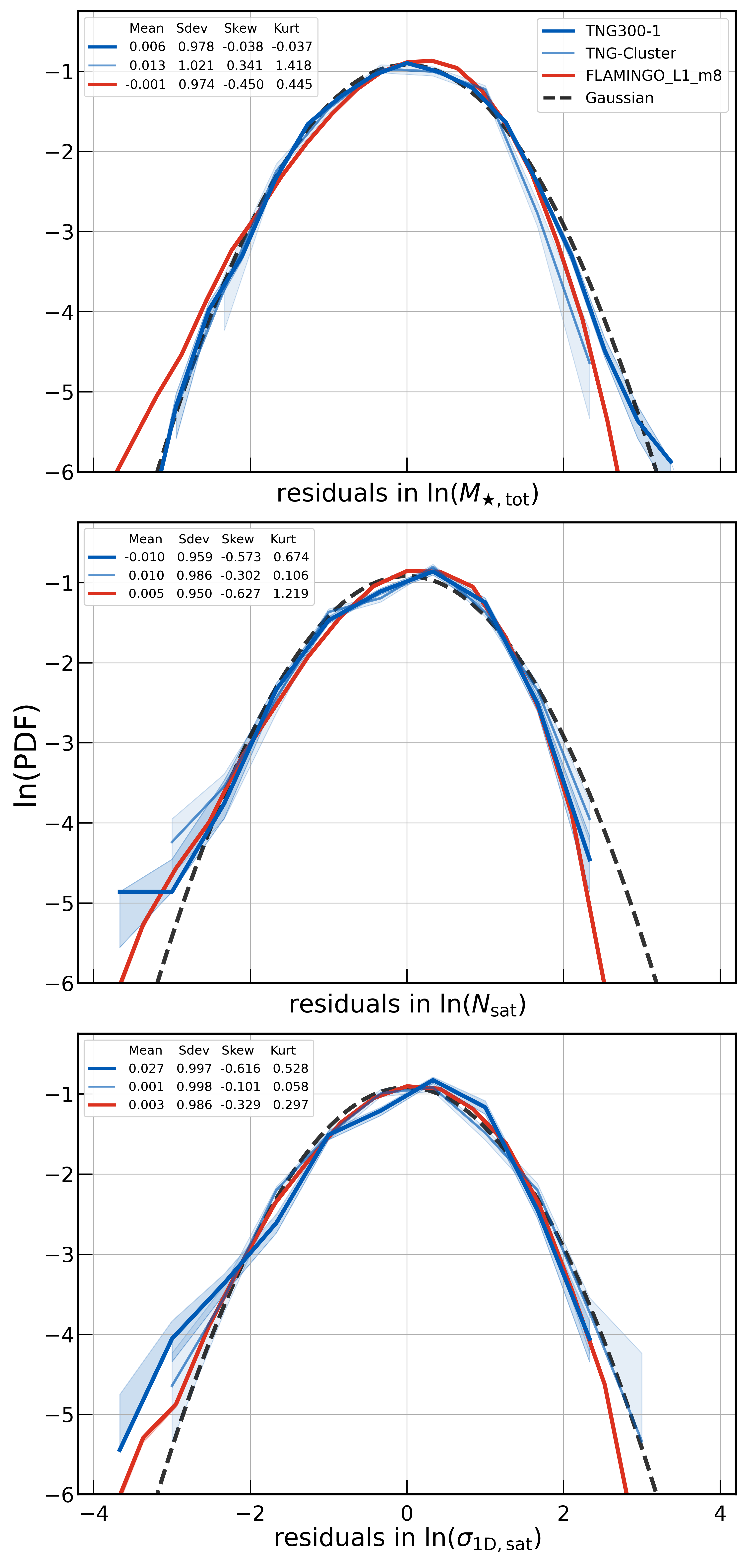}
    \caption{Probability density functions of normalized deviations from mean KLLR behavior at $z=0$, formed by marginalizing across halo mass, for $\Mstartot$ (top panel), $\Nsat$ (middle) and $\sigmaSat$ (bottom) in TNG300-1 (dark blue), TNG-Cluster (light blue) and FLAMINGO (red). The dashed black line is a reference normal distribution. A KLLR kernel width of $0.1$ dex is used to reduce finite kernel width effects on low-order moment estimates given in the panel insets.    
}
    \label{fig:M-star-N-sat-vel-disp-kernel-shapes}
\end{figure}

Pairs for which good agreement among simulations is seen, including ICL expectations of Figure~\ref{fig:M-star-ICL-correlations}, offer particularly tempting targets for empirical follow-up.  In addition, the consistently strong anti-correlation between BCG stellar mass and the mass in satellite galaxies could be leveraged to test models for satellite galaxy membership in massive clusters. 

%%%%%%%%%%%%%%%%%%%%%%%%%%%%%%%%%%%%%%%%%%%%%%%%%%%%%%%%%%%%%%%%%%%%%%%%%%%%%%%%%%
%%%%%%%%%%%%%%%%%%%%%%%%%%%%%%%%%%%%%%%%%%%%%%%%%%%%%%%%%%%%%%%%%%%%%%%%%%%%%%%%%%

\section{Conditional Likelihood Shapes}\label{sec:kernel-shapes}

\begin{figure*}
    \includegraphics[width = 0.75\textwidth]{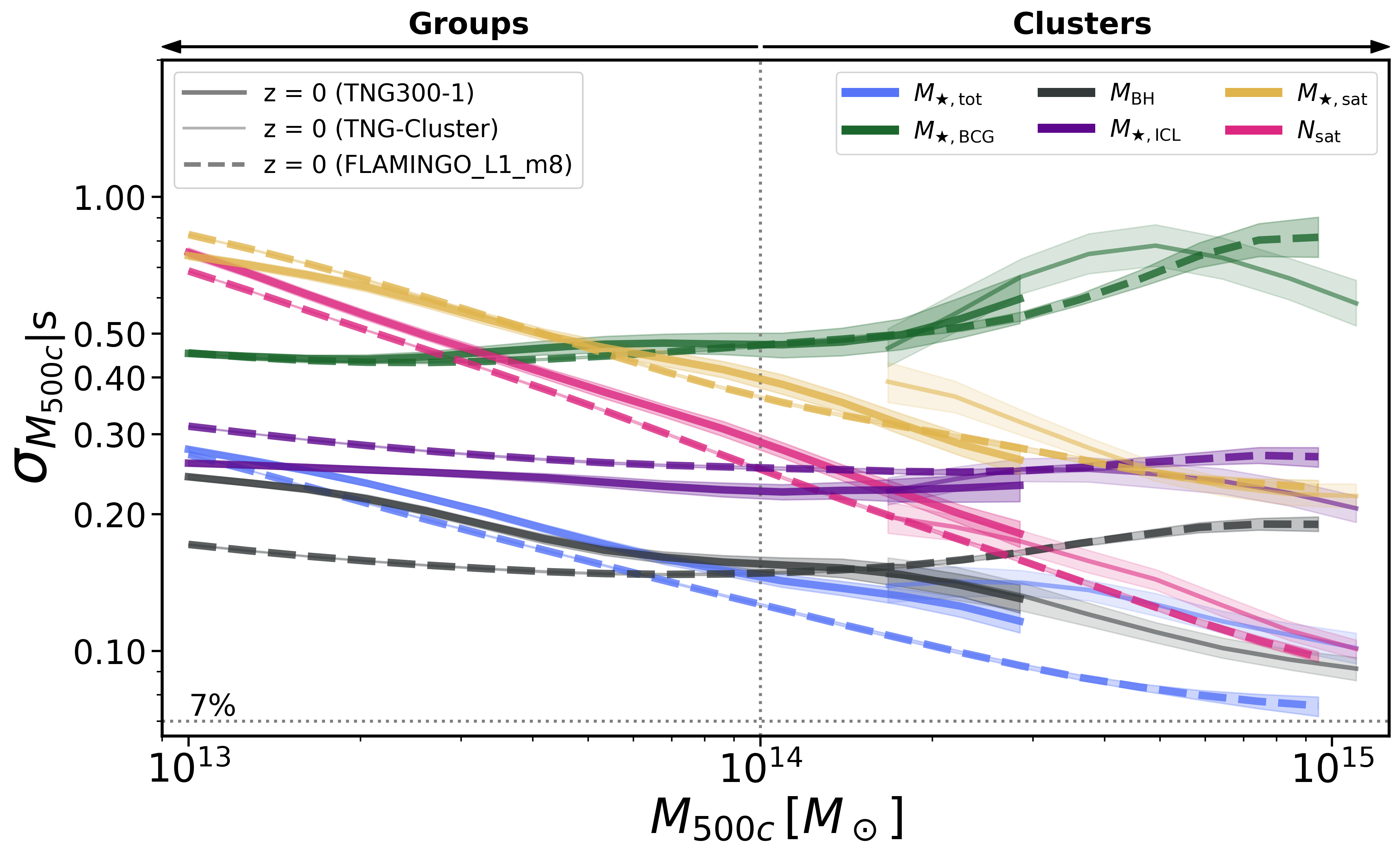}
    \caption{The mass proxy quality---the implied halo mass scatter,  equation~\eqref{eq:mass-proxy-quality}---for stellar, BH, and satellite properties at redshift $z = 0$ in TNG300-1 (thick solid lines), TNG-Cluster (thin solid lines), and {\FLAMLm} (dashed lines).  Analogous to Figure 1 of Paper 1, we approximate the division between group and cluster regimes at low redshift by a halo mass of $10^{14}\msol$. The relative ordering of the mass proxy quality is broadly consistent among the simulations, with the total stellar mass ($\Mstartot$, blue) and total BH mass ($\Mbh$, black) serving as the best mass proxies across most of the mass range. The remaining stellar, BH, and satellite properties exhibit consistency in their implied mass scatter across the simulations.  We omit $\sigmaSat$ and $\Mstargap$ for clarity; both provide $30$--$35\%$ MPQ nearly independent of mass. 
}
    \label{fig:MPQindividualzero}
\end{figure*}

The detailed form of the mass-conditioned likelihood, $\Pr(S|M,z)$, plays a central role in cluster population science \citep{Shaw2010, Erickson2011, Angulo2012, Costanzi2019, Mantz2019, Kugel2025, Aljamal2025}. Previous simulation studies have shown that the central-limit theorem expectation of a log-normal form is well satisfied for some properties, such as hot gas mass and total stellar mass, while a non-trivial amount of skewness is seen in others \citep{LeBrun2017, Truong2018, Farahi2018, Anbajagane2020stellar, Kugel2024, Chen2024}. In Paper I, we show that $\Mgas$ and $\YSZ$ closely follow log-normal likelihood shapes in all three simulation suites, whereas the soft-band X-ray luminosity exhibits a pronounced high-luminosity tail that departs from log-normality.

Normalized residuals, equation~\eqref{eq:KLLR-normalized-res}, from all halos are combined into an aggregate distribution, with each halo contributing equal weight. Because variations in the local slope and scatter, together with the declining halo mass function, can introduce small biases into the recovered likelihood shape, we compute these distributions using a narrower KLLR kernel width of $0.1$ dex.  With nearly 100 thousand halos at the present epoch, FLAMINGO best supports measurement of a distribution's tails, to nearly $\pm 4\sigma$.  The few-hundred halos in the TNG-Cluster sample offers the least support, roughly $\pm 2 \sigma$.

Figure~\ref{fig:M-star-N-sat-vel-disp-kernel-shapes} presents the normalized conditional likelihood distributions of $\Mstartot$, $\Nsat$, and $\sigmaSat$ at $z=0$.  We report the first four moments of each distribution in the panel insets, and bootstrap resampling provides the shaded uncertainty in PDF shape.  The kurtosis is defined as the excess kurtosis, such that a perfectly Gaussian distribution has a kurtosis of zero.  

All properties adhere closely to a Gaussian distribution near their peaks, but most possess a non-negligible leftward skew, with magnitudes $-0.3$ to $-0.6$ in TNG300-1 and FLAMINGO, respectively. The skewness in TNG-Cluster is slightly lower.  

The residuals in $\ln \Mstartot$ are particularly well described by a Gaussian distribution in TNG300.  The FLAMINGO and TNG-Cluster samples exhibit slight negative skewness and positive excess kurtosis. In both simulations, the high-$\Mstartot$ tail is mildly suppressed relative to a Gaussian, while FLAMINGO additionally exhibits a more extended low-$\Mstartot$ tail.  

The residuals in $\ln \Nsat$ and $\ln \sigmaSat$ show similar behavior. Both are close to Gaussian but exhibit negative skewness and positive excess kurtosis, arising from an extended low-value tail and a corresponding suppression of the high-value tail. These deviations are most pronounced in TNG300-1 and FLAMINGO, whereas TNG-Cluster remains closer to Gaussian, with smaller skewness and kurtosis in both properties.

The corresponding distributions for the remaining stellar and BH properties are shown in Appendix~\ref{sec:fgas-fstar}, where more noticeable departures from Gaussianity are evident for some observables.  Future work should address questions of causality: how do halos occupying the low-value tails differ in formation history or galaxy/gas assembly from a typical halo? 

%%%%%%%%%%%%%%%%%%%%%%%%%%%%%%%%%%%%%%%%%%%%%%%%%%%%%%%%%%%%%%
%%%%%%%%%%%%%%%%%%%%%%%%%%%%%%%%%%%%%%%%%%%%%%%%%%%%%%%%%%%%%%
\section{Mass Proxy Quality}\label{sec:MPQ}

We close by returning to the opening topic of Paper I, the mass proxy quality. For a given property, the conditional scatter in true halo mass generally depends on property magnitude (primarily) and redshift (secondarily). In Paper I, we find that thermal energy ($\YSZ$) and hot gas mass ($\Mgas$) are the highest quality mass proxies at the cluster scale, the latter reaching a scatter of only ${\sim}4\%$ on a per-halo basis.  We present here results for stellar and BH properties at $z = 0$ here; behavior at higher redshifts is provided in Appendix~\ref{sec:MPQhighz}.  In Paper III, we provide a full summary of our findings for all properties, including the halo mass precision achieved by joining multiple properties. 

The mass scatter conditioned on single property $s_a$ (recall $\mu \propto \ln M$) is
\begin{equation}
\sigma_{\mu \vert a}(\mu, z) = \frac{\sigma_a(\mu, z)}{|\alpha_a(\mu, z)|} .
\label{eq:mass-proxy-quality}
\end{equation}
where $\alpha_a$ and $\sigma_a$ are the local MPR slope and scatter, respectively. The explicit scale dependence in the above is expressed in terms of total mass, but it could also be expressed in terms of some other property using the appropriate MPR.

Figure~\ref{fig:MPQindividualzero} presents KLLR estimates of the logarithmic halo mass scatter for all but two of the properties listed in Table~\ref{tab:prop-definitions}.  We omit $\sigmaSat$ and $\Mstargap$ for clarity; these properties both have nearly constant MPQ values with amplitudes in the range $0.30$ to $0.35$.  

A striking result is the high degree of agreement among the three simulation suites in the mass proxy quality of many stellar and BH properties, despite their differences in MPR normalizations. The best MPQs, the total stellar mass and the total BH mass, are exceptions that exhibit some quantitative disagreement between the TNG and FLAMINGO expectations.  For $\Mstartot$, the mass scatter is slightly lower in FLAMINGO than in both TNG300-1 and TNG-Cluster, reaching ${\sim} 7\%$, compared to ${\sim} 10\%$ in TNG-Cluster at the cluster scale.

The MPQ of the satellite occupation, $\Nsat$, exhibits strong mass dependence due to the near-Poissonian behavior of its scatter.  It becomes competitive with $\Mstartot$ in the most massive halos, reaching values near $9\%$.

The remaining stellar properties also display consistency. The mass proxy quality of $\MstarICL$ is nearly independent of halo mass over the full two decades in mass considered, with a scatter of $25$-$30\%$, while $\MstarBCG$ remains approximately constant at the group scale before increasing by nearly a factor of two toward cluster masses in both TNG-Cluster and FLAMINGO. 

Finally, the MPQ of the total mass in black holes exhibits some degree of divergence. At the group scale, FLAMINGO predicts a lower mass scatter than that of the TNG300-1 sample. The total BH mass achieves the lowest mass scatter in both TNG300-1 and FLAMINGO for low-mass groups. At the cluster scale, however, the mass scatter in $\Mbh$ becomes comparable to $\Mstartot$ in both TNG300-1 and TNG-Cluster samples, while $\Mstartot$ emerges as the superior mass proxy in FLAMINGO. 

%%%%%%%%%%%%%%%%%%%%%%%%%%%%%%%%%%%%%%%%%%%%%%%%%%%%%%%%%%%%%%%%%%%%%%%%%%%%%%%%%%%%%%%
%%%%%%%%%%%%%%%%%%%%%%%%%%%%%%%%%%%%%%%%%%%%%%%%%%%%%%%%%%%%%%%%%%%%%%%%%%%%%%%%%%%%%%%

\section{Conclusions}\label{sec:conclusion}

This is the second in a series of papers that uses halo samples from modern cosmological simulations to investigate the statistical likelihoods of bulk intrinsic properties of groups and clusters of galaxies.  Under a locally power-law and log-normal conditional likelihood assumption, the scatter in true halo mass at fixed property, which we refer to as the mass-proxy quality (MPQ), can be directly estimated from KLLR-derived mass-property relations (MPRs). Paper I addressed hot gas signatures used for selecting cluster samples based on X-ray or sub-mm observations.  

In this paper, we continue investigating both MPR and MPQ scaling expectations for stellar and BH measures in massive ($\mfiveh \ge 10^{13} \msol$) halo samples identified in TNG300, TNG-Cluster and FLAMINGO simulations, at multiple redshifts ranging from zero to two.

\begin{itemize}
    \item The normalizations of stellar properties such as total and BCG stellar mass differ more dramatically across simulations than is the case for gas properties.  Differences in overall baryon fractions of $10$ to $20\%$ in clusters (Figure~\ref{fig:baryon-fraction}) are found. Critically assessing this difference empirically requires systematic error control at the few percent level for mean population estimates.

    \item As in the hot gas case, we find that the mass-dependent slopes, property variance, and implied mass proxy quality are generally in good qualitative and quantitative agreement across the simulated samples.  This concordance could motivate changes to survey data analysis, such as imposing simulation-informed priors on slopes and (co-)variances \citep{Vikhlinin2009, Benson2013, deHaan2016SPT, Farahi2016, Chiu2023eFEDScosmo, Zhang2024ImpactCovWL}, and be used to improve forward modeling approaches for simulation-based inference \citep{Cerardi2025, Rose2025DREAMS, Regamey2026}.

    \item For many properties, residuals follow a log-normal shape in the conditional likelihood.  In the largest sample of $94,000$ halos from FLAMINGO at $z=0$, a mildly negative skew and positive kurtosis are often seen.  In all samples, the BCG stellar mass likelihood demonstrates significant negative skewness, a finding previously reported for TNG300-1 and the Magneticum simulations by \citet{Anbajagane2020stellar}.   

    \item  The scaling relations of satellite galaxy velocity dispersion are in excellent agreement across the samples. All simulations show local slopes slightly above the self-similar value of $1/3$ with mass-independent scatter of $10\%$ that increases mildly with redshift.  

    \item The correlation structure of property deviations about their mean behaviors is generally complex but there are some common expectations, including ICL correlations with other stellar properties (Figure~\ref{fig:M-star-ICL-correlations}), a strong anti-correlation between BCG stellar mass and the stellar mass and counts of satellite galaxies, and anti-correlated behavior of stellar mass measures with hot gas mass (Figure~\ref{fig:6-prop-corrs}) in clusters.  

    \item A parameter change enhancing the star formation rate at high densities in TNG-Cluster relative to TNG300-1 creates significant statistical shifts. Stellar and BH normalizations shift in opposite directions, the shape of the galaxy stellar mass function is distorted toward higher-mass galaxies in TNG-Cluster, and several terms in the property correlation matrix change sign. For most properties, however, the local slopes and scatter remain consistent, implying similar MPQ expectations (Figure~\ref{fig:MPQindividualzero}). Future work is needed to understand how enhanced star formation effectively causes reduced black hole growth and re-partitions the stellar mass budget among the BCG, satellite, and ICL components in the TNG model framework.

     \item At cluster scales, the total stellar mass of a halo, the number of satellite galaxies, and the total BH mass are the best proxies for total mass, with halo mass precision between 7 and 15\%. 
\end{itemize}

These conclusions should be interpreted in light of several limitations. We consider only two hydrodynamical simulation frameworks with different subgrid, feedback, and subhalo-finding models. (A fourth paper in this series will extend our analysis to the physics variations provided by the lower-resolution FLAMINGO-L1\_m9 simulation.)
Our sample sizes are also limited, particularly in TNG300, which contains few halos with $\Mfivehc > 10^{14.5}\msun$, and in TNG-Cluster, which contains only $352$ halos in total. These limitations restrict the mass range of the analysis and the dynamic range of the conditional likelihood distributions. Finally, our analysis is restricted to four discrete redshifts, $z \leq 2$, and therefore does not capture the continuous evolution between these epochs. 

Most importantly, the properties we consider are entirely intrinsic, measured with no noise within a three-dimensional $\Rfivehc$ aperture.  Future work could focus on equivalent observable measures that incorporate projection effects, non-cluster sources of confusion, and real-world telescope/instrument characteristics.  The intrinsic findings we present here should be considered as ground truth for population analysis based on such synthetic observables.  

\section*{Acknowledgments}
EA and AEE are grateful for support from the National Aeronautics and Space Administration Data Analysis Program (NASA-80NSSC22K0476), as well as the Leinweber Institute for Theoretical Physics. AF acknowledges support from the National Science Foundation under Cooperative Agreement 2421782 and the Simons Foundation award MPS-AI-00010515. DN acknowledges funding from the Deutsche Forschungsgemeinschaft (DFG) through an Emmy Noether Research Group (grant number NE 2441/1-1). We thank Annalisa Pillepich, Joop Schaye, John Helly, Roi Kugel, Matthieu Schaller, Megan Donahue, Ray Wang, Ian McCarthy, Emily Costello, Daisuke Nagai, Elena Rasia, Volker Springel, and Mark Voit for useful discussions. This work used the DiRAC@Durham facility, managed by the Institute for Computational Cosmology on behalf of the STFC DiRAC HPC Facility (\url{www.dirac.ac.uk}). The equipment was funded by BEIS capital funding via STFC capital grants ST/K00042X/1, ST/P002293/1, ST/R002371/1, and ST/S002502/1, Durham University, and STFC operations grant ST/R000832/1. DiRAC is part of the National e-Infrastructure. Computational resources were additionally provided by the Max Planck Computing and Data Facility (MPCDF) of the Max Planck Society (\url{https://www.mpcdf.mpg.de/}).

All analysis in this work was enabled by the following {\sc{Python}} packages:{\sc{illustris\_python}} (\url{https://github.com/illustristng/illustris_python}), {\sc{numpy}} \citep{harris2020array}, {\sc{pandas}} \citep{Mckinney2011}, {\sc{scipy}} \citep{2020SciPy-NMeth}. {\sc{sci-kit Learn}} \citep{scikit-learn}, {\sc{pymc}} \citep{pymc2023}, {\sc{h5py}} 
\citep{collette_python_hdf5_2014}, {\sc{joblib}} (\url{https://github.com/joblib/joblib}), {\sc{swiftsimio}} \citep{Borrow2020}, and {\sc{matplotlib}} \citep{Hunter:2007}.

%%%%%%%%%%%%%%%%%%%%%%%%%%%%%%%%%%%%%%%%%%%%%%%%%%

\section*{Data Availability}
The data calatalogues used in this work, and the code used to analyze it, will be available on Eddie Aljamal's github: \texttt{https://github.com/ealjamal/MPQScaling} \href{https://github.com/ealjamal/MPQScaling}{\faGitSquare}.



\bibliographystyle{mnras}
\bibliography{MPQ-star-properties} % if your bibtex file is called example.bib

%%%%%%%%%%%%%%%%%%%%%%%%%%%%%%%%%%%%%%%%%%%%%%%%%%

%%%%%%%%%%%%%%%%% APPENDICES %%%%%%%%%%%%%%%%%%%%%

\appendix

\section{Total Satellite Stellar Mass}\label{M-star-sat-MPRs}
\begin{figure}
    \centering
    \includegraphics[width = \linewidth]{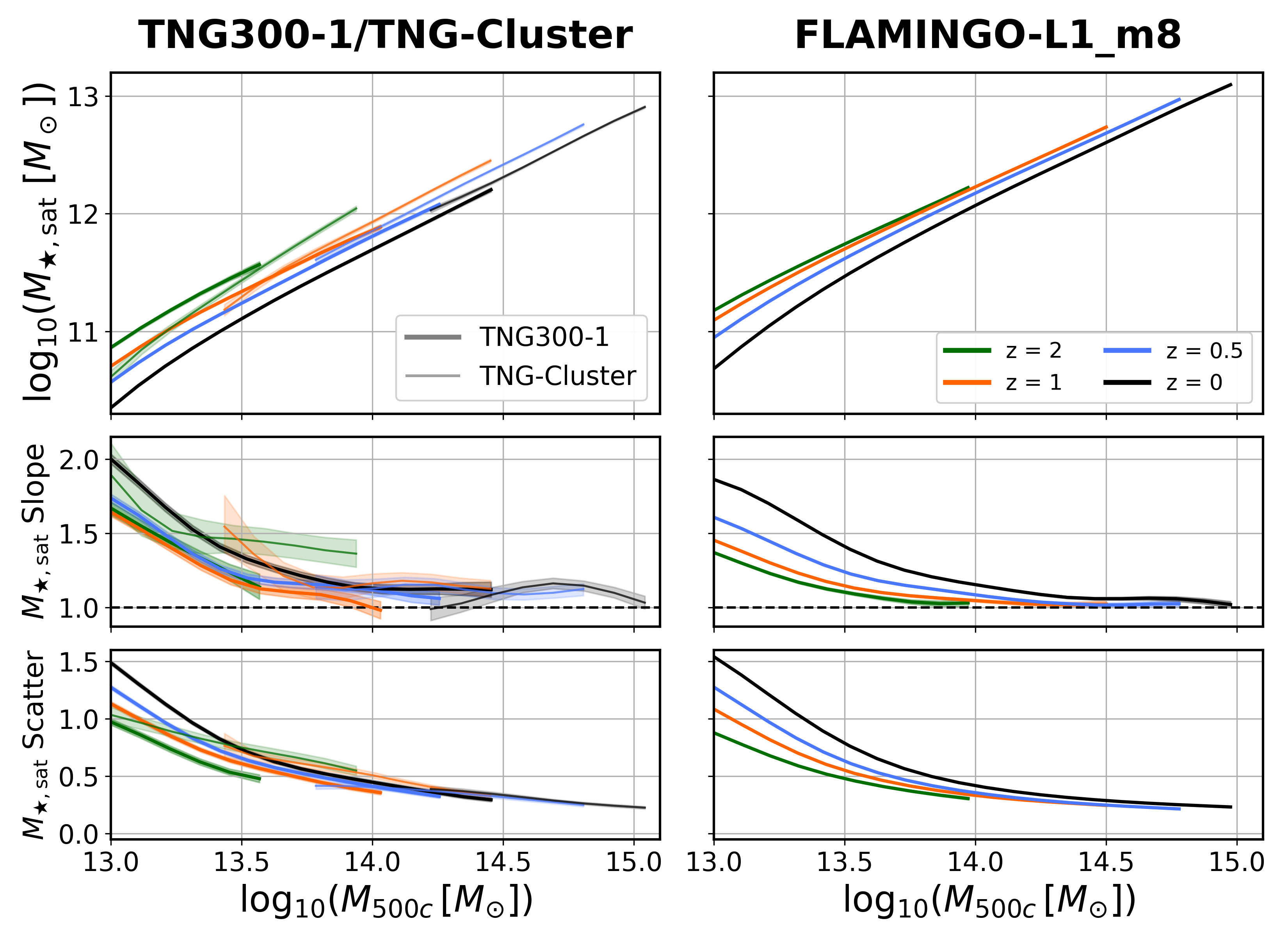}
    \caption{The Total satellite stellar mass scaling parameters: normalization (top panel), slope (middle), and intrinsic scatter (bottom). The format is identical to that of Figure~\ref{fig:N-sat-kllr-params}.
}
    \label{fig:M-star-sat-kllr-params}
\end{figure}

In Figure~\ref{fig:M-star-sat-kllr-params}, we show the KLLR scaling of the total satellite galaxy stellar mass, $\Mstarsat$. In both simulations, the normalization increases at higher redshift, particularly for $z \geq 1$. In FLAMINGO, this strong evolution nearly cancels the redshift evolution of the other two components of the total stellar mass, $\MstarBCG$ and $\MstarICL$, resulting in an approximately redshift-independent $\Mstartot$--$\Mfivehc$ relation. At $z=0$ and $\Mfivehc = 10^{14.5}\msun$, the normalization of $\Mstarsat$ is approximately twice as large in FLAMINGO as in the TNG simulations, consistent with the larger total stellar mass in FLAMINGO. The two TNG simulations show closer agreement, with $\Mstarsat$ in TNG-Cluster approximately $14\%$ higher than in TNG300-1.

The slope of the $\Mstarsat$--$\Mfivehc$ relation is also broadly consistent across the three simulations, exhibiting a decrease from ${\sim}1.9$ for the least massive halos toward the self-similar value of unity for halos with $\Mfivehc > 10^{14}\msun$ at all redshifts. At these higher halo masses, the slope in FLAMINGO remains approximately constant at unity with little redshift dependence, whereas the lower-mass behavior exhibits more pronounced evolution. In particular, the slope decreases continuously with increasing redshift in FLAMINGO, while in the TNG simulations most of the evolution occurs between $z=0$ and $z=0.5$, after which the slope remains broadly unchanged.

The intrinsic scatter shows a similarly strong dependence on halo mass, with broad agreement among the simulations but comparatively weaker evolution with redshift. It decreases with increasing halo mass, reaching approximately $23\%$ for the most massive halos at $z=0$.

\section{Stellar mass discrepancy in TNG}\label{sec:TNG-discrepancy}
\begin{figure}
    \centering
    \includegraphics[width = \linewidth]{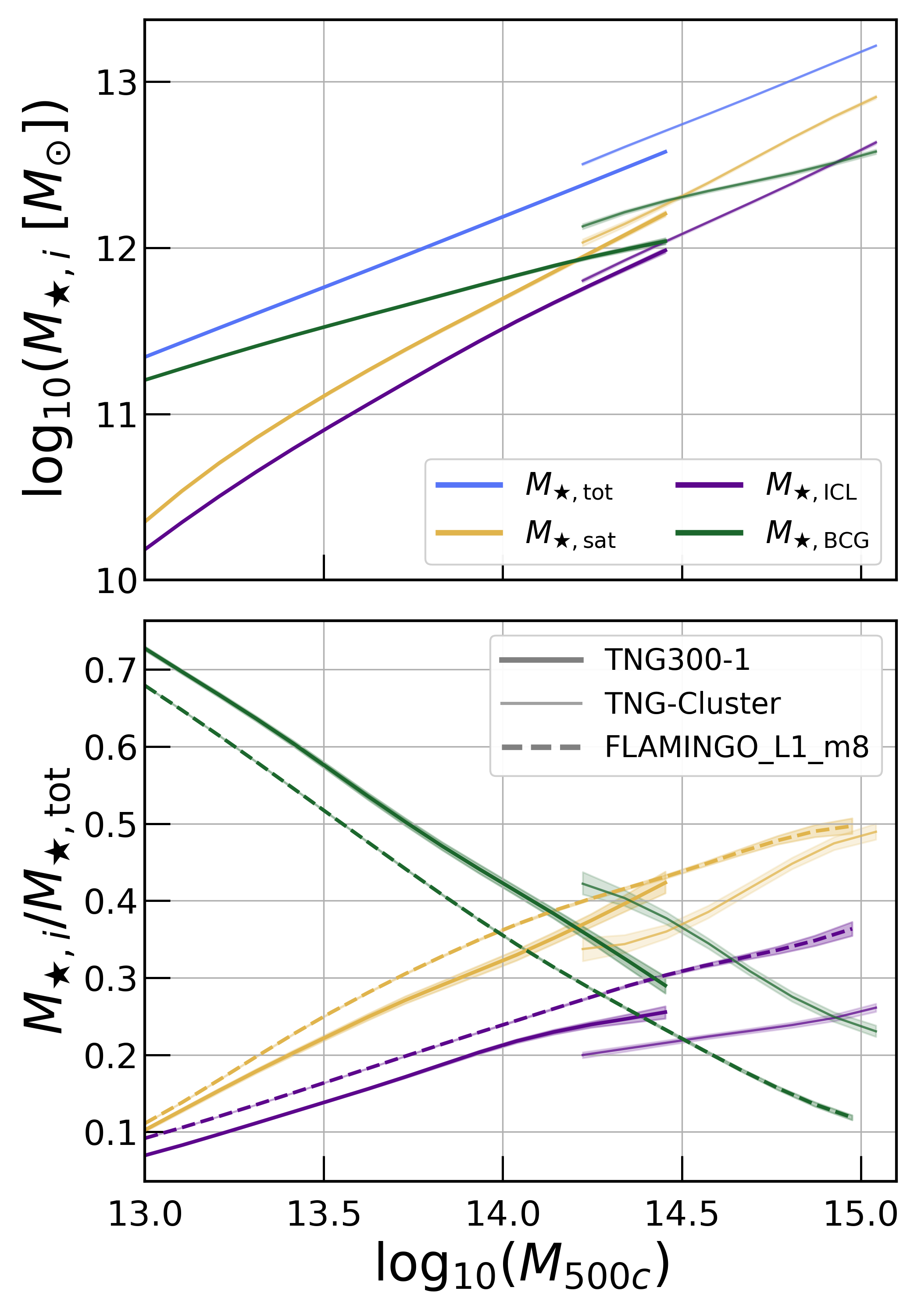}
    \vspace{-12 truept}
    \caption{\textbf{Top.} Present-day scaling relations of the total stellar mass, $\Mstartot$ (blue), and its three components: satellite stellar mass, $\Mstarsat$ (gold), ICL stellar mass, $\MstarICL$ (purple), and BCG stellar mass, $\MstarBCG$ (green), for TNG300-1 (thick lines) and TNG-Cluster (thin lines). We omit FLAMINGO (dashed lines) from the top panel for clarity, as its results are already presented in \S\ref{sec:mpr-star-bh-norm}.  \textbf{Bottom.} Fraction of the total stellar mass contained in each component, $M_{\star,i}/\Mstartot$, as a function of halo mass at $z=0$.
}
    \label{fig:Mstartot-components-scaling-fractions}
\end{figure}

We further examine the discrepancies in the stellar mass contents of TNG300-1 and TNG-Cluster halo samples here, and explore possible physical interpretations of these differences.

\subsection{$\Mstartot$ Scaling Relation}\label{sec:Mstartot-discrepancy}
In Section~\ref{sec:Mstartot-scaling}, we found that at fixed halo mass, the total stellar mass in TNG-Cluster is approximately $35\%$ larger than in TNG300-1 at $z=0$. To further investigate this discrepancy, the upper panel of Figure~\ref{fig:Mstartot-components-scaling-fractions} breaks the total stellar mass into the stellar mass components contributing to it. 

\begin{figure}
    \centering
    \includegraphics[width = \linewidth]{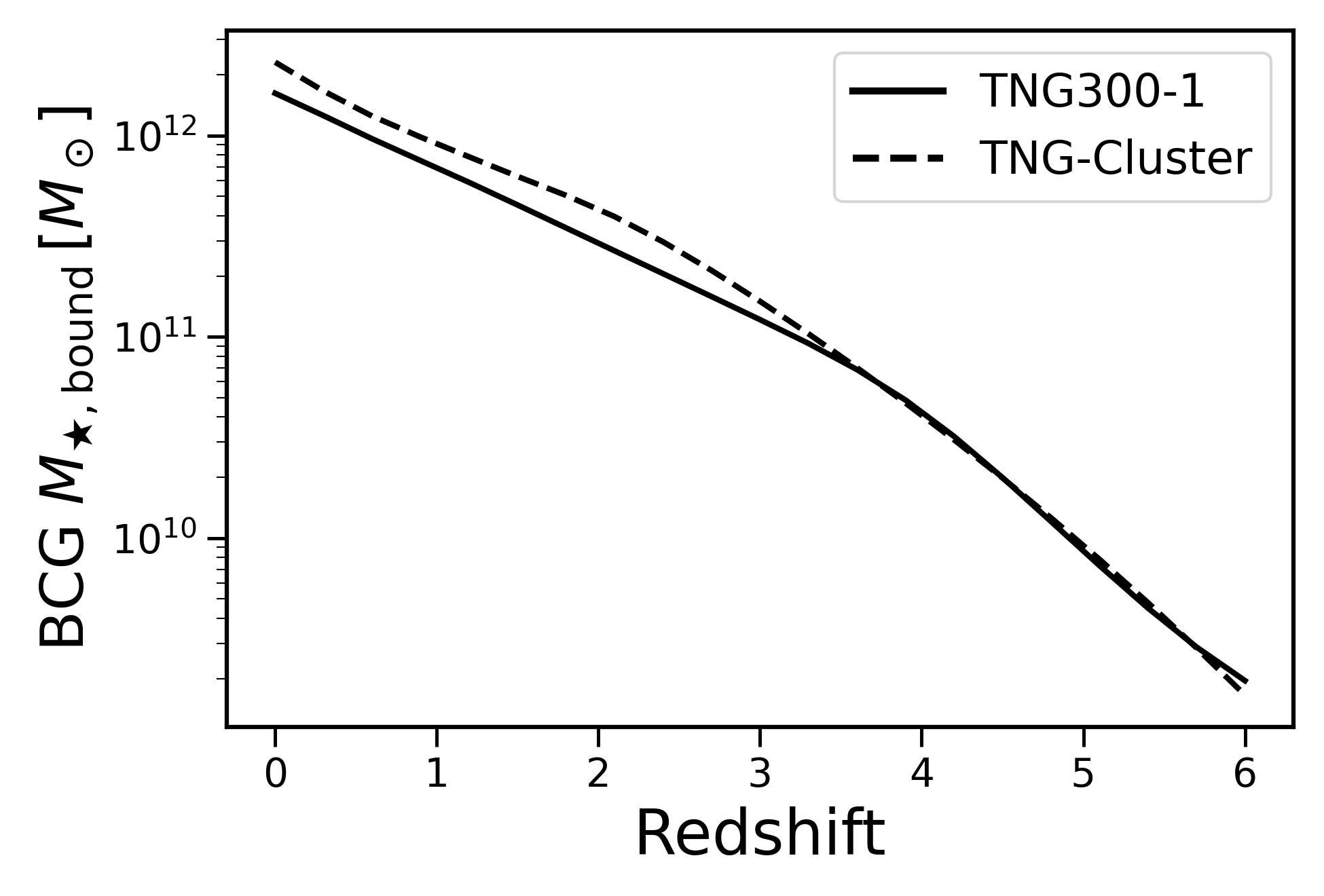}
    \caption{The geometric mean bound stellar mass of BCGs as a function of redshift for TNG300-1 (solid) and TNG-Cluster (dashed). The BCGs are selected from $50$ randomly chosen halos at $z=0$ within the overlapping halo mass range of the two simulations, $10^{14.15}\msun < \Mfivehc < 10^{14.40}\msun$, and are traced back to $z=6$ along their main progenitor branches.
}
    \label{fig:BCG-history}
\end{figure}

The largest difference is seen in the BCG stellar mass, $\MstarBCG$, which is higher in TNG-Cluster by $0.25$ dex ($76\%$) at $\Mfivehc=10^{14.5}\msun$.  Satellite and ICL values are also larger, but by only ${\sim}0.06$ dex (${\sim}14\%$). The mean fraction of the total stellar mass contributed by each component is shown in the lower panel, where FLAMINGO behavior is also included. Although both $\Mstarsat$ and $\MstarICL$ are slightly enhanced in TNG-Cluster, their fractional contributions are lower than in TNG300-1 because their increases are smaller than the increase in the total stellar mass. In contrast, the BCG stellar mass fraction, $\MstarBCG/\Mstartot$, is approximately $10\%$ larger in TNG-Cluster, demonstrating that the excess total stellar mass is driven primarily by the enhanced growth of the BCGs.  
In the most massive halos of TNG-Cluster, the BCG holds nearly $25\%$ of all stars, a factor of two more than the BCGs of FLAMINGO.  

The stellar component partitioning in FLAMINGO and TNG300-1 follows similar, but slightly offset, mass-dependent behaviors. The enhancement to the star formation rate in the TNG-Cluster simulation enlarges this offset, with the exception of the mass in satellite galaxies in halos near $10^{15}\msol$.

Since the excess stellar mass in TNG-Cluster is primarily concentrated in the BCGs—the densest stellar environments—rather than being distributed proportionally among all satellite galaxies and the ICL, it is reasonable to attribute this enhancement to the modified star formation prescription (Section~\ref{sec:starFormCal}) in TNG-Cluster. 

To further investigate the BCG stellar mass enhancement in TNG-Cluster relative to TNG300, we randomly select $50$ halos from each simulation within the overlapping halo mass range, $10^{14.15}\msun < \Mfivehc < 10^{14.40} \msun$. We identify their BCGs at $z=0$, trace their main progenitors back to $z=6$, and compute the mean logarithmic bound stellar mass of the BCGs as a function of redshift. The result, shown in Figure~\ref{fig:BCG-history}, demonstrates that the mean BCG stellar masses are initially similar and grow in tandem until $z\sim4$. At later times, however, the BCGs in TNG-Cluster grow more rapidly than those in TNG300-1, becoming more massive by $z=0$.

\begin{figure}
    \centering
    \includegraphics[width = \linewidth]{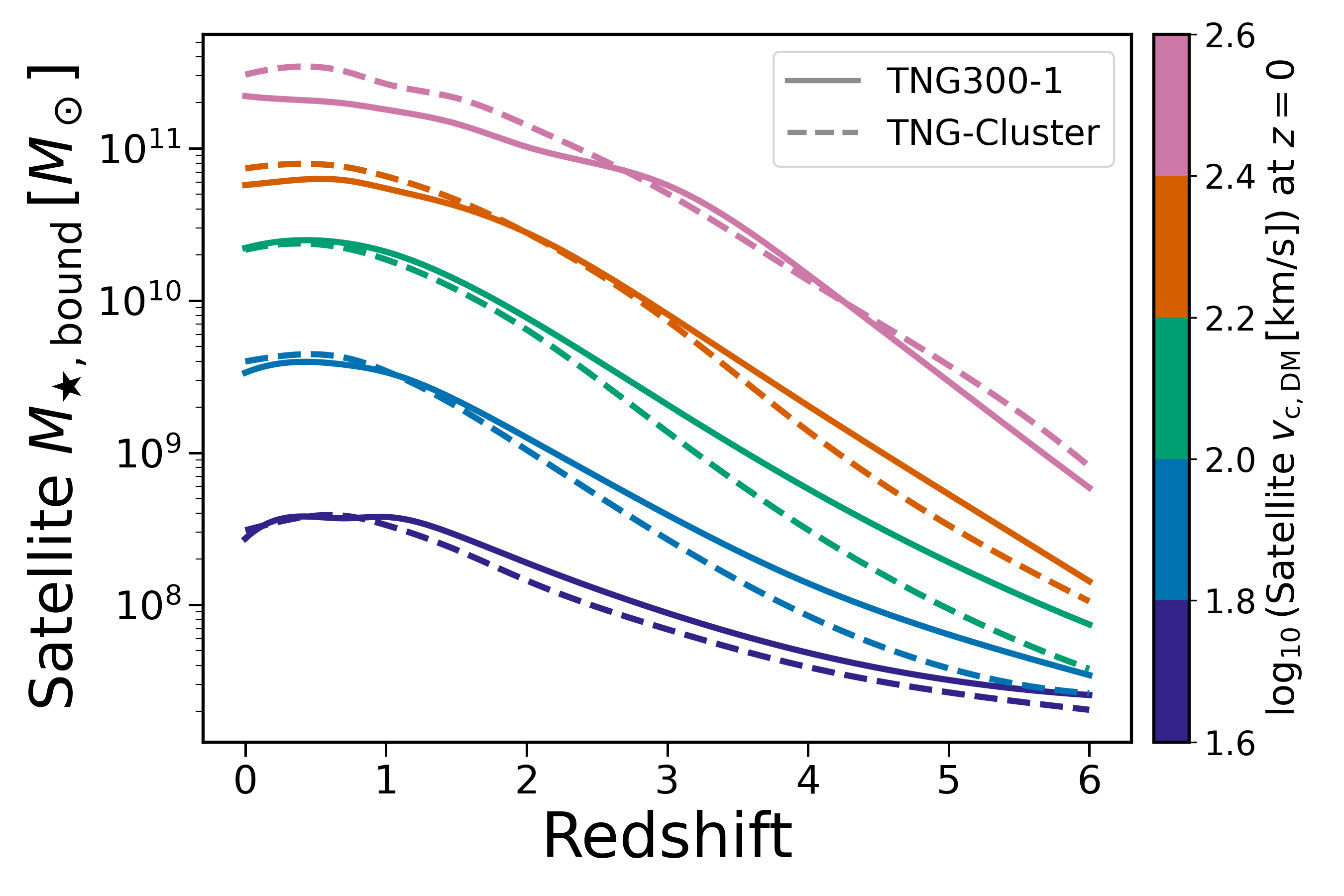}
    \caption{The geometric mean bound stellar mass of satellite galaxies in TNG300-1 (solid) and TNG-Cluster (dashed) as a function of redshift, binned by $\vcircdm$ (Equation~\eqref{eq:v-circ-def}). The sample consists of satellite galaxies within $50$ randomly selected halos at $z=0$ in the overlapping halo mass range $10^{14.15}\msun < \Mfivehc < 10^{14.40}\msun$. For each halo, we select up to $10$ satellites with the largest $\vcircdm$ and trace their bound stellar masses back to $z=6$ along their main progenitor branches.
}
    \label{fig:satellite-history}
\end{figure}

\subsection{Satellite Occupation}\label{sec:Nsat-discrepancy}
We showed in Section~\ref{sec:Nsatstats} that TNG300-1 contains more satellite galaxies than TNG-Cluster at fixed halo mass, despite TNG-Cluster having a larger total stellar mass. As discussed above, however, the stellar mass in satellites is only slightly larger in TNG-Cluster. Together with the S-GSMF (Section~\ref{sec:GSMF}), which shows fewer low-mass satellites and more high-mass satellites in TNG-Cluster, this naturally explains its lower satellite occupation.

To investigate the origin of the more massive satellite population in TNG-Cluster, we follow the evolution of satellite galaxies in the two simulations. Using the same $50$ halos within the overlapping halo mass range of TNG300-1 and TNG-Cluster used above, we identify sets of subhalos within $\Rfivehc$ defined by their dark matter circular velocities, 
\begin{equation}\label{eq:v-circ-def}
\vcircdm = \sqrt{G\frac{M_{\rm DM}(<2r_{\star,1/2})}{2r_{\star,1/2}}},
\end{equation}
where $G$ is the gravitational constant, $M_{\rm DM}(<2r_{\star,1/2})$ is the enclosed dark matter mass within twice the stellar half-mass radius, and $r_{\star,1/2}$ is the stellar half-mass radius. We then trace the mean stellar mass evolution of these subhalos in bins of $\log \vcircdm$. 

We find that low-mass satellite subhalos, with $\vcircdm < 10^{2.2}\,\mathrm{km\,s^{-1}}$, exhibit little difference in their bound stellar masses at late times. At earlier epochs, however, satellites in TNG-Cluster are systematically less massive but subsequently experience more rapid stellar mass growth. In contrast, the most massive satellite subhalos, with $\vcircdm > 10^{2.4},\mathrm{km,s^{-1}}$, contain on average approximately $60\%$ more bound stellar mass in TNG-Cluster than in TNG300-1 by $z=0$. 

\begin{figure}
    \centering
    \includegraphics[width = \linewidth]{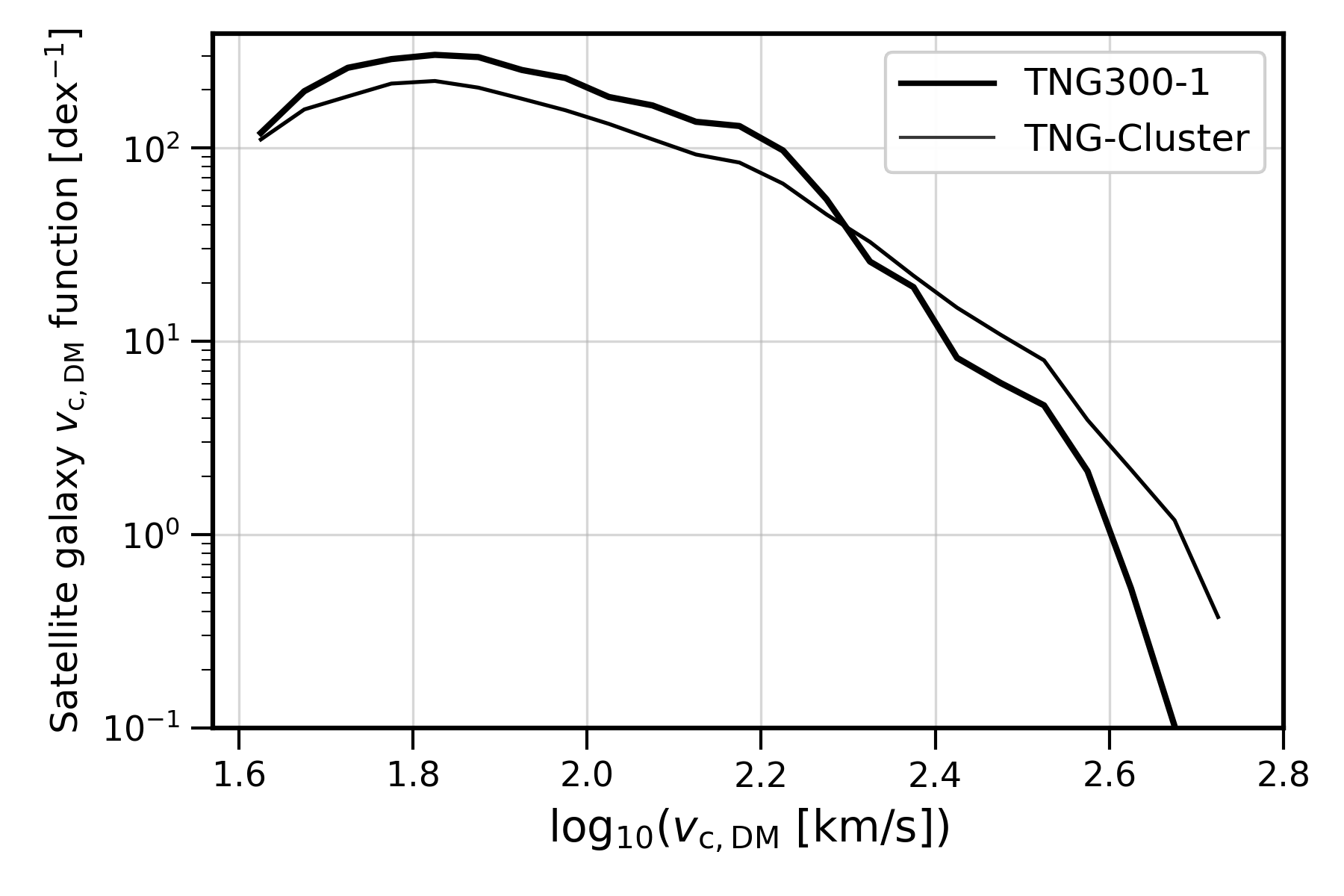}
    \caption{The satellite galaxy $\vcircdm$ function at $z = 0$ for satellite galaxies residing in halos with $\Mfivehc \geq 10^{14}\msun$, analogous to the S-GSMF in Figure~\ref{fig:S-GSMF} but using the satellite galaxies' DM circular velocity from Eq.~\eqref{eq:v-circ-def} instead of bound stellar mass. It uses the same weighting scheme as the S-GSMF and is evaluated at the pivot halo mass of $\Mfivehc = 10^{14.5}\msun$. This shows that TNG-Cluster has fewer galaxies with low $\vcircdm$ and more with high $\vcircdm$ compared to TNG300-1.}
    \label{fig:satellite-Vcirc-function}
\end{figure}

The agreement in bound stellar mass between TNG300-1 and TNG-Cluster galaxies with $\vcircdm < 10^{2.2}$ km/s, together with the underabundance of low-stellar-mass galaxies in TNG-Cluster seen in Figure~\ref{fig:S-GSMF}, can be further understood by examining the satellite galaxy $\vcircdm$ function in Figure~\ref{fig:satellite-Vcirc-function}. This function is constructed using the same methods utilized for the S-GSMF, such that the satellite populations are compared at fixed halo-mass distributions for halos with $\Mfivehc > 10^{14},\msun$. At a representative halo mass of $\Mfivehc \simeq 10^{14.5},\msun$, TNG-Cluster contains fewer satellites at both low stellar masses and low $\vcircdm$, while exhibiting an excess of satellites with both higher stellar masses and higher $\vcircdm$ relative to TNG300-1. Thus, the difference between the two simulations extends beyond a simple change in the stellar mass distribution: the satellite populations also differ in their dark-matter structure and dynamical properties. In particular, the joint shift toward higher bound stellar mass and $\vcircdm$ indicates that the TNG-Cluster satellites are, on average, associated with more massive or more centrally concentrated dark-matter subhalos.

These evolutionary trends explain the differences in the satellite galaxy stellar mass function between the two simulations. Enhanced late-time stellar growth in the most massive satellites shifts stellar mass toward the high-mass end of the satellite population, producing fewer low-mass satellites and more massive satellites in TNG-Cluster despite the similar total satellite stellar mass at fixed halo mass. Consequently, a larger fraction of the satellite stellar mass is contained within a relatively small number of massive satellites, naturally leading to the lower satellite occupation in TNG-Cluster.

\section{Gas and Stellar Fractions}\label{sec:fgas-fstar}
\begin{figure}
    \centering
    \includegraphics[width = 0.85\linewidth]{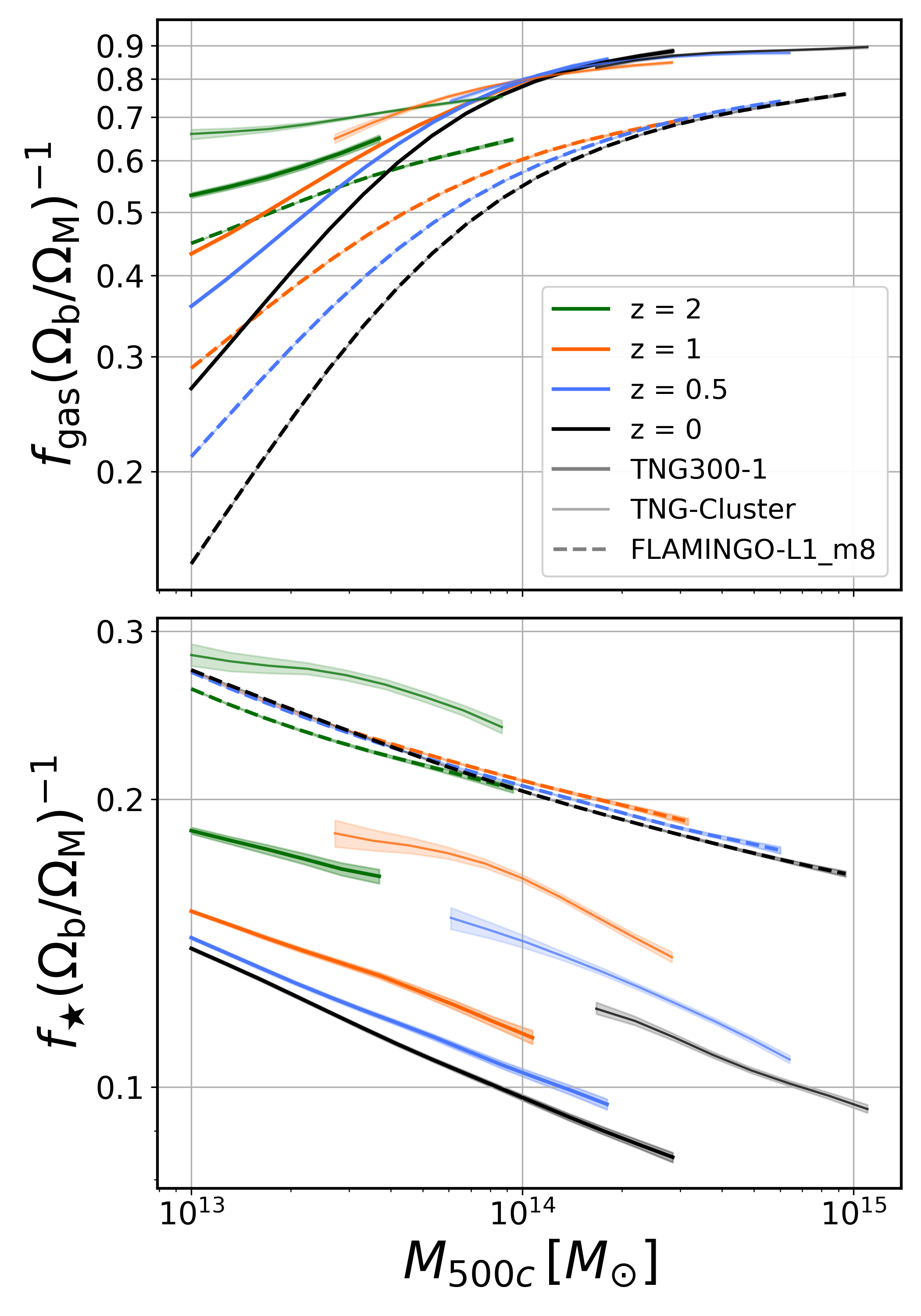}
    \caption{Mean gas fractions, $\fgas$ (top), and stellar fractions, $\fstar$ (bottom), within $\Rfivehc$, normalized by the cosmic baryon fraction, for halo populations in TNG300-1 (thick solid line), TNG-Cluster (thin solid line), and {\FLAMLm} (dashed line).
}
    \label{fig:fgasfstar}
\end{figure}

Expanding on the overall baryon fractions presented in Figure~\ref{fig:baryon-fraction}, Figure~\ref{fig:fgasfstar} shows the gas (top) and stellar mass (bottom) fractions separately.

\begin{figure*}
    \centering
    \includegraphics[width = 0.85\textwidth]{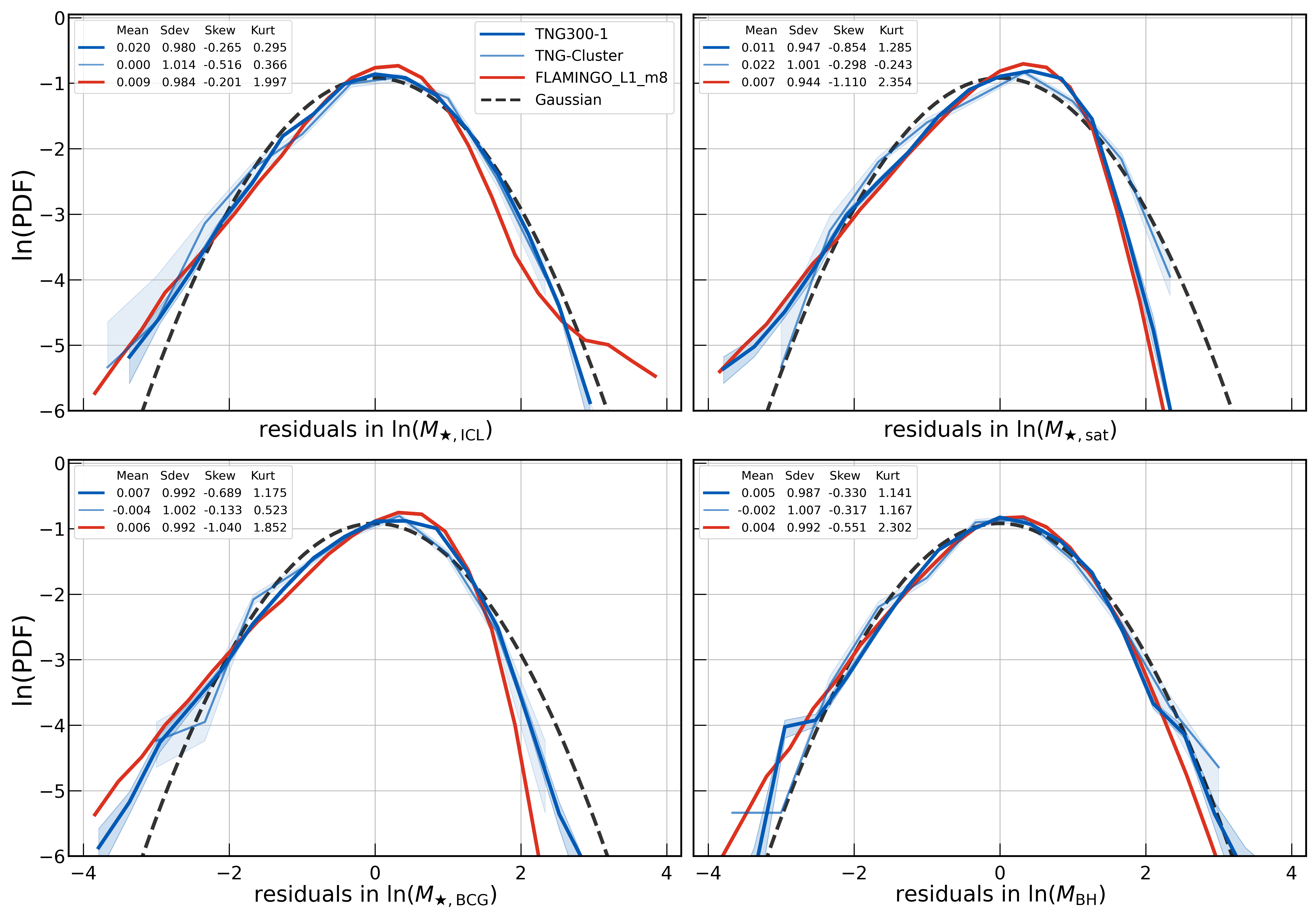}
    \caption{Mass-conditioned likelihood forms at $z=0$ of the ICL stellar mass $\MstarICL$ (top left), the satellite stellar mass $\Mstarsat$ (top right), the BCG stellar mass $\MstarBCG$ (bottom left), and the total BH mass $\Mbh$ (bottom right). We report the first four moments of each distribution in the panel insets, where a Gaussian distribution has an excess kurtosis of zero.   
}
    \label{fig:M-ICL-M-sat-BCG-star-M-bh-kernel-shapes}
\end{figure*}

The dominant baryonic component of gas exhibits trends similar to those of the total baryon fraction. More massive halos retain a larger fraction of their gas within $\Rfivehc$, while at fixed halo mass the gas fraction increases with redshift for halos with $\Mfivehc \lesssim 10^{14}\msun$ and becomes nearly redshift independent at higher masses. At the highest halo masses, both TNG300-1 and TNG-Cluster retain approximately $90\%$ of their baryons in the gaseous component, compared to only ${\sim}77\%$ in FLAMINGO. The gas fractions of the two TNG simulations are in close agreement for $z\leq1$, indicating that the systematically higher baryon fraction in TNG-Cluster is not driven by differences in the gas content.

At $z=2$, low-mass halos in TNG-Cluster contain on average approximately $10\%$ more gas than their counterparts.  As seen in the lower panel, those halos also contain more stars, and the combination yields the large difference in overall baryon fraction seen in the solid green lines of Figure~\ref{fig:baryon-fraction}.  

The systematically higher stellar mass in TNG-Cluster relative to TNG300-1 is evident in $\fstar$ across the full halo mass range at all redshifts. FLAMINGO shows redshift-independent evolution in stellar mass fraction. In contrast, TNG300-1 and TNG-Cluster exhibit a modest increase in $\fstar$ with redshift, becoming more pronounced at $z=2$, particularly in TNG-Cluster, whose stellar fraction exceeds even that of FLAMINGO. 
This enhanced stellar fraction allows the baryon fraction in TNG-Cluster to remain close to the cosmic value despite only ${\sim}70\%$ of the baryons residing in the gaseous component. Similarly, at $z=2$ the larger gas fraction in TNG300-1 relative to FLAMINGO is nearly offset by the higher stellar fraction in FLAMINGO, resulting in comparable total baryon fractions in the two simulations.

\section{Additional Likelihood Shapes}

Extending the analysis of the three representative residual distributions presented in Figure~\ref{fig:M-star-N-sat-vel-disp-kernel-shapes}, Figure~\ref{fig:M-ICL-M-sat-BCG-star-M-bh-kernel-shapes} presents the mass-conditioned likelihoods at $z=0$ for four additional properties: the ICL stellar mass $\MstarICL$, the total satellite stellar mass $\Mstarsat$, the BCG stellar mass $\MstarBCG$, and the total BH mass $\Mbh$.

While all are broadly consistent with Gaussian shapes, each exhibits departures from normality. The normalized residuals of $\MstarICL$ are mildly negatively skewed and displays extended negative tails in all three simulations, while a positive tail is apparent only in FLAMINGO, giving rise to a relatively high kurtosis. The residual distributions of both $\MstarBCG$ and $\Mstarsat$ exhibit the strongest departures from Gaussianity, with pronounced negative skewness, extended low-value tails, and comparatively weak support at high residual values. In contrast, the $\Mbh$ residual distribution is nearly Gaussian and closely resembles that of $\Mstartot$, exhibiting only a slight negative skewness, a modest low-value tail, and substantial support at high residual values.

\section{Central BH correlations}\label{sec:cent-BH-corrs}
In Section~\ref{sec:corr-divergent}, we identified the largest discrepancies between TNG300-1 and TNG-Cluster in the correlations involving the total BH mass, particularly $\Mstartot-\Mbh$ and $\MstarBCG-\Mbh$. At $z=1$, these correlations are strongly positive in TNG300-1 but negative in TNG-Cluster. We further showed that the discrepancies in the $\Mbh-\Mstarsat$ and $\Mbh-\Nsat$ correlations can be understood as a consequence of these differences. Here, we demonstrate that replacing the total BH mass with the central BH mass, $\Mbhcent$, defined as the mass of the most massive BH within $30\kpc$ of the BCG center, largely removes these discrepancies in both the sign and, in some cases, the magnitude of the correlations.

\begin{figure}
    \centering
    \includegraphics[width = 0.6\linewidth]{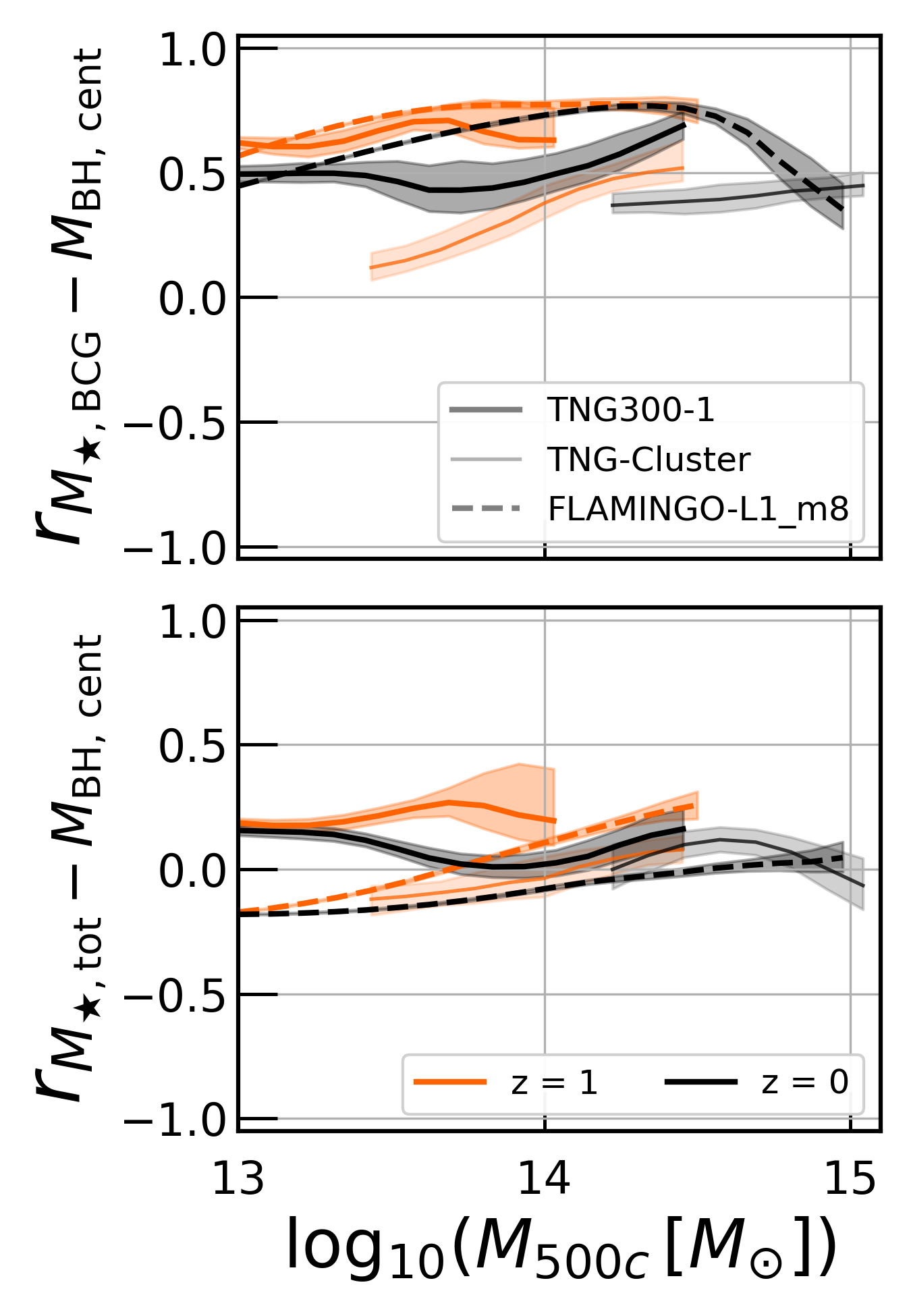}
    \caption{Correlations involving the central BH mass, $\Mbhcent$, defined as the mass of the most massive BH within $30\,\kpc$ of the BCG center. (\textbf{Top}) Correlation between $\Mbhcent$ and the BCG stellar mass, $\MstarBCG$. (\textbf{Bottom}) Correlation between $\Mbhcent$ and the total stellar mass, $\Mstartot$.
}
    \label{fig:Mbh-cent-corrs}
\end{figure}

As shown in Figure~\ref{fig:Mbh-cent-corrs}, all three simulations predict a positive BCG stellar mass--central BH mass correlation at all redshifts, with close agreement in both the sign and magnitude of the correlation. The correlation coefficient lies between $0.2$ and $0.7$, making it slightly stronger than the corresponding correlation with the total BH mass. The most striking difference is found in TNG-Cluster. At $z=0$, the total BH mass is essentially uncorrelated with the BCG stellar mass (Figure~\ref{fig:6-prop-corrs}(l)), whereas the central BH mass exhibits a moderately strong positive correlation. At $z=1$, the contrast is even more pronounced: the total BH mass is moderately anti-correlated with the BCG stellar mass, while the central BH mass remains positively correlated, in agreement with both TNG300-1 and FLAMINGO.

Whereas the $\Mstartot-\Mbh$ correlation shown in Figure~\ref{fig:6-prop-corrs}(g) spans a wide range of values, from $-0.2$ to $0.6$, for halos with $\Mfivehc > 10^{13.5}\msun$, all three simulations instead show little evidence for a correlation between the total stellar mass and the central BH mass over the same halo mass range and at all redshifts.

\section{Redshift dependence of MPQ}\label{sec:MPQhighz}

\begin{figure}
    \includegraphics[width = 0.9\linewidth]{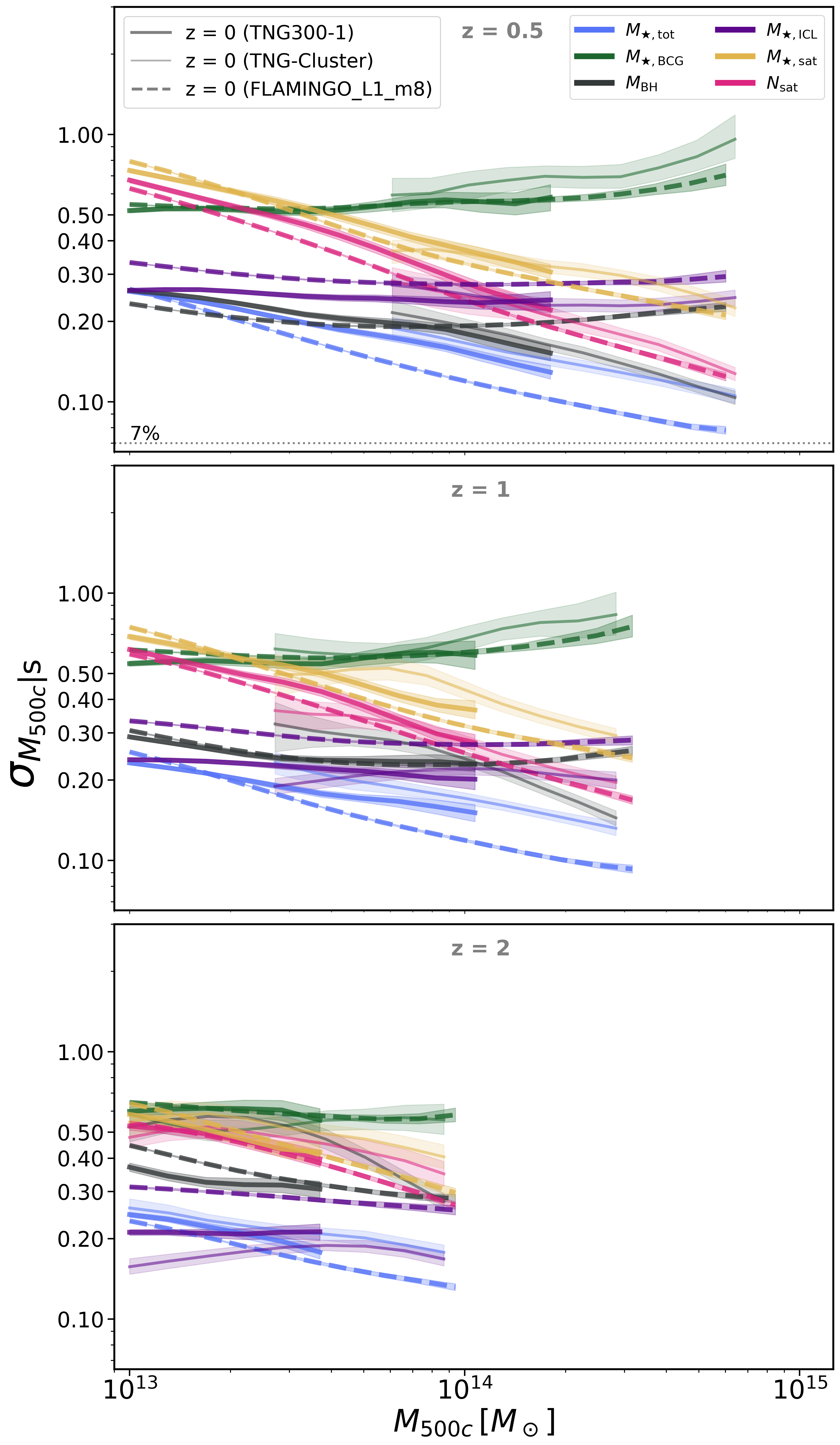}
    \caption{Halo mass dependence of the MPQ at $z=0.5$ (top), $1$ (middle), and $2$ (bottom), displayed using the same format as Figure~\ref{fig:MPQindividualzero}. At each redshift, the curves extend only to the halo mass of the $21^{\rm st}$ most massive halo in each simulation.}
    \label{fig:MPQ-redshifts}
\end{figure}

Figure~\ref{fig:MPQ-redshifts} shows the halo mass scatter at redshifts $z=0.5$, $1$, and $2$ (top to bottom). The color scheme and axis scales are identical to those in Figure~\ref{fig:MPQindividualzero}.

Overall, properties with strong mass dependence in their MPQ ($\Mstartot$, $\Mstarsat$, and $\Nsat$) exhibit relatively weak redshift evolution, whereas properties whose MPQ is nearly independent of halo mass ($\MstarBCG$, $\MstarICL$, $\Mbh$, and $\Mstargap$) show more noticeable redshift evolution.

For $\Mstartot$, the implied mass scatter in TNG300-1 and FLAMINGO decreases by approximately $5\%$ between successive redshift bins for halos with $\Mfivehc \lesssim 10^{13.5}\msun$, while remaining nearly unchanged at higher masses. Owing to its higher mass range, TNG-Cluster samples only halos above this scale at $z\le1$ and correspondingly shows little redshift evolution in the MPQ. Although TNG300-1 and FLAMINGO exhibit similar evolution in the implied mass scatter, the underlying causes differ. In TNG300, the improvement is driven primarily by a steepening of the $\Mstartot$ scaling relation, whereas in FLAMINGO it results mainly from a decrease in the intrinsic scatter. The MPQ of $\Mstarsat$ follows a similar trend as $\Mstartot$, with decreasing implied mass scatter for halos with $\Mfivehc < 10^{14}\msun$ and redshift independence at larger scales.

In contrast, the MPQ of $\MstarBCG$ exhibits stronger redshift evolution than the previously discussed properties. In both TNG300-1 and FLAMINGO, the implied mass scatter increases primarily between $z=0$ and $z=0.5$, with comparatively little evolution at higher redshifts. This evolution is strongest for halos with $\Mfivehc \lesssim 10^{14}\msun$, whereas TNG-Cluster shows only weak redshift dependence over its halo mass range. 

%%%%%%%%%%%%%%%%%%%%%%%%%%%%%%%%%%%%%%%%%%%%%%%%%%%%%%%%%%%%%%%%%%%%%%%%%%%%%%%%%%%%%%%%%%%%%%%%%%%

% Don't change these lines
\bsp	% typesetting comment
\label{lastpage}
\end{document}